\documentclass[zpreprint,zbstnp]{zeus_paper}
\usepackage[english]{babel}
\chardef\usc=95
\chardef\til=126
\catcode`\@=11 
\DeclareRobustCommand\xdotspace{\futurelet\@let@token\@xdotspace}
\def\@xdotspace{%
  \ifx\@let@token.\else
  \ifx\@let@token\bgroup.\else
  \ifx\@let@token\egroup.\else
  \ifx\@let@token\/.\else
  \ifx\@let@token\ .\else
  \ifx\@let@token~.\else
  \ifx\@let@token!.\else
  \ifx\@let@token,.\else
  \ifx\@let@token:.\else
  \ifx\@let@token;.\else
  \ifx\@let@token?.\else
  \ifx\@let@token/.\else
  \ifx\@let@token'.\else
  \ifx\@let@token).\else
  \ifx\@let@token-.\else
  \ifx\@let@token\@xobeysp.\else
  \ifx\@let@token\space.\else
  \ifx\@let@token\@sptoken.\else
   .\space
   \fi\fi\fi\fi\fi\fi\fi\fi\fi\fi\fi\fi\fi\fi\fi\fi\fi\fi}
\catcode`\@=12 

\newcommand{\stru}[2]{%
   \relax\ifmmode\hbox{\vrule height#1 depth#2 width0pt}%
   \else\vrule height#1 depth#2 width0pt\fi}

\newcommand{\Ronum}[1]{\uppercase\expandafter{\romannumeral#1}}
\newcommand{\ronum}[1]{\expandafter{\romannumeral#1}}
\DeclareRobustCommand{\LaTeXZ}{%
  \LaTeX\kern-.05em4\kern-.1em
  {\raisebox{-0.2ex}{$\scriptstyle\text{ZEUS}$}}\xspace}

\DeclareMathAlphabet{\mathbf}{OT1}{cmr}{bx}{sl}
\newcommand{\eVdist}{\kern-0.06667em}

\newcommand{\Gev}{{\text{Ge}\eVdist\text{V\/}}}

\newcommand{\slashfrac}[2]{%
  \raisebox{0.5ex}{\ensuremath #1}\kern-0.12em/\kern-0.08em
  \raisebox{-.8ex}{\ensuremath #2}}

\newcommand{\sqr}[3]{%
    {\vcenter{\hrule height.#3ex\hbox{\vrule width.#2ex height#1ex
     \kern#1ex\vrule width.#3ex}\hrule height.#2ex}}}

\catcode`\@=11 
\newcommand{\parenbar}{\mathpalette\p@renb@r}
\def\p@renb@r#1#2{\vbox{%
  \ifx#1\scriptscriptstyle \dimen@.7em\dimen@ii.2em\else
  \ifx#1\scriptstyle \dimen@.8em\dimen@ii.25em\else
  \dimen@1em\dimen@ii.4em\fi\fi \offinterlineskip
  \ialign{\hfill##\hfill\cr
    \vbox{\hrule width\dimen@ii}\cr
    \noalign{\vskip-.3ex}%
    \hbox to\dimen@{$\mathchar300\hfil\mathchar301$}\cr
    \noalign{\vskip-.3ex}%
    $#1#2$\cr}}}
\catcode`\@=12 

\newcommand{\IP}{{\rm I$\kern-0.01667em$P}\xspace}

\mathchardef\qsm=63
\mathchardef\pls=43
\mathchardef\mns=512
\mathchardef\plm=518
\mathchardef\eql=61
\mathchardef\smallleft=300
\mathchardef\smallright=301
\mathchardef\les=316
\mathchardef\gre=318
\mathchardef\leq=532
\mathchardef\grq=533

\catcode`\@=11 
\newcounter{pict@width}
\newcounter{pict@height}
\newlength{\pict@scale}
\newcommand{\psfigadd}[4]{%
\setcounter{pict@width}{1*\ratio{#2+\pict@scale/2}{\pict@scale}}
\setcounter{pict@height}{1*\ratio{#3+\pict@scale/2}{\pict@scale}}
\setlength{\unitlength}{\pict@scale}
\hbox to #2{\hspace{-\fill}\begin{picture}(\thepict@width,\thepict@height)
\put(0,0){\psfig{figure=#1,width=#2,height=#3,clip=}}
\SetScale{0.283466457}
\SetWidth{1.763889}
{#4}
\end{picture}}
}
\newcounter{pict@widthfst}
\newcounter{pict@widthscd}
\newcounter{pict@widthtot}
\newcommand{\psfigaddtwo}[7]{%
\setcounter{pict@widthfst}{1*\ratio{#2+\pict@scale/2}{\pict@scale}}
\setcounter{pict@widthscd}{1*\ratio{#2+#4+\pict@scale/2}{\pict@scale}}
\setcounter{pict@widthtot}{1*\ratio{#2+#4+#6+\pict@scale/2}{\pict@scale}}
\setcounter{pict@height}{1*\ratio{#3+\pict@scale/2}{\pict@scale}}
\setlength{\unitlength}{\pict@scale}
\hbox{\hspace{-\fill}\begin{picture}(\thepict@widthtot,\thepict@height)
\put(0,0){\psfig{figure=#1,width=#2,height=#3,clip=}}
\put(\thepict@widthscd,0){\psfig{figure=#5,width=#6,height=#3,clip=}}
\SetScale{0.283466457}
\SetWidth{1.763889}
{#7}
\end{picture}}
}
\newcommand{\psfigror}[4]{%
\setcounter{pict@width}{1*\ratio{#2+\pict@scale/2}{\pict@scale}}
\setcounter{pict@height}{1*\ratio{#3+\pict@scale/2}{\pict@scale}}
\setlength{\unitlength}{\pict@scale}
\hbox{\begin{picture}(\thepict@width,\thepict@height)
\put(0,\thepict@height){\psfig{figure=#1,width=#3,height=#2,clip=,angle=270}}
\SetScale{0.283466457}
\SetWidth{1.763889}
{#4}
\end{picture}}
}
\newcommand{\psfigrol}[4]{%
\setcounter{pict@width}{1*\ratio{#2+\pict@scale/2}{\pict@scale}}
\setcounter{pict@height}{1*\ratio{#3+\pict@scale/2}{\pict@scale}}
\setlength{\unitlength}{\pict@scale}
\hbox{\begin{picture}(\thepict@width,\thepict@height)
\put(0,0){\psfig{figure=#1,width=#3,height=#2,clip=,angle=90}}
\SetScale{0.283466457}
\SetWidth{1.763889}
{#4}
\end{picture}}
}
\catcode`\@=12 
\newlength\listtextwidth

\catcode`\@=11 
\newlength{\@tabfninsert}
\newlength{\@tabfnwidth}
\newcommand{\tabfootnote}[2]{%
  \setlength{\@tabfninsert}{0.8em}
  \setlength{\@tabfnwidth}{\textwidth}
  \addtolength{\@tabfnwidth}{-\@tabfninsert}
  \addtolength{\@tabfnwidth}{-0.4em}
  \noindent\makebox[\@tabfninsert][r]{\footnotesize$^{#1}$\hfil}\hfill%
  \parbox[t]{\@tabfnwidth}{\footnotesize #2\hfill}}
\catcode`\@=12 

\def\citeCAL{{\cite{%
nim:a309:77,*nim:a309:101,*nim:a321:356,*nim:a336:23%
}}\xspace}

\begin{document}
\newcommand{\pythia}{{\sc Pythia}}
\newcommand{\herwig}{{\sc Herwig}}
\newcommand{\jimmy}{{\sc Jimmy}}
\newcommand{\etcut}{\mbox{$E_T^{\rm CUT}$}}
\newcommand{\etgap}{\mbox{$E_T^{\rm GAP}$}}
\newcommand{\xg}{\mbox{$x_\gamma^{\rm OBS}$}}
\newcommand{\xp}{\mbox{$x_p^{\rm OBS}$}}
\newcommand{\GeV}{\mbox{\rm ~GeV}}
\newcommand{\pom} {I\hspace{-0.2em}P}
\newcommand{\xpom}{\mbox{$x_{_{\pom}}$}}
\newcommand{\reg} {I\hspace{-0.2em}R}

\prepnum {DESY 05-011}
\title{Study of deep inelastic inclusive and diffractive scattering with the ZEUS forward plug calorimeter}                                                       

                    
\author{ZEUS Collaboration}
\date{January 2005}

\abstract
{Deep inelastic scattering and its diffractive component, $ep \to e^{\prime}\gamma^* p \to e^{\prime}XN$, have been studied at HERA with the ZEUS detector using an integrated luminosity of 4.2 pb$^{-1}$. The measurement covers a wide range in the $\gamma^*p$ c.m. energy $W$ (37 - 245 GeV), photon virtuality $Q^2$ (2.2 - 80 GeV$^2$) and mass $M_X$. The diffractive cross section for $M_X > 2$ GeV rises strongly with $W$; the rise is steeper with increasing $Q^2$. The latter observation excludes the description of diffractive deep inelastic scattering in terms of the exchange of a single Pomeron. The ratio of diffractive to total cross section is constant as a function of $W$, in contradiction to the expectation of Regge phenomenology combined with a naive extension of the optical theorem to $\gamma^*p$ scattering. Above $M_X$ of 8 GeV, the ratio is flat with $Q^2$, indicating a leading-twist behaviour of the diffractive cross section. The data are also presented in terms of the diffractive structure function, $F^{D(3)}_2(\beta,\xpom,Q^2)$, of the proton. For fixed $\beta$, the $Q^2$ dependence of $\xpom F^{D(3)}_2$ changes with $\xpom$ in violation of Regge factorisation. For fixed $\xpom$, $\xpom F^{D(3)}_2$ rises as $\beta \to 0$, the rise accelerating with increasing $Q^2$. These positive scaling violations suggest substantial contributions of perturbative effects in the diffractive DIS cross section.}

\makezeustitle

\def\3{\ss}                                                                                        
\pagenumbering{Roman}                                                                              
                                                   %
\begin{center}                                                                                     
{                      \Large  The ZEUS Collaboration              }                               
\end{center}                                                                                       
  S.~Chekanov,                                                                                     
  M.~Derrick,                                                                                      
  S.~Magill,                                                                                       
  S.~Miglioranzi$^{   1}$,                                                                         
  B.~Musgrave,                                                                                     
  \mbox{J.~Repond},                                                                                
  R.~Yoshida\\                                                                                     
 {\it Argonne National Laboratory, Argonne, Illinois 60439-4815}, USA~$^{n}$                       
\par \filbreak                                                                                     
  M.C.K.~Mattingly \\                                                                              
 {\it Andrews University, Berrien Springs, Michigan 49104-0380}, USA                               
\par \filbreak                                                                                     
  N.~Pavel, A.G.~Yag\"ues Molina \\                                                                
  {\it Institut f\"ur Physik der Humboldt-Universit\"at zu Berlin,                                 
           Berlin, Germany}
% ...
\newpage                                                                                           
\begin{tabular}[h]{rp{15cm}} 
$^{\    1}$ & also affiliated with University College London, UK \\                                  
$^{\    2}$ & retired \\                                                                             
$^{\    3}$ & now at the University of Victoria, British Columbia, Canada \\                         
$^{\    4}$ & PPARC Advanced fellow \\                                                               
$^{\    5}$ & partly supported by Polish Ministry of Scientific Research and Information             
Technology, grant no.2P03B 12625\\                                                                 
$^{\    6}$ & supported by the Polish State Committee for Scientific Research, grant no.             
2 P03B 09322\\                                                                                     
$^{\    7}$ & on leave of absence from Cornell University, Ithaca, NY, USA \\                        
$^{\    8}$ & now at Max-Planck-Institut f\"{u}r Physik, M\"{u}nchen, Germany \\                     
$^{\    9}$ & now at DESY group FEB, Hamburg, Germany \\                                             
$^{  10}$ & now at Dept. Estructura; Constituentes de la Materia, Faculta
de Fisica, Universitat de Barcelona, Spain\\
$^{  11}$ & now at DESY group FLC, Hamburg, Germany \\                                               
$^{  12}$ & now at LAL, Universit\'e de Paris-Sud, IN2P3-CNRS, Orsay, France \\                     
$^{  13}$ & now at Insitut de Fisica d'Altes Energias (IFAE), Universitat
de Barcelona, Spain\\
$^{  14}$ & partly supported by Moscow State University, Russia \\                                   
$^{  15}$ & also affiliated with DESY \\                                                             
$^{  16}$ & now at Baylor University, USA \\                                                         
$^{  17}$ & now at University of Oxford, UK \\                                                       
$^{  18}$ & now at the Department of Physics and Astronomy, University of Glasgow, UK \\             
$^{  19}$ & now at Royal Holloway University of London, UK \\                                        
$^{  20}$ & also at Nara Women's University, Nara, Japan \\                                          
$^{  21}$ & also at University of Tokyo, Japan \\                                                    
$^{  22}$ & Ram{\'o}n y Cajal Fellow \\                                                              
$^{  23}$ & PPARC Postdoctoral Research Fellow \\                                                    
$^{  24}$ & on leave of absence at The National Science Foundation, Arlington, VA, USA \\            
$^{  25}$ & also at Max Planck Institute, Munich, Germany, Alexander von Humboldt                    
Research Award\\                                                                                   
$^{  26}$ & present address: Tokyo Metropolitan University of Health                                 
Sciences, Tokyo 116-8551, Japan\\                                                                  
$^{  27}$ & also at University of Hamburg, Germany, Alexander von Humboldt Fellow \\                 
$^{  28}$ & partially funded by DESY \\                                                              
$^{  29}$ & also at \L\'{o}d\'{z} University, Poland \\                                              
$^{  30}$ & \L\'{o}d\'{z} University, Poland \\                                                      
$^{  31}$ & \L\'{o}d\'{z} University, Poland, supported by the KBN grant 2P03B12925 \\               
$^{  32}$ & supported by the KBN grant 2P03B12725 \\                                                 
\end{tabular}          
                                                 %
\newpage   
                                                           %
                                                           %
\begin{tabular}[h]{rp{14cm}}                                                                       
$^{a}$ &  supported by the Natural Sciences and Engineering Research Council of Canada (NSERC) \\  
$^{b}$ &  supported by the German Federal Ministry for Education and Research (BMBF), under        
          contract numbers HZ1GUA 2, HZ1GUB 0, HZ1PDA 5, HZ1VFA 5\\                                
$^{c}$ &  supported in part by the MINERVA Gesellschaft f\"ur Forschung GmbH, the Israel Science   
          Foundation (grant no. 293/02-11.2), the U.S.-Israel Binational Science Foundation and    
          the Benozyio Center for High Energy Physics\\                                            
$^{d}$ &  supported by the German-Israeli Foundation and the Israel Science Foundation\\           
$^{e}$ &  supported by the Italian National Institute for Nuclear Physics (INFN) \\                
$^{f}$ &  supported by the Japanese Ministry of Education, Culture, Sports, Science and Technology 
          (MEXT) and its grants for Scientific Research\\                                          
$^{g}$ &  supported by the Korean Ministry of Education and Korea Science and Engineering          
          Foundation\\                                                                             
$^{h}$ &  supported by the Netherlands Foundation for Research on Matter (FOM)\\                   
$^{i}$ &  supported by the Polish State Committee for Scientific Research, grant no.               
          620/E-77/SPB/DESY/P-03/DZ 117/2003-2005 and grant no. 1P03B07427/2004-2006\\             
$^{j}$ &  partially supported by the German Federal Ministry for Education and Research (BMBF)\\   
$^{k}$ &  supported by RF Presidential grant N 1685.2003.2 for the leading scientific schools and  
          by the Russian Ministry of Education and Science through its grant for Scientific        
          Research on High Energy Physics\\                                                        
$^{l}$ &  supported by the Spanish Ministry of Education and Science through funds provided by     
          CICYT\\                                                                                  
$^{m}$ &  supported by the Particle Physics and Astronomy Research Council, UK\\                   
$^{n}$ &  supported by the US Department of Energy\\                                               
$^{o}$ &  supported by the US National Science Foundation\\                                        
$^{p}$ &  supported by the Polish Ministry of Scientific Research and Information Technology,      
          grant no. 112/E-356/SPUB/DESY/P-03/DZ 116/2003-2005 and 1 P03B 065 27\\                  
$^{q}$ &  supported by FNRS and its associated funds (IISN and FRIA) and by an Inter-University    
          Attraction Poles Programme subsidised by the Belgian Federal Science Policy Office\\     
\end{tabular}                                                                                      
                                                           %
                                                           %

\pagenumbering{arabic} 
\pagestyle{plain}
\section{Introduction} 
Inclusive deep inelastic lepton-nucleon scattering (DIS) 
has been measured over a wide kinematic range. This has allowed a precise description of the nucleon structure functions obtained through QCD analyses using the DGLAP evolution equations~\cite{sovjnp:15:438, *sovjnp:20:94, *jetp:46:641,*np:b126:298,pr:d67:012007}. It has been established at HERA, that diffraction, where the proton or a low-mass nucleonic system emerge from the interaction with almost the full energy of the incident proton, contributes substantially to the DIS cross section~\cite{pl:b315:481}. Extensive measurements of diffractive DIS have been made by both the ZEUS and H1 collaborations~\cite{zfp:c76:613,epj:c6:43,epj:c25:169,desy-04-131}. 

The diffractive component of DIS is analysed in terms of conditional parton distribution functions (PDFs)~\cite{zfp:c76:613,desy-04-131}. According to the QCD factorisation theorem~\cite{trentadue,qcdf,*qcdf1}, these diffractive PDFs will also undergo QCD evolution as a function of the photon virtuality $Q^2$ in the same way as the inclusive proton PDFs. The dipole model{~\cite{proc:ep:1971,*pr:d3:1382,*pr:d8:1341,zfp:c53:331,*jetp:81:625,*pl:b422:238,pr:d59:014017,*pr:d60:114023,np:b335:115} provides an appealing picture that can be applied in DIS to both inclusive and diffractive scattering. In this model, the virtual photon dissociates into $q\overline{q}$ and $q\overline{q}g$ dipoles which then interact with the proton target, predominantly through gluon exchange. The size of the dipole is given by $Q^2$ and, in the kinematic range of HERA, varies from a typical hadron size down to much smaller values.

For hadron-hadron collisions, a large body of data on total, elastic and diffractive cross sections~\cite{prep:74:1,*prep:101:169} has been 
parameterised in Regge phenomenology by the exchange of the Pomeron trajectory. An early seminal suggestion to combine Regge phenomenology with 
QCD~\cite{pl:b152:256} in a $t$-channel picture introduced the idea of a 
Pomeron structure function. Assuming that diffraction can be described by the exchange of the Pomeron, its partonic structure can be determined in diffractive DIS. Such an approach depends on the validity of Regge factorisation, which implies a Pomeron flux that is independent of $Q^2$.

For scattering of on-shell particles, such as $\gamma p \to \gamma p$, the optical theorem relates the imaginary part of the forward elastic amplitude to the total $\gamma p$ cross section. Similarly, diffractive scattering of $\it virtual$ photons leading to a low-mass hadronic system should also be closely related to the total virtual photon-proton cross section.

This paper reports high statistics results from the ZEUS experiment on $e^-p$ deep inelastic scattering (Fig.~\ref{f:nondifdiag}),
\begin{eqnarray}
e p \to e + {\rm anything},   \nonumber
\end{eqnarray}
with the focus on diffractive production by virtual photon-proton scattering (Fig.~\ref{f:diffdiag}),
\begin{eqnarray}
\gamma^* p \to XN,                   \nonumber
\end{eqnarray}
(where $N$ is a proton or a low-mass nucleonic state) and a comparison with the total $\gamma^* p$ cross section.

In comparison to previous ZEUS measurements of diffraction~\cite{epj:c6:43}, the detector configuration was improved in the following way. The installation of a forward plug calorimeter (FPC) in the beam hole of the forward uranium calorimeter extended the forward rapidity coverage. As a result, the measurable range in the mass of the system $X$ was increased by a factor of 1.7. At the same time, the contribution of nucleon dissociation was limited to masses $M_N \le 2.3$ GeV. The rear beam hole in the detector was decreased in size by moving the calorimeter modules above and below closer to the beams. This increased the acceptance for low $Q^2$ and large $W$ events. These measures substantially improved the precision and kinematic coverage in comparison to previous HERA measurements~\cite{zfp:c76:613,epj:c6:43,epj:c25:169}. 

This paper is organized as follows. The experimental setup is described in Section~\ref{sec-exp}. Reconstruction of event kinematics and event selection are described in Section~\ref{sec-reconkinem}. Models for inclusive and diffractive DIS are presented in Section~\ref{sec-models}. Extraction of the diffractive contribution is discussed in Section~\ref{sec-lnmxmethod}. Evaluation of the total and diffractive cross sections is described in Section~\ref{sec-evalsig}. Section~\ref{sec-f2sigtot} presents the results on the proton structure function and the total $\gamma^*p$ cross section. The diffractive cross section is presented in terms of $M_X$,  $W$ and $Q^2$, and compared to the total cross section in Section~\ref{sec-sigdiff}. The diffractive structure function of the proton is discussed in Section~\ref{sec-difff2d}.

\section{Experimental set-up}
\label{sec-exp}
The data used for this measurement were taken at the HERA $ep$ collider using the ZEUS detector in 1998-1999 when electrons of 27.5 GeV collided with protons of 920 GeV. The data correspond to an integrated luminosity of 4.2 pb$^{-1}$.
 
A detailed description of the ZEUS detector can be found elsewhere~\cite{bluebook,pl:b293:465}. A brief outline of the components that are most relevant for this analysis is given below.

Deep inelastic scattering events were identified using information from the uranium-scintillator calorimeter (CAL), the forward plug calorimeter (FPC), the central tracking detector (CTD), the small angle rear tracking detector (SRTD) and the rear part of the hadron-electron separator (RHES).

Charged particles are tracked in the CTD~\cite{nim:a279:290,*npps:b32:181,*nim:a338:254}.
The CTD consists of 72 cylindrical drift chamber layers, organized in nine superlayers covering the polar-angle~\footnote{The ZEUS coordinate system is a right-handed Cartesian system, with the $Z$ axis pointing in the proton direction, referred to as the ``forward direction'', and the $X$ axis pointing left towards the centre of HERA. The coordinate origin is at the nominal interaction point.} region $15^0 < \theta < 164^0$. The CTD operates in a magnetic field of 1.43 T provided by a thin solenoid. The transverse-momentum resolution for full-length tracks is $\sigma(p_T)/p_T = 0.0058p_T \oplus 0.0065 \oplus 0.0014/p_T$, with $p_T$ in GeV.

The CAL~\citeCAL consists
of three parts: the forward (FCAL), the barrel (BCAL) and the rear (RCAL)
calorimeters. Each part is subdivided transversely into towers and
longitudinally into one electromagnetic section (EMC) and either one (in RCAL)
or two (in BCAL and FCAL) hadronic sections (HAC). The smallest subdivision of
the calorimeter is called a cell.  The CAL energy resolutions, as measured under
test-beam conditions, are $\sigma(E)/E=0.18/\sqrt{E}$ for electrons and
$\sigma(E)/E=0.35/\sqrt{E}$ for hadrons ($E$ in $\Gev$).
The CAL covers 99.7\% of the total solid angle.
The beam hole in the RCAL was $20 \times 8$ cm$^2$~\cite{epj:c21:443}.

The position of electrons scattered at small angles to the electron-beam direction was determined including the information from the SRTD~\cite{nim:a401:63,epj:c21:443}. The SRTD is attached to the front face of the RCAL and consists of two planes of scintillator strips, 1 cm wide and 0.5 cm thick, arranged in orthogonal orientations. Ambiguities in SRTD hits were resolved with the help of the RHES~\cite{nim:a277:176}, which consists of a layer of approximately 10,000 ($2.96 \times 3.32$ cm$^2$) silicon-pad detectors inserted in the RCAL at a depth of 3.3 radiation lengths. Electrons scattered at higher $Q^2$ were also measured in the CTD.

The FPC~\cite{nim:a450:235} was used to measure the energy of particles in the pseudorapidity range $\eta \approx 4.0 - 5.0$. It was a lead-scintillator sandwich calorimeter read out by wavelength-shifter (WLS) fibers and photomultipliers (PMT). It was installed in the $20 \times 20$ cm$^2$ beam hole of FCAL. The FPC had outer dimensions of 192 $\times$ 192 $\times$ 1080 mm$^3$ and a hole of 3.15 cm radius for the passage of the beams. The minimum angle for particle detection was 12 mrad which corresponds to a pseudorapidity of 5.1. In the FPC, 15 mm-thick lead plates alternated with 2.6-mm thick scintillator layers. The WLS fibers of 1.2 mm diameter passed through 1.4 mm holes located on a 12 mm grid in the lead and scintillator layers. The FPC was subdivided longitudinally into an electromagnetic (10 layers) and a hadronic section (50 layers) representing a total of 5.4 nuclear absorption lengths. The scintillator layers consisted of tiles forming towers which are read out individually. The cell cross sections were $24 \times 24$ mm$^2$ in the electromagnetic and 48$ \times 48$ mm$^2$ in the hadronic section. The FPC was tested and calibrated at CERN with electron, muon and hadron beams. The measured energy resolution for electrons was $\sigma_E/E = (0.41 \pm 0.02)/\sqrt{E} \oplus 0.062 \pm 0.002$, ($E$ in GeV). When installed in the FCAL, the energy resolution for pions was $\sigma_E/E = (0.65 \pm 0.02)/\sqrt{E} \oplus 0.06 \pm 0.01$ ($E$ in GeV) and the $e/h$ ratio was close to unity. The relative calibration of the FPC cells was regularly adjusted using measurements from a $^{60}$Co source, resulting in an average energy scale uncertainty of 4\% (3\%) for the EMC (HAC) cells~\cite{goebel:2001} as determined with DIS events at high $Q^2$~\footnote{Throughout the running period, DIS neutral current events, $ep \to eX$, with $Q^2 > 80$ GeV$^2$, were selected without using information from the FPC. The average energy deposited by the hadronic system in the individual FPC cells was used to monitor the energy calibration of each cell during the data-taking period.}. 

\section{Reconstruction of kinematics and event selection}
\label{sec-reconkinem}
This section describes event reconstruction and selection common to DIS inclusive and diffractive data samples.

The reaction $e^- (k) \; p(P) \to e^- (k^{\prime}) + \rm anything$ at fixed squared centre-of-mass (c.m.) energy, $s = (k+P)^2$, is described in terms of $Q^2 \equiv -q^2 = -(k -k^{\prime})^2$, Bjorken $x = Q^2/(2 P \cdot q)$ and $s \approx 4 E_e E_p$, where  $E_e$ and $E_p$ denote the electron and proton beam energies, respectively. For this data set, $\sqrt{s} = 318$ GeV. The fractional energy transferred to the proton in its rest system is $y \approx Q^2/(sx)$. The c.m. energy of the total hadronic system, $W$, is given by $W^2 = [p+(k-k^{\prime})]^2 = m^2_p+Q^2(1/x-1)\approx Q^2/x = ys$, where $m_p$ is the mass of the proton.

Diffraction, $e^- (k) \; p(P) \to e^- (k^{\prime}) + N(N) + X$, is described in terms of the mass $M_X$ of the system $X$, and the mass $M_N$ of the system $N$. Since $t$, the four-momentum transfer squared, between the incoming proton and the outgoing system $N$, $t = (p-N)^2$, was not measured, the results presented are integrated over $t$.  

The diffractive structure function was analyzed in terms of the momentum fraction of the proton carried by the Pomeron,
$\xpom = [(P  -  N) \cdot q]/(P \cdot q ) \approx (M^2_X+Q^2)/(W^2 + Q^2)$, and
the fraction of the Pomeron momentum carried by the struck quark, $\beta = Q^2/[2(P  -  N) \cdot q ]\approx Q^2 / (M^2_X+Q^2)$.
The variables $\xpom$ and $\beta$ are related to the Bjorken scaling variable, $x$, via $x = \beta \xpom$.

The events studied are of the type
\begin{eqnarray}
ep \to e^{\prime} {\cal X} \; + \; {\rm rest},
\label{eq:xobs}
\end{eqnarray}
where ${\cal X}$ denotes the hadronic system observed in the detector and `rest' the particle system escaping detection through the forward and/or rear beam holes.
 
Scattered electrons were identified with an algorithm based on a neural network~\cite{nim:a365:508}. The direction and energy of the scattered electron were determined from the combined information given by CAL, SRTD, RHES and CTD. The impact point of the electron on the face of the RCAL had to lie outside an area of $26.6{\;\rm cm} \times 17{\;\rm cm}$ (box cut) centred on the beam axis. Further fiducial cuts on the impact point were imposed to ensure reliable measurement of the electron energy. 

The value of $Q^2$ was reconstructed from the measured energy $E^{\prime}_e$ and scattering angle $\theta_e$, of the electron, $Q^2 = 2 E_e E^{\prime}_e (1 + \cos \theta_e)$. The hadronic system was reconstructed from energy-flow objects (EFO)~\cite{goebel:2001,gennady} which combine the information from CAL and FPC clusters and from CTD tracks, and which were not assigned to the scattered electron (hadronic EFOs). 

The value of $W$ was determined using the weighted average of the values given by the electron and the hadron measurements (see Appendix~\ref{asec-reconw}).

The mass of the system ${\cal X}$ was determined by summing over all hadronic EFOs,
\begin{eqnarray}
M^2_{\cal X} = (\sum P_h)^2,                           \nonumber
\label{eq:mxcalc}
\end{eqnarray}
where $P_h$ is the four-momentum vector of each EFO $h$. All kinematic variables used to describe inclusive and diffractive scattering were derived from $M_{\cal X}$, $W$ and $Q^2$.

The coordinates $X_{\rm vtx}, Y_{\rm vtx}, Z_{\rm vtx}$ of the event vertex were determined with tracks reconstructed in the CTD. The average $X_{\rm vtx}, Y_{\rm vtx}$ values varied by $\pm 0.1$ cm and $\pm 0.03$ cm, respectively, over the data-taking period. Since the variations were small, and the transverse size of the beams were smaller than the resolution, the average $X_{\rm vtx}, Y_{\rm vtx}$ values were used. The distribution of $Z_{\rm vtx}$ was approximately Gaussian with an r.m.s. of $\pm 11$~{\rm cm}. The value of $Z_{\rm vtx}$ was taken from the reconstructed event vertex. For events without a measured primary vertex, the average $Z$ vertex for each data run was used.  

If a scattered-electron candidate was found, the following criteria were imposed to select the DIS events:
\begin{itemize}
\item the scattered-electron energy $E_e^{\prime}$ be at least 10 GeV;
\item the total measured energy of the hadronic system be at least 400 MeV;
\item $y^{\rm FB}_{\rm JB} > 0.004$, where $y^{\rm FB}_{\rm JB} = \sum_h (E_h - P_{Zh})/(2E_e)$, summed over all hadronic EFOs in FCAL plus BCAL; or at least 400 MeV be deposited in the BCAL or in the RCAL outside of the ring of towers closest to the beamline;
\item $-54 < Z_{\rm vtx} < 50$ cm;
\item $46 < \sum_{i=e,h} (E_i - P_{Zi}) < 64$ GeV, where the sum runs over both the scattered electron and all hadronic EFOs. This cut reduces the background  from photoproduction and beam-gas scattering and removes events with large initial-state QED radiation; 
 
\item candidates for QED-Compton events, consisting of a scattered electron candidate and a photon candidate with mass $M_{e \gamma}$ less than 0.25 GeV and total transverse momentum less than $\approx$ 1.5 GeV, were removed.
\end{itemize}

The contamination from electron (proton) beam-gas scattering was measured using non-colliding electron (proton) bunches and found to be negligible.

About 800,000 events passed the selection cuts. The kinematic range for inclusive and diffractive events was chosen taking into account detector resolution and statistics. About 612,000 events were retained which satisfied $37 < W < 245$ GeV and $2.2 < Q^2 < 80$ GeV$^2$. 

The resolutions of the reconstructed kinematic variables were estimated using Monte Carlo (MC) simulation of diffractive events of the type $\gamma^*p \to {X} N$ (see Section~\ref{sec-models}). For the $M_{X}$, $W$ and $Q^2$ bins considered in this analysis, the resolutions are approximately $\frac{\sigma (W)}{W} = \frac{1}{W^{1/2}}$, $\frac{\sigma(Q^2)}{Q^2} = \frac{0.25}{(Q^2)^{1/3}}$ and $\frac{\sigma (M_X)}{M_X} = \frac{c}{M_X^{1/3}}$, where $c = 0.6$ GeV$^{1/3}$ for $M_X <$ 1 GeV and $c = 0.4$ GeV$^{1/3}$ for $M_X \ge 1$ GeV, with $M_X, W$ in units of GeV and $Q^2$ in GeV$^2$.

Results are presented for seven bins in $W$, seven bins in $Q^2$ and six bins in $M_X$, as shown in Table~\ref{t:binning}. The QED-Born-level cross sections and structure functions are determined as averages over these intervals and transported (see Section~\ref{sec-evalsig}) to the reference values ($M_{X\;{\rm ref}}, W_{\rm ref}, Q^2_{\rm ref}$) listed in Table~\ref{t:binning}.\\

\section{Monte Carlo simulations}
\label{sec-models}
The data were corrected for detector acceptance and resolution with suitable combinations of several MC models. Events from inclusive DIS, including radiative effects, were simulated using the HERACLES 4.6.1~\cite{cpc:69:155-tmp-3cfb28c9} program with the DJANGOH 1.1~\cite{cpc:81:381} interface to the hadronisation programs and using the CTEQ4D next-to-leading-order PDFs~\cite{pr:d55:1280}. In order to improve the description of the existing measurements at $Q^2 < 2$ GeV$^2$, a parametrisation~\cite{nokind:dhaidt:2002} of the measured $F_2$ data was used to reweight the generated non-diffractive events. In HERACLES, $O(\alpha)$ electroweak corrections are included. The colour-dipole model of ARIADNE 4~\cite{cpc:71:15}, including boson-gluon fusion, was used to simulate the $O(\alpha_S)$ plus leading-logarithmic corrections to the quark-parton model. The Lund string model as implemented in JETSET 7.4~\cite{cpc:82:74} was used by ARIADNE for hadronisation.

Diffractive DIS in which the proton does not dissociate, $ep \to eXp$  (including the production of $\omega$ and $\phi$ mesons via $ep \to e V^0 p$, $V = \omega, \phi$ but excluding $\rho^0$ production), were simulated with SATRAP~which is based on a saturation model~\cite{pr:d59:014017,*pr:d60:114023} and is interfaced to the RAPGAP 2.08 
framework~\cite{cpc:86:147}. The QED radiative effects were simulated with HERACLES. The QCD parton showers were simulated with LEPTO 6.5~\cite{cpc:101:108}. The production of $\rho^0$ mesons, $ep \to e \rho^0 p$, was simulated with JETSET 7.4 interfaced to the module ZEUSVM~\cite{zeusvm:1996} which uses a parametrisation of the measured $\rho^0$ cross sections as well as of the production and decay angular distributions~\cite{epj:c6:603,epj:c12:393}. 

The diffractive process in which the proton dissociates, $ep \to eXN$, was simulated with SATRAP interfaced to a module called SANG~\cite{helim:2002}. SANG includes the production of $\rho^0$ mesons. The mass spectrum of the system $N$ was generated according to $d\sigma /dM^2_N \propto (1/M^2_N)^{n}$ with $n = 1$. The fragmentation of the system $N$ was simulated using JETSET 7.4. The reweighting procedure used to match the generated $M_N$ spectrum with that of the data is described in Section~\ref{sec:lim_pdiss}.

All DIS processes except for those simulated by ZEUSVM were generated starting at $Q^2 = 0.5$ GeV$^2$; events from ZEUSVM were generated starting at $Q^2 = 0.7$ GeV$^2$, since the contribution from lower values of $Q^2$ was negligible. 
Since the diffractive events in data and MC showed different $W$ and $\beta$ dependences, events generated by SATRAP, which were the bulk of MC diffractive events, were reweighted to match the data.

In order to test for a possible contribution from Reggeon exchange to the final state reaction $\gamma^* p \to XN$, events were generated with RAPGAP in accordance with the analysis of the Regge contribution given in Appendix~\ref{asec-lnmxreggeon}. The background from photoproduction was estimated with events generated by PYTHIA 5.7~\cite{cpc:82:74}.

The ZEUS detector response was simulated using a program based on GEANT 3.13~\cite{tech:cern-dd-ee-84-1}. The generated events were passed through the detector and trigger simulation and processed by the same reconstruction and analysis programs as the data. The $Z_{\rm vtx}$ distribution used in the MC was reweighted to agree with the data.

The simulation of the measured total hadronic energy was checked with the balance of the measured transverse momenta of the scattered electron and that of the observed total hadronic system. For both MC and data, an average transverse momentum balance was achieved by increasing the measured hadronic energies by a factor of 1.065. The mass $M_{\cal{X}}$ reconstructed from the energy-corrected EFOs, in the $M_{\cal X}$ region analyzed, required an additional correction factor of 1.10 which was determined from MC simulation~\footnote{The hadrons produced in diffractive events, on average, have lower momenta than those for hadrons from non-peripheral events, and their fractional energy loss in the material in front of the calorimeter is larger.}.

Good agreement between data and simulated event distributions was obtained for both the inclusive and diffractive samples. More details on the event simulation can be found elsewhere~\cite{goebel:2001,helim:2002}. 

\section{Determination of the diffractive contribution}
\label{sec-lnmxmethod}
\subsection{The $M_X$ method}
The diffractive contribution was extracted from the data using the $M_X$ method~\cite{epj:c6:43,zfp:c70:391}.

The virtual photon-proton collision can be described as (see Eq.~(\ref{eq:xobs})) $\gamma^*p \to \cal{X} \; + \; {\rm rest}$,
where $\cal{X}$ represents all particles measured in the detector and `rest'
all particles that escape through the beam holes.
In the QCD picture of non-peripheral DIS, $\cal{X}$ is related to the struck quark and `rest' to the proton remnant, both of which are coloured states.
The final-state particles are expected to be uniformly emitted in rapidity along the $\gamma^*p$ collision axis leading to final-state particles which populate uniformly the rapidity gap between the struck quark and the proton remnant~\cite{feynman:1972:photon}. In this case, it can be shown from general arguments (see Appendix~\ref{asec-lnmx}) that the mass, $M_{\cal X}$, is distributed as 
\begin{eqnarray}
\frac{d{\cal N}^{\rm non-diff}}{d \ln M^2_{\cal X}} = c \cdot \exp(b \cdot \ln M^2_{\cal X}),
\label{eq:lnmxsquare}
\end{eqnarray}
where $b$ and $c$ are constants. DJANGOH predicts, for non-peripheral DIS, $b \approx 1.9$.

The diffractive reaction, $\gamma^*p \to X N$, on the other hand, has different characteristics. The incoming proton undergoes a small perturbation and emerges either intact, or as a low-mass nucleonic state, carrying a large fraction, $x_L$, of the incoming proton momentum. Diffractive scattering shows up as a peak near $x_L = 1$, the mass of the system $X$ being limited by kinematics to $M^2_X/W^2 \stackrel{<}{\sim} 1 - x_L$. Moreover, the distance in rapidity between the outgoing nucleon system $N$  and the system $X$ is $\Delta \eta \approx \ln(1/(1-x_L))$~\cite{Barone}, becoming large when $x_L$ is close to one. Combined with the limited values of $M_X$ and the peaking of the diffractive cross section near $x_L = 1$, this leads to a large separation in rapidity between $N$ and any other hadronic activity in the event. For the vast majority of diffractive events, the decay particles from the system $N$ leave undetected through the forward beam hole. For a wide range of $M_X$ values, the particles of the system $X$ are emitted entirely within the acceptance of the detector and the measured system $\cal X$ can be identified with $X$. Monte Carlo studies show that $X$ can be reliably reconstructed over the full $M_X$ range of this analysis: Fig.~\ref{f:mxresol} shows the distribution of the measured versus the generated value of $\ln M^2_X$ for the lowest and highest $W$ bins at two different $Q^2$ values. The horizontal bars indicate the maximum values of $\ln M^2_X$ up to which the diffractive contribution was extracted. There is a close correlation between the measured and the generated $\ln M^2_X$ value. From this point on, the distinction between $\cal X$ and $X$ will be omitted.

Regge phenomenology predicts the shape of the $M_X$ distribution for peripheral processes (see Appendix~\ref{asec-lnmx}). Diffractive production by Pomeron exchange in the $t$-channel, which dominates $x_L$ values close to unity, leads to an approximately constant $\ln M^2_X$ distribution ($b \approx 0$). Figure~\ref{f:lnmxsel} shows the distribution of $\ln M^2_X$ for the lowest and highest $W$ bins at low and high $Q^2$ for the data together with the expectations from MC simulation for non-peripheral DIS (DJANGOH) and for diffractive processes (SATRAP + ZEUSVM and SANG). The observed distributions agree well with the expectation for a non-diffractive component giving rise to an exponentially growing $\ln M^2_X$ distribution, and for a diffractive component producing an almost constant distribution in a large part of the $\ln M^2_X$ range.

The recent ZEUS measurement of diffraction in DIS with the leading proton spectrometer (LPS)~\cite{desy-04-131} allows the Reggeon exchange contribution (Fig.~\ref{f:gptripreg}) to be estimated, as discussed in Appendix~\ref{asec-lnmxreggeon}. Figure~\ref{f:lnmxselreggeon} compares, for the same ($W,Q^2$) bins as in Fig.~\ref{f:lnmxsel}, the $\ln M^2_X$ distributions for the data with those expected from Reggeon exchange. In this case, the $\ln M^2_X$ distribution increases exponentially with increasing $\ln M^2_X$ with a slope value of $b \approx 1.3$. 
  
The exponential rise of the $\ln M^2_X$ distribution for non-diffractive processes permits the subtraction of this component and, therefore, the extraction of the diffractive contribution without assumptions about its exact $M_X$ dependence. The distribution is of the form:
\begin{eqnarray}
\frac{dN}{d\ln M^2_X} = D + c \cdot \exp(b \; \ln M^2_X), \; \ln M^2_X < \ln W^2 - \eta_0, 
\label{eq:lnmxshape}
\end{eqnarray}
where $D$ is the diffractive contribution and the second term represents the non-diffractive contributions. The quantity $(\ln W^2 - \eta_0)$ specifies the maximum value of $\ln M^2_X$ up to which the exponential behaviour of the non-diffractive contribution holds. A value of $\eta_0 = 2.2$ was found from the data. Equation~(\ref{eq:lnmxshape}) was fitted to the data in the limited range $\ln W^2 - 5.6 < \ln M^2_X < \ln W^2 - \eta_0$ in order to determine the parameters $b$ and $c$. The diffractive contribution is expected to be a slowly varying function of $\ln M^2_X$ when $M^2_X > Q^2$, and to approach, for large $M^2_X$, an approximately constant $\ln M^2_X$ distribution~\cite{zfp:c53:331,*jetp:81:625,*pl:b422:238,pl:b191:309,np:b303:634}. Therefore, $D$ was assumed to be constant over the fit range. However, the diffractive contribution was not taken from the fit but was obtained from the observed number of events after subtracting the non-diffractive contribution determined using the fitted values of $b$ and $c$.

The non-diffractive contribution in the ($M_X, W, Q^2$) bins was measured in two steps. In the first step, the slope $b$ was determined as an average of the values obtained from the fits to the data for the intervals with $134 < W < 245$ GeV and $2.2 < Q^2 < 10$ GeV$^2$. The fits yielded $b_{\rm nom} = 1.63 \pm 0.07$. In the second step, the fits were repeated for all ($W,Q^2$) intervals, using $b = b_{\rm nom}$ as a fixed parameter and assuming $D$ to be constant. Good fits with $\chi^2$ per degree of freedom (dof) of about unity were obtained. The statistical error of the diffractive contribution includes the uncertainty of $b_{\rm nom}$.

Figures~\ref{f:lnmxsel} and~\ref{f:lnmxselreggeon} show the results from the fit according to Eq.~(\ref{eq:lnmxshape}) for the non-diffractive and the sum of the non-diffractive and diffractive contributions. Figure~\ref{f:lnmxselreggeon} shows the $M_X$ distribution of the RAPGAP Reggeon simulation described in Section~\ref{sec-models}. As discussed above, the Reggeon contribution to the $M_X$ spectra is similar in slope to the non-peripheral contribution described by Eq.~(\ref{eq:lnmxsquare}), and is always smaller than the nondiffractive contributions~\footnote{A recent determination of the diffractive contribution based on the presence of a leading proton~\cite{desy-04-131} has been limited to the region $\xpom < 0.01$ to exclude contributions from Reggeon exchange. The fact that the $M_X$ method excludes the Reggeon contribution allows the diffractive component to be extracted also in the region $\xpom > 0.01$.}. The same conclusions hold for all other ($W,Q^2$) bins considered in this analysis.

Finally, a MC event sample was prepared which consisted of the sum of the contributions from diffraction (SATRAP+ZEUSVM+SANG) and non-peripheral scattering (DJANGOH). The MC event sample was subjected to the same analysis procedure as the data and the diffractive contribution was extracted for all $(M_X, W, Q^2)$ bins. For all bins, the accuracy of the determination of the fraction of diffractive events in the MC sample was better than the statistical precision of the data. The same test was repeated with the RAPGAP Reggeon sample included in the summed MC event sample. The normalisation of the Reggeon sample was increased by a factor of two - with respect to that found in the study of Appendix C - in order to test the robustness of the $M_X$-method extraction of the diffractive component against the large uncertainties of the Reggeon contribution. The result showed that the diffractive component was extracted with accuracy similar to that of the extraction without the Reggeon component. 
 
For the final analysis of the diffractive cross section and structure function, only ($M_X, W, Q^2$) bins where the non-diffractive background was less than 50\% were kept.

\subsection{Contribution from diffractive proton dissociation}\label{sec:lim_pdiss}
In addition to single dissociation, $\gamma^* p \to Xp$, processes where the proton also dissociates, $\gamma^* p \to XN$, can contribute to the diffractive event sample. Events from double dissociation can be grouped into those events where $N$ has a low mass and disappears in the forward beam hole without energy deposition in the calorimeters FPC or CAL, and into those where decay particles deposit energy in the calorimeters. The probability of depositing energy in the calorimeters depends on the mass $M_N$.
On average, in events where $N$ has a mass below 2.3 GeV, the system $N$ disappears in the forward beam hole without energy deposition in the FPC or the CAL, while for those events with $M_N > 2.3$ GeV, the system $N$ deposits energy in the calorimeters. In the latter case, the reconstructed mass of the total hadronic system is larger than the mass of $X$. Such events lead to a distortion of the $\ln M^2_X$ distribution at high $M_X$ values. In order to study this effect, double dissociative events were generated using SANG. 

The parameters of SANG, in particular those determining the shape of the $M_N$ spectrum and the overall normalization, were checked with the subset of the data dominated by the contribution from double dissociation. Events in this subset were required to have a minimum rapidity gap $\Delta \eta > \eta_{\rm min}$ between at least one EFO and its neighbours. Good sensitivity for double dissociation was obtained with $\eta_{\rm min}$ = 3.0 for $W = 55 - 99$ GeV and $\eta_{\rm min}$ = 4.0 for $W = 99 - 245$ GeV. The study was performed with four event samples for the kinematic regions shown in Figs.~\ref{f:mffcalsang} and~\ref{f:mffcal}.

The mass of the hadronic system reconstructed from the energy deposits in FPC+FCAL, $M_{\rm FFCAL}$, depends approximately linearly on the mass $M^{\rm gen}_N$ of the generated system $N$. The distribution of $M_{\rm FFCAL}$ at low $M_{\rm FFCAL}$ is dominated by double dissociation. After reweighting the $M_N$ distribution generated for the process $\gamma^*p \to XN$, good agreement was obtained between the number of events measured and the number of events predicted from the sum of the simulated $Xp$, $\rho^0 p$, $X N$ and non-diffractive processes. 

Figure~\ref{f:mffcalsang} demonstrates the sensitivity of the $M_{\rm FFCAL}$ distribution to the shape of the $M_N$-spectrum: it shows, for the four ($Q^2,W$) regions, the distribution of $M_{\rm FFCAL}$ as predicted by SANG, and after reweighting SANG for $M_N > M_{N0} = 2.3$ GeV by $(\frac{M_N}{M_{N0}})^{+1}$, or by $(\frac{M_N}{M_{N0}})^{-1}$. In the first case, the event rate increases (e.g. near $M_{\rm FFCAL} = 1$ GeV by roughly a factor of two); in the second case, the event rate decreases (near $M_{\rm FFCAL} = 1$ GeV by about a factor of 1.5). To achieve agreement with the data (see Fig.~\ref{f:mffcal}) SANG was reweighted for $M_N \le 4$ GeV by a factor of $0.89 \sqrt{M_N/4}$ ($M_N$ in GeV), and for $M_N > 4$ GeV by a factor of $(2.5/M_N)^{0.25}$. In this exercise, the diffractive contribution for $M_N > 2.3$ GeV is assumed not to change with $W$. The good description of the data obtained from this simulation supports this assumption (see below).

The data distributions of $M_{\rm FFCAL}$ at low $M_{\rm FFCAL}$ and the reweighted MC predictions are compared in Fig.~\ref{f:mffcal} for the four ($Q^2,W$) regions. The sum of the contributions calculated for $Xp$, $\rho^0 p$ and the non-diffractive component are shown, as well as the $XN$ contribution which dominates the region of low $M_{\rm FFCAL}$ values. The sum of the four contributions reproduces the data well. Double dissociation, ($\gamma^*p \to XN$), accounts for more than 80\% of the events predicted by MC: for Fig.~\ref{f:mffcal}a when $M_{\rm FFCAL} < 2.5$ GeV, for Fig.~\ref{f:mffcal}b when $M_{\rm FFCAL} < 2$ GeV, for Fig.~\ref{f:mffcal}c when $M_{\rm FFCAL} < 5$ GeV and for Fig.~\ref{f:mffcal}d when $M_{\rm FFCAL} < 3$ GeV.

This study showed that, approximately, events generated with $M_N < 2.3$ GeV deposit less than 1 GeV of energy in the FPC, while events with $M_N \ge 2.3$ GeV deposit more than 1 GeV. No information could be gained from the data on the contribution from double dissociation with $M_N > 2.3$ GeV. Figure~\ref{f:lnmxxn} shows the $\ln M^2_X$ spectra for the same $W$ and $Q^2$ regions as in Fig.~\ref{f:lnmxsel} together with the expected contribution from double dissociation, $\gamma^* p \to XN$, for those events with $M_N \ge 2.3$ GeV. This contribution is of the order of 6\% (18\%, 36\%) for $M_X/W < 0.05$ ($M_X/W =$ 0.1,0.14). Since this contribution can affect the determination of the slope $b$ for the non-diffractive contribution, it has been subtracted, using the MC simulation, from the data as a function of $M_X$, $W$ and $Q^2$. The systematic error (see Section~\ref{subsec-sys}) allows for a $30\%$ increase/decrease of the number of events removed. The diffractive cross section presented later is therefore the sum of the contributions from the $Xp$ and $XN$ ($M_N < 2.3$ GeV) final states. 

Also shown in Fig.~\ref{f:lnmxxn} is the expected contribution from photoproduction which is negligible, except at high $W$.

\subsection{The $\ln M^2_X$ distributions}
The $\ln M^2_X$ spectra for all ($W,Q^2$) bins studied in this analysis are displayed in Fig.~\ref{f:lnmxall}. The data distributions, from which the contributions from double dissociation ($M_N < 2.3$ GeV) and from photoproduction background have been subtracted, are shown. They are compared with the MC predictions for the contributions from non-peripheral and diffractive production. It can be seen that the events at low and medium values of $\ln M^2_X$ originate exclusively from diffractive production. The MC simulations are in good agreement with the data.

\section{Evaluation of cross sections and systematic uncertainties}
\label{sec-evalsig}
The total and diffractive cross sections for $ep$ scattering 
in a given ($W,Q^2$) bin were determined from the number of observed events, corrected for background, acceptance and smearing, and corrected to the QED Born level.

The cross sections and structure functions are presented at the reference values $W_{\rm ref}$, $Q^2_{\rm ref}$, and $M_{X {\rm ref}}$. This was achieved as follows: first, the cross sections and structure functions were determined at the weighted average of each ($M_X$, $W$, $Q^2$) bin. They were then transported to the reference position using a parametrisation~\cite{nokind:dhaidt:2002} in the case of the proton structure function $F_2$, and the result of the BEKW(mod) fit (see Section~\ref{sec-bekw}) for the diffractive cross sections and structure functions. The resulting changes to the cross section and structure function values from the average to those at the reference positions were 5 - 15\%.  

\subsection{Systematic uncertainties} 
\label{subsec-sys}
A study of the main sources contributing to the systematic uncertainties of the measurements were performed. The systematic uncertainties were calculated by varying the cuts or modifying the analysis procedure and repeating the full analysis for every variation. The size of the variations of cuts and the changes of the energy scales were chosen commensurate with the resolutions or the uncertainties of the relevant variables:
\begin{itemize}
\item the acceptance at low values of $Q^2$ depends critically on the position measurement of the scattered electron. The vertical separation of the upper and lower halves of the SRTD was increased (decreased) by 0.2 cm in the data (systematic uncertainties 1a,b) while their positions in the MC were left unchanged. The resulting deviations of the cross sections were typically 5 - 7 \%;
\item  the box cut was changed from $26.6{\;\rm cm} \times 17{\;\rm cm}$ to $27.6{\;\rm cm} \times 18{\;\rm cm}$ (systematic uncertainty 2). This affected the low-$Q^2$ region. Changes of 5 - 15\% were observed, mainly for $W < 100$ GeV; 
\item the measured energy of the scattered electron was increased (decreased) by 2\% in the data, but not in the MC (systematic uncertainties 3a,b). In most cases the changes were smaller than, or of the order of, the statistical error;
\item the lower cut for the energy of the scattered electron was lowered to 8 GeV (raised to 12 GeV) (systematic uncertainties 4a,b). This produced changes of 0 - 2\%; 
\item to estimate the systematic uncertainties due to the uncertainty in the hadronic energy, the analysis was repeated after increasing (decreasing) the hadronic energy measured by the CAL by 2\% in the data but not in MC (systematic uncertainties 5a,b). The typical changes were below 5\%;
\item the energies measured by the FPC were increased (decreased) by 10\% in the data but not in MC (systematic uncertainties 6a,b). The effect was negligible; 
\item the minimum hadronic energy cut of 400 MeV was increased by 50\% (systematic uncertainty 7). This led to changes at the 1-3\% level;
\item in order to check the simulation of the hadronic final state, the selection on $\sum_{i=e,h}(E_i - P_{Zi})$ was changed from (46 to 64) to (43 to 64) GeV (systematic uncertainty 8), leading to changes at the level of 20 to 30\% of the statistical uncertainty except for one measurement at low $Q^2$ and high $W$. Also in this case, the change was small compared to the total systematic uncertainty;
\item the reconstructed mass $M_X$ of the system ${\cal }X$ was increased (decreased) by 5\% in the data but not in the MC (systematic uncertainties 9a,b). Changes of the order of 5 - 10\% were observed for the lowest $M_X$ bin while, for the higher $M_X$ bins, the changes were much smaller; 
\item a substantial fraction of events at $W > 164$ GeV, low $M_X$ and low $Q^2$ have no measured primary vertex. The position of the vertex affects the measurement of the polar angle of the scattered electron and so of $Q^2$. For these events, the average vertex for each data-taking run was used. The fraction of events with no vertex found in the data agrees well with the MC predictions~\footnote{For four ($M_X,W,Q^2$) bins - all of which have $M_X = 1.2$ GeV and $W = 180$ or 220 GeV, the fraction of events in the data without a vertex is above 35\%. The fraction of events predicted by the MC for these bins is the same to within 7\%.}. The difference between the observed fraction of events and that predicted by MC was used as a systematic uncertainty (systematic error 10) which amounted to 5 - 10\% for low $M_X$ and $Q^2$, and to much smaller values elsewhere; 

\item the contribution from double dissociation with $M_N > 2.3$ GeV was determined with the help of the reweighted SANG simulation and was subtracted from the data. The diffractive cross section was redetermined by increasing (decreasing) the predicted contribution from SANG by 30\% (systematic uncertainties 11a,b). The resulting changes in the diffractive cross section were well below the statistical uncertainty except for four data points, where they were of similar magnitude.
\end{itemize}

In order to evaluate the uncertainties arising from the form of the $\ln M^2_X$ distribution assumed for the diffractive contribution ($D =$ constant, see Eq.~(\ref{eq:lnmxshape})), the fits were repeated with the form $D = d_0(1-\beta)[\beta(1-\beta)+d_1(1-\beta)^{\kappa}]$, where $d_0, d_1, \kappa$ are fit parameters~\cite{zfp:c53:331,*jetp:81:625,*pl:b422:238,pl:b191:309,np:b303:634,ejp:c7:443}. Negligible changes were found. 

The total systematic error for each bin was determined by adding the individual contributions in quadrature.

In the fits reported below, except for the BEKW(mod) fit, the fits were performed (a) to the nominal values, (b) to every data set (j) obtained by shifting the measured values by the amount given by the systematic uncertainty (j). The statistical uncertainties were included in each fit. The fit parameters quoted are those given by the fit to the nominal values; the systematic uncretainties were obtained as the square root of the sum of the squares of the differences between the fit parameters obtained with the nominal set and those obtained with the systematic shifts. In the case of conjugated uncertainties (labelled as (a,b) above) the averages of the squares of the two uncertainties were taken. For the BEKW(mod) fit, the statistical and systematic uncertainties of the measured values of the diffractive structure function were added in quadrature.

\section{Proton structure function $F_2$ and the total $\gamma^*p$ cross section}
\label{sec-f2sigtot}
A prime goal of this analysis is the study of the $W$ and $Q^2$ dependences of the diffractive cross section as a function of $M_X$, and the comparison with the {\it total} cross section. As a first step, the total cross section was determined for the same bins in $W$ and $Q^2$ as for the diffractive cross section and using an identical analysis procedure. This minimises systematic uncertainties and allows the most direct comparison of the two cross sections.

The differential cross section for inclusive $ep$ scattering mediated by virtual photon exchange is given in terms of the structure functions $F_i$ of the proton by
\begin{eqnarray}
\frac{d^2 \sigma^{e^- p}}{dx dQ^2} = \frac{2 \pi \alpha^2}{x Q^4} [Y F_2(x,Q^2) - y^2 F_L(x,Q^2)](1 + \delta_r(x,Q^2)),
\end{eqnarray}
where $Y = 1 + (1-y)^2$, $F_2$ is the main component of the cross section which in the DIS factorisation scheme corresponds to the sum of the momentum densities of the quarks and antiquarks weighted by the squares of their charges, $F_L$ is the longitudinal structure function and $\delta_r$ is a term accounting for radiative corrections. In the $Q^2$ range considered in this analysis, $Q^2 < 80$ GeV$^2$, the contributions from $Z^0$ exchange and $Z^0$ - $\gamma$ interference are well below 1\% and were ignored. The contribution of $F_L$ to the cross section relative to that from $F_2$ is given by $(y^2/Y) \cdot (F_L/F_2)$. For the determination of $F_2$, the $F_L$ contribution was taken from the QCD fits to the structure function data obtained by ZEUS~\cite{pr:d67:012007} and H1~\cite{epj:c21:33}, which may be approximated by $F_L = 0.2 F_2$. The contribution of $F_L$ to the cross section in the highest $y$ (= lowest $x$) bin of this analysis was 3.8\%, decreasing to 1.5\% for the next highest $y$-bin. For the other bins, the $F_L$ contribution is below 1\%. The uncertainties of the $F_L$ corrections were estimated to be below 20\%; the resulting uncertainties on $F_2$ are below 1\%.
 
The measured $F_2$ values are listed in Table~\ref{t:f2tab} and shown in Fig.~\ref{f:fpcf2}. The data are compared to the predictions of the ZEUS QCD fit~\cite{pr:d67:012007} obtained from the previous ZEUS $F_2$ measurements~\cite{epj:c21:443}. The fit describes the data well. The proton structure function, $F_2$, rises rapidly as $x \to 0$ for all values of $Q^2$, the slope increasing as $Q^2$ increases.

The form:
\begin{eqnarray}
F_2 = c \cdot x^{-\lambda},
\label{eq:f2fitlambda} 
\end{eqnarray}
was fitted for every $Q^2$ bin to the $F_2$ data. Here $\lambda$ is related to the intercept of the Pomeron trajectory, $\lambda = \alpha_{\pom}(0) -1$. For later comparison with the diffractive results, these $\alpha_{\pom}$ values will be referred to as $\alpha^{\rm tot}_{\pom}$. The resulting values for $c$ and $\alpha^{\rm tot}_{\pom}(0)$ are listed in Table~\ref{t:f2fit}. Within errors, $c$ is independent of $Q^2$. Figure~\ref{f:slopef2dif} shows that $\alpha^{\rm tot}_{\pom}(0)$ lies above the `soft Pomeron' value of $1.096^{+0.012}_{-0.009}$ deduced from hadron-hadron scattering data
~\cite{np:b244:322,*pl:b296:227,pl:b395:311}
; it rises approximately linearly with $\ln Q^2$ from $\alpha^{\rm tot}_{\pom}(0) = 1.155 \pm 0.011({\rm stat.})^{+0.007}_{-0.011}({\rm syst.})$ at $Q^2 = 2.7$ GeV$^2$, to $1.307 \pm 0.019({\rm stat.})^{+0.027}_{-0.037}({\rm syst.})$ at $Q^2 = 55$ GeV$^2$, in agreement with previous observations~\cite{epj:c7:609,pl:b520:183}. A Pomeron intercept which changes with $Q^2$ violates the assumption of single Pomeron exchange plus Regge factorisation of the vertex functions.
 
The total cross section for virtual photon-proton scattering, $\sigma^{\rm tot} _{\gamma^{\ast} p} \equiv \sigma_T(x,Q^2) + \sigma_L(x,Q^2)$,
was extracted from the measurement of $F_2$ using the relation
\begin{eqnarray}
\sigma^{\rm tot} _{\gamma^{\ast} p} = \frac{4 \pi^2 \alpha}{Q^2(1-x)} F_2(x,Q^2),
\end{eqnarray}
which is valid for $4m^2_p x^2 \ll Q^2$~\cite{pr:129:1834,anphy:28:18,pr:167:1365}. The total cross section values are listed in Table~\ref{t:sigtottab} for fixed $Q^2$ as a function of $W$. The total cross section multiplied by $Q^2$, shown in Fig.~\ref{f:sigtot}, exhibits a strong rise with $W$, becoming steeper as $Q^2$ increases. This behaviour of $\sigma^{\rm tot}_{\gamma^*p}$ reflects the $x$ dependence of $F_2$ as $x \to 0$, viz. $\sigma^{\rm tot}_{\gamma^*p} \propto W^{2(\alpha^{\rm tot}_{\pom}(0)-1)}$.

\section{Diffractive cross section}
\label{sec-sigdiff}
The cross section for diffractive scattering via $ep \to eXN$ can be expressed in terms of the transverse (T) and longitudinal (L) cross sections, $\sigma^{\rm diff}_T$ and $\sigma^{\rm diff}_L$, for $\gamma^* p \to XN$ as\\
\begin{eqnarray}
\frac{d\sigma^{\rm diff}_{\gamma^* p \to XN}(M_X,W, Q^2)}{dM_X} \equiv \frac{d(\sigma^{\rm diff}_T + \sigma^{\rm diff}_L)}{dM_X} \approx \frac{2 \pi}{\alpha}\frac{Q^2}{(1-y)^2+1}\frac{d\sigma^{\rm diff}_{ep \to eXN}(M_X,W,Q^2)}{dM_X d\ln W^2 dQ^2} .
\end{eqnarray}
Here, a term $(1 - y^2 /[1+(1-y)^2])\sigma^{\rm diff}_L/(\sigma^{\rm diff}_T + \sigma^{\rm diff}_L)$
multiplying $(\sigma^{\rm diff}_T + \sigma^{\rm diff}_L)$ has been neglected~\cite{pr:129:1834,anphy:28:18,pr:167:1365}. Since $y \approx W^2/s$, this approximation reduces the cross section by less than 4\% for $W < 200$ GeV, and by less than $ 8\%$ in the highest $W$ bin, 200 - 245 GeV, if $\sigma^{\rm diff}_L \le \; \sigma^{\rm diff}_T$~\footnote{The relative contribution to diffractive production by longitudinal photons is expected to be small~\cite{zfp:c53:331,ejp:c7:443} except for the production of vector mesons, $\gamma^*p \to VN$. The processes $\gamma^*p \to VN$, $V = \rho^0, \omega, \phi$, contribute about 40 - 60\% of the diffractive cross section measured in the lowest $M_X$ bin ($0.28 < M_X < 2$ GeV) and are dominated by longitudinal photons. Assuming that these were the only contributions from longitudinal photons, extrapolation of the cross sections for $\gamma^*p \to Vp$ measured at $W < 150$ GeV~\cite{epj:c6:603,epj:c12:393,pl:b487:273} to higher $W$ gives an effect of the order of 3$\%$ at $7 < Q^2 < 27$ GeV$^2$ and $W$ = 180 GeV. This estimate assumed the same fraction of nucleon dissociation for $\gamma^*p \to VN$ as for inclusive diffraction. The measured data on $J/\Psi$ production indicate that this process contributes less than 10\% of the diffractive cross section in the bin $M_X = 2 - 4$ GeV and therefore even smaller corrections are expected for this $M_X$ bin.}.

The diffractive cross section $d\sigma^{\rm diff}/dM_X$ for $\gamma^*p \to XN$, where $M_N < 2.3$ GeV, is presented in Tables~\ref{t:dsigdmx1} -~\ref{t:dsigdmx6} and Fig.~\ref{f:dsigdmx}, after transporting the measured cross sections to the reference values ($M_X, W, Q^2$) using the BEKW(mod) fit (see Section~\ref{sec-bekw}). 

These diffractive cross sections do not include contributions from deeply virtual Compton scattering, $\gamma^* p \to \gamma p$ (DVCS). The DVCS cross sections in the region $5 < Q^2 < 30$ GeV$^2$, $40 < W < 140$ GeV have been measured by the ZEUS collaboration~\cite{pl:b573:46} and are between 2 and 4\% of $\sigma^{\rm diff}_{\gamma^* p \to XN} (0.28 < M_X < 2$ GeV).

\subsection{$W$ dependence of the diffractive cross section}
\label{subsec-wdepdiff}
As seen in Fig.~\ref{f:dsigdmx}, for the bin $M_X = 1.2$ GeV, the diffractive cross section, $d\sigma^{\rm diff}/dM_X$, shows only a modest increase with $W$. For higher $M_X$ values, a rise with $W$ is observed for $Q^2 \ge 4$ GeV$^2$. The $W$ dependence was quantified by fitting the form
\begin{eqnarray}
\frac{d\sigma^{\rm diff}_{\gamma^* p \to XN}}{dM_X} = h \cdot(W/W_0)^{a^{\rm diff}},
\end{eqnarray}
to the data for each $(M_X,Q^2)$ bin with $M_X < 15$ GeV; here $W_0 = 1$ GeV and $h$, $a^{\rm diff}$ are free parameters. 
Under the assumption that the diffractive cross section can be described by the exchange of a single Pomeron, the parameter $a^{\rm diff}$ is related to the Pomeron trajectory averaged over $t$: $\overline{\alpha_{\pom}} = 1+ a^{\rm diff}/4$. In the framework of Regge phenomenology, the cross section for diffractive scattering can be written as~\cite{Collins:1977:regge}, 
\begin{eqnarray}
d\sigma/dt = f(t) \cdot e^{2(\alpha_{\pom}(t)-1)\cdot \ln (s/s_0)},
\end{eqnarray}
where $t$ is the four-momentum-transfer squared from $\gamma^*$ to $X$, $f(t)$ characterises the $t$-dependences of the ($\gamma^* \pom \gamma^*$) and ($p \pom N$) vertices, and $s_0 = 1$ GeV$^2$. In the present measurement, the diffractive cross section is integrated over $t$, providing $t$-averaged values $\overline{\alpha_{\pom}}$. Assuming $d\sigma/dt \propto e^{A \cdot t}$ and $\alpha_{\pom}(t) = \alpha_{\pom}(0) + \alpha_{\pom}^{\prime} \cdot t$, leads to $\alpha_{\pom}(0) = \overline{\alpha_{\pom}} + \alpha_{\pom}^{\prime}/A$. Taking $A = 7.9 \pm 0.5({\rm stat.})^{+0.9}_{-0.5}({\rm syst.})$ GeV$^{-2}$, as measured by this experiment with the leading proton spectrometer~\cite{desy-04-131}}~\footnote{This value of $A$ has been determined for $\xpom < 0.01$, where diffraction is dominant in the ZEUS LPS data. It is assumed that $A$ for the diffractive contribution remains the same in the region $0.01 < \xpom < 0.022$.}, and $\alpha_{\pom}^{\prime} = 0.25$ GeV$^{-2}$~\cite{np:b244:322,*pl:b296:227}, gives $\alpha_{\pom}(0) \approx \overline{\alpha_{\pom}} +0.03 = 1.03 + a^{\rm diff}/4$. The $\alpha_{\pom}(0)$ values deduced from diffractive cross sections are denoted as $\alpha^{\rm diff}_{\pom}(0)$.

The resulting $\alpha^{\rm diff}_{\pom}(0)$ values are listed in Table~\ref{t:adiff} and shown in Fig.~\ref{f:difslope} as a function of $Q^2$ for different $M_X$ intervals. For $M_X$ below 2 GeV, $\alpha^{\rm diff}_{\pom}(0)$ is compatible with the soft Pomeron. For larger $M_X$, $\alpha^{\rm diff}_{\pom}(0)$ lies above the soft-Pomeron result, the difference increasing with $Q^2$.  

Figure~\ref{f:slopef2dif} compares the $Q^2$ dependence of $\alpha^{\rm tot}_{\pom}(0)$ with $\alpha^{\rm diff}_{\pom}(0)$ deduced from the diffractive cross section for $2< M_X < 15$ GeV. Both sets of results lie above the soft-Pomeron result and show a rise with $Q^2$. The $\alpha^{\rm diff}_{\pom}(0)$ values lie, however, consistently below those obtained from $\sigma^{\rm tot}_{\gamma^*p}$, or equivalently $F_2$, with $[{\alpha^{\rm diff}_{\pom}(0)-1}]/[{\alpha^{\rm tot}_{\pom}(0)-1}] \approx 0.5 - 0.7$. Thus, the $W$ dependences of the total and diffractive cross sections yield different Pomeron trajectories. 

\subsection{Combined $W$ and $Q^2$ dependence of the diffractive cross section}

The assumption of Regge factorisation requires that the Pomeron trajectory be independent of $Q^2$ if diffractive scattering is to be described by the exchange of a single Pomeron. As a consequence, the $W$ dependence of the diffractive cross section should also be independent of $Q^2$. In order to test this hypothesis with the full body of data, the form
\begin{eqnarray}
\frac{d\sigma^{\rm diff}_{\gamma^*p}}{dM_X} = c(M_X,Q^2) \cdot (\frac{W}{W_0})^{4(\overline{\alpha_{\pom}}(Q^2)-1)} ,
\end{eqnarray}
was fitted to the diffractive cross section; here $W_0 = 1$ GeV. The values $\overline{\alpha_{\pom}}(Q^2)$ and the constants $c(M_X,Q^2)$ were determined from the fit. In this way, the $Q^2$ dependence of the $W$ dependence of the diffractive cross section was tested independently of its ($M_X, Q^2$) dependence. Since the diffractive cross section for $M_X < 2$ GeV receives a substantial contribution from the process $\gamma^* p \to \rho^0 p$, which is dominated by longitudinal photons, and since the $W$ range covered for $M_X > 15$ GeV is rather limited, the fitting was done for the data with $2 < M_X < 15$ GeV. In total, 126 diffractive cross section measurements were included. There are 25 free parameters: four $\overline{\alpha_{\pom}}$ values for four bins of $Q^2$, and 21 constants $c(M_X,Q^2)$ for three $M_X$ bins (2 - 4, 4 - 8, 8 - 15 GeV) and the corresponding seven $Q^2$ bins.

The results obtained are presented in Table~\ref{t:alphapomvsq2} and shown in Fig.~\ref{f:alpomvsq2}. Within errors, $\alpha^{\rm diff}_{\pom}(0)$ is constant for $Q^2$ between 2.7 and 20 GeV$^2$ ($\langle Q^2 \rangle$ = 7.8 GeV$^2$) but has a substantially larger value for $Q^2$ between 20 and 80 GeV$^2$ ($\langle Q^2\rangle$ = 34.6 GeV$^2$). 

The statistical significance of the rise of $\alpha^{\rm diff}_{\pom}(0)$ with $Q^2$  was determined by a fit with the following free parameters: the normalisation constants for the four bins in $Q^2$, a single value of $\overline{\alpha^{\rm diff}_{\pom}(0)}$ averaged over $2.7 < Q^2 < 20$ GeV$^2$ and the difference $\Delta\alpha_{\pom} \equiv \alpha^{\rm diff}_{\pom}(0,\langle Q^2 \rangle  = 34.6 \; {\rm GeV}^2) - \alpha^{\rm diff}_{\pom}(0, \langle Q^2 \rangle = 7.8 \; {\rm GeV}^2)$. Considering all systematic uncertainties and their correlations, the fit yielded: 
\begin{eqnarray}
\alpha^{\rm diff}_{\pom}(0,\langle Q^2 \rangle = 7.8 \; {\rm GeV}^2) = 1.1220 \pm 0.0046({\rm stat.})^{+0.0132}_{-0.0114}({\rm syst.})
\end{eqnarray}
and
\begin{eqnarray}
 \Delta\alpha^{\rm diff}_{\pom}  = 0.0714 \pm 0.0140({\rm stat.})^{+0.0047}_{-0.0100}({\rm syst.}).
\end{eqnarray}
The addition of the statistical and systematic uncertainties in quadrature gives 
\begin{eqnarray}
\Delta\alpha^{\rm diff}_{\pom} = 0.0714^{+0.0147}_{-0.0172}.
\end{eqnarray}
The result {\rm establishes} the rise of $\alpha^{\rm diff}_{\pom}(0)$ with $Q^2$, with a significance of 4.2 standard deviations. Assuming single Pomeron exchange, this observation contradicts Regge factorisation.  

This experiment, using the LPS~\cite{desy-04-131}, obtained for the kinematic region $\xpom < 0.01$, $0.03 < Q^2 < 39$ GeV$^2$, the value $\alpha^{\rm diff}_{\pom}(0) = 1.16 \pm 0.02 (\rm stat.) \pm 0.02 ({\rm syst.})$. Restricting the data in the present analysis to $\xpom < 0.01$ gives $\alpha^{\rm diff}_{\pom}(0, \langle Q^2 \rangle  = 7.8 \; {\rm GeV}^2) = 1.1209 \pm 0.0051({\rm stat.})^{+0.0136}_{-0.0122}({\rm syst.})$ and $\Delta\alpha^{\rm diff}_{\pom} = 0.0578 \pm 0.0178({\rm stat.})^{+0.0081}_{-0.0118}(\rm syst.)$ (see Table~\ref{t:alphapomvsq2lt01} and Fig.~\ref{f:alpomvsq2}). The results are consistent with the fit to the full data set and also in agreement with the LPS result.

\subsection{$M_X$ and $Q^2$ dependences of the diffractive cross section at fixed $W$}
The $M_X$ and $Q^2$ dependences of the diffractive cross section for $W = 220$ GeV are shown in Fig.~\ref{f:dsigdmxvsmx}. The highest-$W$ region is used since it covers the largest range in $M_X$. The cross section has been multiplied by a factor of $Q^2$, since a leading-twist behaviour would give approximate $Q^2$ independence. For low and medium $Q^2$, the $M_X$ spectrum is dominated by the production of states with $M_X < 3$ GeV (Fig.~\ref{f:dsigdmxvsmx}a). The cross section for these states decreases rapidly for higher $Q^2$, consistent with a predominantly higher-twist behaviour. Above $M_X = 11$ GeV, little dependence on $Q^2$ is observed (Fig.~\ref{f:dsigdmxvsmx}b), corresponding to a leading twist behaviour. 

\subsection{Diffractive contribution to the total cross section}
The relationship between the total and diffractive cross sections can be derived under certain assumptions. For instance, the imaginary part of the amplitude for elastic scattering, $A_{\gamma^* p \to \gamma^* p}(t,W,Q^2)$, at $t = 0$ can be assumed to be linked to the total cross section by a generalisation of the optical theorem to virtual photon scattering. Assuming that $\sigma^{\rm tot}_{\gamma^*p} \propto W^{2 \lambda}$ and that the elastic and inclusive diffractive amplitudes at $t=0$ are purely imaginary and have the same $W$ and $Q^2$ dependences, then $A_{\gamma^* p \to \gamma^* p}(t=0,W,Q^2)$ is proportional to $W^{2 \lambda}$. Neglecting the real part of the scattering amplitudes, the rise of the diffractive cross section with $W$ should then be proportional to $W^{4 \lambda}$, so that the ratio of the diffractive cross section to the total $\gamma^* p$ cross section,
\begin{eqnarray}
r^{\rm diff}_{\rm tot} \equiv \frac{\sigma^{\rm diff}}{\sigma^{\rm tot}} = \frac{\int^{M_b}_{M_a} dM_X d\sigma^{\rm diff}_{\gamma^* p \to XN, M_N < 2.3 {\rm GeV}}/dM_X}{\sigma^{\rm tot}_{\gamma^* p}}
\end{eqnarray}
should behave as $r^{\rm diff}_{\rm tot}  \propto W^{2 \lambda}$. 

The ratio $r^{\rm diff}_{\rm tot}$
was determined for all $M_a < M_X < M_b$ bins, with the $\sigma^{\rm tot}_{\gamma^* p}$ values taken from this analysis. The ratio $r^{\rm diff}_{\rm tot}$ is shown in Tables~\ref{t:rdiftot1} -~\ref{t:rdiftot6} and in Fig.~\ref{f:rdiftot}. The observed near constancy with $W$ is explained by the dipole saturation model~\cite{pr:d59:014017,*pr:d60:114023,jp:g28:1057}.

For $M_X < 2$ GeV, $r^{\rm diff}_{\rm tot}$ decreases with increasing $Q^2$, while, for $M_X > 4$ GeV, this decrease becomes weaker, and almost no $Q^2$ dependence is observed for $M_X > 8$ GeV. Here, the diffractive cross section has approximately the same $W$ and $Q^2$ dependences as the total cross section, in agreement with the conclusion drawn from Fig.~\ref{f:slopef2dif}. 

The ratio $\sigma^{\rm diff}(0.28 < M_X < 35 {\rm \; GeV}, M_N < 2.3 {\rm \; GeV})/\sigma^{\rm tot}$ was evaluated as a function of $Q^2$ for the highest $W$ bin ($200 < W < 245$ GeV) which provides the best coverage in $M_X$. The ratio is given in Table~\ref{t:sumrdiftot}. At $Q^2 = 4$ GeV$^2$, $\sigma^{\rm diff}/\sigma^{\rm tot}$ reaches 15.8$^{+1.2}_{-1.0}$ \%. It decreases slowly with $Q^2$, reaching 9.6$^{+0.7} _{-0.7}$ \% at $Q^2 =27$ GeV$^2$. Diffractive processes thus account for a substantial part of the total deep inelastic cross section.

\section{Diffractive structure function of the proton}
\label{sec-difff2d}
The diffractive structure function of the proton, $F^{D(3)}_2(\beta,\xpom,Q^2)$, is related to the diffractive cross section for $W^2 \gg Q^2$ as follows:
\begin{eqnarray}
\frac{1}{2M_X} \frac{d\sigma^{\rm diff}_{\gamma^*p \to XN}(M_X,W,Q^2)}{dM_X} = \frac{4 \pi^2 \alpha}{Q^2(Q^2+M^2_X)} \xpom F^{D(3)}_2(\beta,\xpom,Q^2).
\end{eqnarray}
If $F^{D(3)}_2$ is interpreted in terms of quark densities, it specifies the probability to find, in a proton undergoing a diffractive reaction, a quark carrying a fraction $x = \beta \xpom$ of the proton momentum.

The measurements of $\xpom F^{D(3)}_2$ are given in Tables~\ref{t:f2d301} -~\ref{t:f2d304} as a function of $\beta, \xpom$ and $Q^2$. Figure~\ref{f:f2d3} shows $\xpom F^{D(3)}_2$ as a function of $\xpom$ for different values of $\beta$ and $Q^2$.

\subsection{Comparison with other measurements}
\label{sec-compother}
The measurements of $\xpom F^{D(3)}_2$ obtained from this analysis for $M_N < 2.3$ GeV are consistent with those determined previously by this experiment with the $M_X$ method for $M_N < 5.5$ GeV~\cite{epj:c6:43}.

The measurements of $\xpom F^{D(3)}_2$ from this analysis (FPC) have been compared with those from this experiment determined with the leading proton spectrometer (LPS)~\cite{desy-04-131} and from the H1 experiment~\cite{zfp:c76:613}. Since the three analyses quote the values of $\xpom F^{D(3)}_2$ at different $(\beta, Q^2)$ points, the values of this analysis (FPC) were transported to the $(\beta, Q^2)$ points of the other measurements using the BEKW(mod) fit, provided that the $(\beta, Q^2)$ values of the corresponding FPC measurement satisfied the conditions $0.8 < Q^2_{FPC}/Q^2 < 1.2$, $|\beta_{FPC} - \beta|/\beta < 0.5$. 
The LPS analysis measures the reaction $\gamma^* p \to Xp$ while the FPC analysis includes the contribution from proton dissociation, $\gamma^* p \to XN$, $1.08 < M_N < 2.3$ GeV. The LPS data for $x_{\pom} < 0.005$ - where Reggeon contributions to the LPS data are negligible - were used to estimate the fractional contribution $f_{\rm pdissoc}$ from proton dissociation to the FPC results, assuming $f_{\rm pdissoc}$ to be independent of $\beta, x_{\pom}$ and $Q^2$. The relative normalisation of the two data sets was determined using the result of the BEKW(mod) fit to the FPC data. This yielded $1 - f_{\rm pdissoc} = 0.70 \pm 0.03$, which shows that about 30$\%$ of the diffractive cross section in the FPC analysis comes from nucleon dissociation with masses $M_N$ between 1.08 and 2.3 GeV. Figure~\ref{f:f2d3fpclps} shows the LPS data together with the FPC data multiplied by a factor of 0.70. The LPS data for $x_{\pom} < 0.01$ agree well with those of the current analysis.

Figure~\ref{f:f2d3fpch1} shows a comparison of the measurements of this analysis with that of the H1 collaboration~\cite{zfp:c76:613} which includes the contribution from nucleon dissociation for $M_N < 1.6$ GeV. No correction was applied to the FPC data to account for the possible difference in $\xpom F^{D(3)}_2$ for $M_N < 2.3$ GeV (this analysis) and $M_N < 1.6$ GeV (H1 analysis). Qualitative agreement between the present data and the H1 measurements is observed, with the possible exception of the region of $\xpom > 0.01$, where the  H1 data include contributions from Reggeon exchange.

\subsection{Discussion of the $\xpom F^{D(3)}_2$ results from this analysis}
The diffractive structure function presented in Fig.~\ref{f:f2d3} is a function of $\xpom$ for fixed $M_X$ (or, equivalently fixed $\beta$) and $Q^2$. For the lowest $M_X$ region - which corresponds to large $\beta$ values - little dependence on $\xpom$ is observed. This is in contrast to the regions with smaller $\beta$ where $\xpom F^{D(3)}_2$ rises strongly as $\xpom \to 0$, reflecting the rapid increase of the diffractive cross section $d\sigma^{\rm diff}/dM_X$ with $W$ for $M_X > 2$ GeV.

For the following study of $\xpom F^{D(3)}_2$ as a function of $\xpom$ and $\beta$ for fixed values of $(\xpom, Q^2)$ and $(\beta, Q^2)$, respectively, bin centering was done by using the BEKW(mod) fit (Section~\ref{sec-bekw}).

The $Q^2$ dependence of $\xpom F^{D(3)}_2$ is displayed in Fig.~\ref{f:f2d3vsq2} for different values of $\beta$ and $\xpom$. Different regions in the $\beta - \xpom$  space show markedly different behaviours with $Q^2$. For $\beta = 0.9$, the region dominated by diffractive production of states with $M_X < 2$ GeV, $\xpom F^{D(3)}_2$ is constant or slowly decreasing with $Q^2$. For $\beta \le 0.7$, $\xpom F^{D(3)}_2$ increases with increasing $Q^2$ provided $\beta \xpom < 2 \cdot 10^{-3}$. This $Q^2$ dependence is similar to the scaling violations of the proton structure function $F_2$. By noting that $\beta \xpom = x$, it can be seen from Fig.~\ref{f:f2d3vsq2} that the behaviour of the scaling violations with $Q^2$ depends primarily on $x$ rather than on $\beta$ and $\xpom$ separately. Disregarding the region $\beta = 0.9$, positive scaling violations dominate at low $x < 0.002$. The fact that the $Q^2$ dependence of $\xpom F^{D(3)}_2$ for fixed $\beta$ changes with $\xpom$ shows again that the data are inconsistent with the Regge-factorisation hypothesis, a concept which implies that, for given $\beta$ and $Q^2$, the same Pomeron structure is probed, independently of $\xpom$.
 
Figure~\ref{f:f2d3vsbeta} shows $\xpom F^{D(3)}_2$ as a function of $\beta$ for fixed $(\xpom, Q^2)$. For those $(\xpom, Q^2)$ values where the measurements cover a wide range in $\beta$, $\xpom F^{D(3)}_2$ is observed to have a broad maximum around $\beta = 0.5$, a dip near $\beta = 0.1$ and a rise as $\beta \to 0$.

The data of Fig.~\ref{f:f2d3vsbeta} can be better visualized by plotting them for fixed $\xpom$. Figure~\ref{f:f2d201} shows the results obtained by using the $\xpom F^{D(3)}_2$ measurements with $0.5 x_0 < \xpom < 1.5 x_0$, where $x_0 = 0.01$. For each measurement, the $\xpom F^{D(3)}_2$ value measured at $\xpom_{\rm meas}$ was transported to $\xpom = x_0$ using the BEKW(mod) fit. On average, the difference between measured and transported $x_0 F^{D(3)}_2(\beta,x_0,Q^2)$ value was of the order of 5\%. Finally, for every ($\beta,Q^2$) point, the weighted average of the selected measurements was made.  

In a model where diffraction proceeds by the exchange of a Pomeron, the diffractive structure function factorises into the flux of Pomerons and the structure function of the Pomeron, $\xpom F^{D(3)}_2(\beta,\xpom,Q^2) = \Phi(\xpom) \cdot F^{\pom}_2(\beta,Q^2)$. Up to a normalisation constant, $x_0 F^{D(3)}_2(\beta,x_0,Q^2)$, would represent the structure function of the Pomeron, $F^{\pom}_2(\beta,Q^2) = x_0 F^{D(3)}_2(\beta,x_0,Q^2)$. In such a model, however,  the flux is independent of $Q^2$, which is at variance with the data from this analysis, as shown in Fig.~\ref{f:f2d3vsq2}. 

The  resulting measurements of $x_0 F^{D(3)}_2(\beta,x_0,Q^2)$ are presented in Table~\ref{t:f2d2tab} and Fig.~\ref{f:f2d201}. Several aspects are noteworthy. Firstly, $x_0 F^{D(3)}_2(\beta,x_0,Q^2)$ has a maximum near $\beta = 0.5$, consistent with a $\beta(1-\beta)$ variation. Secondly, in the region of high $\beta$, $x_0 F^{D(3)}_2(\beta,x_0,Q^2)$ tends to decrease as $Q^2$ increases from 14 to 27 GeV$^2$. Finally, for $\beta < 0.1$, $x_0 F^{D(3)}_2$ rises as $\beta \to 0$, the rise increasing with increasing $Q^2$. 

The $\beta (1-\beta)$ dependence is explained in dipole models of diffraction by $\gamma^* \to q \overline{q}$ splitting{~\cite{proc:ep:1971,*pr:d3:1382,*pr:d8:1341,zfp:c53:331,*jetp:81:625,*pl:b422:238,pr:d59:014017,*pr:d60:114023,np:b335:115} and two gluon exchange. The rise of $x_0 F^{D(3)}_2(\beta,x_0,Q^2)$ as $\beta \to 0$ and its increase as $Q^2$ increases is reminiscent of the logarithmic scaling violations of the proton structure function $F_2$ at low $x$, which are ascribed to the contribution from the sea.

The data are consistent with the idea that diffractive DIS probes the diffractive PDFs of the proton; their dependence on $\beta$ and $Q^2$ is similar to the different $Q^2$ dependence of the proton inclusive PDFs at different values of $x$~\cite{hep-ph-0406225}. 
The positive scaling violations observed for $x = \beta \xpom < 2 \times 10^{-3}$ suggest substantial perturbative effects such as gluon emission. 

\subsection{Comparison with the BEKW model}
\label{sec-bekw}
The BEKW model~\cite{ejp:c7:443} provides a general parametrisation for inclusive diffraction in DIS and allows the identification of certain subprocesses by their characteristic behaviour in $\beta$ and $Q^2$. In the model, the incoming virtual photon fluctuates into a $q\overline{q}$ or $q\overline{q}g$ dipole which interacts with the target proton via two-gluon exchange. The $\beta$ spectrum and the scaling behaviour in $Q^2$ are derived from the wave functions of the incoming transverse (T) or longitudinal (L) photon on the light cone in the non-perturbative limit. The $\xpom$ dependence of the cross section is not predicted by the model but is to be determined by experiment. Specifically
\begin{eqnarray}
\xpom F^{D(3)}_2(\beta,\xpom,Q^2) & = & c_T \cdot F^T_{q\overline{q}} + c_L \cdot F^L_{q\overline{q}} + c_g \cdot F^T_{q\overline{q}g},
\label{eq:bekw}
\end{eqnarray}
where 
\begin{eqnarray}
F^T_{q\overline{q}} & = & (\frac{x_0}{\xpom})^{n_T(Q^2)} \cdot \beta(1 - \beta), \\
\label{eq:bekwqqT}
F^L_{q\overline{q}} & = & (\frac{x_0}{\xpom})^{n_L(Q^2)} \cdot \frac{Q^2_0}{Q^2+Q^2_0} \cdot [\ln(\frac{7}{4} + \frac{Q^2}{4 \beta Q^2_0})]^2 \cdot \beta^3 (1 - 2\beta)^2, \\
\label{eq:bekwqqL}
F^T_{q\overline{q}g} & = & (\frac{x_0}{\xpom})^{n_g(Q^2)} \cdot \ln(1+\frac{Q^2}{Q^2_0})\cdot (1-\beta)^{\gamma}.
\label{eq:bekwqqg}
\end{eqnarray}
The contribution from longitudinal photons coupling to $q \overline{q}$ is limited to $\beta$ values close to unity. The $q \overline{q}$ contribution from transverse photons is expected to have a broad maximum around $\beta = 0.5$, while the $q \overline{q} g$ contribution becomes important at small $\beta$, provided $\gamma$ is large. The original BEKW model also includes a higher-twist term for $q \overline{q}$ produced by transverse photons. The present data are insensitive to this term, and therefore it has been neglected.

For $F^{L}_{q \overline{q}}$, the term $(\frac{Q^2_0}{Q^2})$ provided by BEKW was replaced by the factor $(\frac{Q^2_0}{Q^2+Q^2_0})$ to avoid problems as $Q^2 \to 0$. The powers $n_{T,L,g}(Q^2)$ were assumed by BEKW to be of the form 
$n(Q^2) =  n_0 + n_1 \cdot \ln [1 + \ln(\frac{Q^2}{Q_0^2})]$. The rise of $\alpha_{\pom}(0)$ with $\ln Q^2$ observed in the present data suggested using the form $n(Q^2) = n_0 + n_1\ln(1+\frac{Q^2}{Q_0^2})$. The modified BEKW form will be referred to as BEKW(mod). Taking $x_0 = 0.01$ and $Q^2_0 = 0.4$ GeV$^2$, the BEKW(mod) form gives a good description of the data, viz. $\chi^2 = 112$ for 188 ${\rm dof}$. According to the fit, all the coefficients $n_0$ and $n_1$ for the longitudinal component can be set to zero, and the powers $n_T$, $n_g$ are the same, within errors, for the $q\overline{q}$ and $q\overline{q}g$ components produced by transverse photons. This leads to:
$c_T = 0.112 \pm 0.003$, $c_L = 0.154 \pm 0.012$, $c_g = 0.0091 \pm 0.0003$, $n_1^{T,g} = 0.067 \pm 0.004$ and $\gamma = 8.62 \pm 0.55$
with $\chi^2$ = 114 for 193 ${\rm dof}$. The value of the power $\gamma$ is considerably larger than the value of about three expected by BEKW. Results from a similar analysis of the LPS data can be found elsewhere~\cite{desy-04-131}.

Figures~\ref{f:f2d3vsq2},~\ref{f:f2d3vsbeta} and~\ref{f:f2d3.bekw} compare the measurement of $\xpom F^{D(3)}_2(\beta,\xpom,Q^2)$ as a function of $\xpom$, $\beta$ and $Q^2$ with the BEKW(mod) fit. The fit describes the data well. The weak rise of $\xpom F^{D(3)}_2$ as $\xpom \to 0$ observed for $M_X = 1.2$ GeV, and the strong rise for $M_X \ge 3$ GeV, are explained by the BEKW fit as follows: the high $\beta$ region ($\beta > 0.9$) receives substantial contributions from longitudinal photons with a weak dependence on $\xpom$, while transverse photons dominate at lower $\beta$ and lead to a strong rise as $\xpom \to 0$. The observed increase of the rise of $\xpom F^{D(3)}_2$ as $\xpom \to 0$ with increasing $Q^2$ is accommodated in the model by assuming that the power $n_T(Q^2)$ increases with $Q^2$. In the BEKW model\footnote{Although the BEKW(mod) fit gives an excellent description of the data from this analysis, its prediction for the contribution from longitudinal photons at low $M_X$ and low $Q^2$ is at variance with existing data on vector meson production. For $M_X < 2$ GeV, the contribution from longitudinal photons accounts for at least $\approx 20\%$ of $x_{\pom}F^{D(3)}_2$. The BEKW(mod) fit curves for the longitudinal photon contribution at $M_X < 2$ GeV (dotted lines in Fig.~\ref{f:f2d3.bekw}) are in broad agreement with the data for $Q^2 \ge 6$ GeV$^2$, but are too low at lower $Q^2$.}, the broad maximum seen in the $\beta$ distribution around $\beta = 0.5$ is a result of the dominance of the $q \overline{q}$ configuration at medium $\beta$, and the rise towards small $\beta$ is a result of the $(q \overline{q} g)$ configuration. 
The good agreement of the BEKW fit with the data for $M_X > 2$ GeV lends strong support to the dipole picture.

\section{Conclusions}
A simultaneous measurement of the proton structure function $F_2$, the diffractive $\gamma^*p$ cross section and the diffractive structure function has been made. The kinematic range of the measurement was $2.2 < Q^2 < 80$ GeV$^2$, $37 < W < 245$ GeV and $0.28 < M_X < 35$ GeV. The forward plug calorimeter (FPC) was used to extend the range of $M_X$ compared to previous measurements. The $M_X$ method was used to extract the diffractive cross section; the method is shown to exclude non-peripheral as well as Reggeon contributions to the cross section.

The results for the proton structure function $F_2(x,Q^2)$ are in good agreement with previous measurements of the ZEUS collaboration. The $F_2$ data were analysed in the framework of Regge phenomenology. The intercept of the Pomeron trajectory of these data is significantly higher than that measured in hadron-hadron collisions (`soft Pomeron', $\alpha_{\pom}(0) = 1.096^{+0.012}_{-0.009}$) and is a strong function of $Q^2$: $\alpha^{\rm tot}_{\pom}(0) = 1.155 \pm 0.011 ({\rm stat.})^{+0.007}_{-0.011}({\rm syst.})$ at $Q^2 = 2.7$ GeV$^2$ and $\alpha^{\rm tot}_{\pom}(0) = 1.307 \pm 0.019 ({\rm stat.})^{+0.027}_{-0.037}({\rm syst.})$ at $Q^2 = 55$ GeV$^2$. The $Q^2$ dependence of the Pomeron intercept corresponds to the rise of $F_2$ towards low $x$, which increases with $Q^2$ and which has been observed previously at HERA. In a Regge approach, this shows that $F_2$, and the total $\gamma^*p$ cross section, cannot be interpreted in terms of the exchange of a single Pomeron combined with the assumption of Regge factorisation.

The measured diffractive DIS cross section is in good agreement with previous ZEUS measurements when proton dissociation is taken into account. The diffractive cross section for $0.28 < M_X < 2$ GeV shows a weak dependence on $W$ but a much stronger decrease than $1/Q^2$, characteristic of a higher twist behaviour. For $M_X > 8$ GeV, the cross section decreases as $1/Q^2$, indicating a leading twist behaviour. The diffractive cross section was also analysed in terms of Regge phenomenology. The excess of the intercept of the Pomeron trajectory above unity is about half of that extracted from the $F_2$ data, but still significantly higher than that of the soft Pomeron. The Pomeron intercept rises by $\Delta\alpha^{\rm diff}_{\pom} = 0.0714\pm 0.0140({\rm stat.})^{+0.0047}_{-0.0100}({\rm syst.})$ between $Q^2$ of 7.8 and 27 GeV$^2$. This establishes a $Q^2$ dependence of the Pomeron intercept and shows that the diffractive DIS as well as the inclusive DIS cross sections cannot be interpreted as resulting from single Pomeron exchange combined with the assumption of Regge factorisation.

The ratio of the diffractive to the total $\gamma^*p$ cross section was studied. For fixed $M_X$ and $Q^2$, the ratio is flat as a function of $W$ in the kinematic range of these measurements. For $0.28 < M_X < 35$ GeV, $W = 220$ GeV, the ratio is $15.8 ^{+1.2}_{-1.0}$ \% at $Q^2 = 4$ GeV$^2$ and $9.6^{+0.7}_{-0.7}$\% at $Q^2$ = 27 GeV$^2$.

The diffractive cross section was also analysed in terms of the diffractive structure function of the proton $F^{D(3)}_2(\beta,\xpom,Q^2)$. The $\beta$ and $Q^2$ dependences of $\xpom F^{D(3)}_2$ are a function of $\xpom$; this is expected in view of the Regge factorisation breaking observed for the diffractive cross section. The pattern of scaling violations depends primarily on the proton variables $x$ and $Q^2$; for $\beta < 0.9$, positive scaling violations are observed when $x < 0.002$. The analysis of $\xpom F^{D(3)}_2(\beta,\xpom,Q^2)$ for $\xpom = x_0 = 0.01$ exhibits several remarkable properties: $x_0 F^{D(3)}_2(\beta,x_0,Q^2)$ shows a broad maximum near $\beta = 0.5$ consistent with a $\beta (1-\beta)$ variation as expected by dipole models for $\gamma^* \to q \overline{q}$ splitting; for $\beta < 0.1$, $x_0 F^{D(3)}_2$ rises as $\beta \to 0$, the rise increasing with increasing $Q^2$. The positive scaling violations observed for $\beta < 0.1$ suggest that diffraction in DIS receives substantial contributions from perturbative effects.  

The results of this paper show that Regge phenomenology cannot give a good description of the diffractive and total DIS cross section without extensive modifications that would undermine the simplicity of the Regge approach. The large fraction of the DIS cross section that is diffractive even at high $Q^2$, and the leading twist nature of the diffractive cross section at higher $M_X$ may mean that some assumptions~\cite{hep-ph-0406225} inherent in the DGLAP analysis of the structure function $F_2$ need to be reexamined.
\\

{\Large \bf Acknowledgements}

We thank the DESY directorate for their strong support and encouragement. The effort of the HERA machine group is gratefully acknowledged. We thank the DESY computing and network services for their support. The construction, testing and installation of the ZEUS detector have been possible by the effort of many people not listed as authors. We would like to thank, in particular, J. Hauschildt and K. L\"{o}ffler (DESY), R. Feller, E. M\"{o}ller and H. Prause (I. Inst. for Exp. Phys., Univ.  Hamburg), A. Maniatis (II. Inst. for Exp. Phys., Univ. Hamburg), N. Wilfert and the members of the mechanical workshop of the Faculty of Physics from the Freiburg University for their work on the FPC. We would like to thank also L. Lindemann for his contributions at the start of this analysis.





\appendix
{

\pagestyle{plain}
{\Large\bf Appendix}

\section{Reconstruction of $W$ with the weighting method}
\label{asec-reconw}
The value of $W$ is reconstructed using the weighted average of the values determined from the electron and the hadron measurements (the `weighting method') denoted by $W_W$. The electron kinematics yields $W_e$:
\begin{eqnarray}
W^2_e & = & (P+k-k^{\prime})^2  \nonumber \\
   & = & 2 P(k-k^{\prime}) + P^2 + (k-k^{\prime})^2  \nonumber \\
   & = & 2 E_p[2E_e -E^{\prime}_e(1 - \cos \theta_e)] +m^2_p -Q^2, \nonumber
\end{eqnarray}
where $m_p$ is the mass of the proton. The measurement accuracy of $W_e$, $\Delta W_e$, depends on the uncertainties with which $E^{\prime}_e$ and $\theta_e$ are measured,
\begin{eqnarray}
\Delta W^2_e = -[2 E_p(1- \cos \theta_e) + 2 E_e (1 + \cos \theta_e)] \Delta E^{\prime}_e + 2(E_p E^{\prime}_e - E_e E^{\prime}_e) \Delta \cos \theta_e \;. \nonumber
\end{eqnarray}
Monte Carlo (MC) studies yielded for the resolutions $\Delta E^{\prime}_e \approx 5/E^{\prime}_e \oplus 0.08$ and $\Delta \theta_e  \approx  0.007$,
where the energies are given in units of GeV and $\Delta \theta_e$ in radians.
The dominant contribution to $\Delta W_e$ comes from the uncertainty in the measurement of the electron energy. This can be serious for low values of $W$ ($W < 60$ GeV) where $E^{\prime}_e$ is close to $E_e$: for instance, at low $Q^2$ and when the measured $E^{\prime}_e > E_e$, $W^2_e$ becomes negative.

The energies $E_h$ and production angles $\theta_h$ of the EFOs provide the hadronic measurement of $W$:
\begin{eqnarray}
W^2_h & \approx  & 2E_p \sum_h E_h(1 - \cos \theta_h), \nonumber
\end{eqnarray}
with an uncertainty of
\begin{eqnarray}
\Delta W^2_h & = & 2 E_P \sum_h (1- \cos \theta_h)\Delta E_h + 2 E_P \sum_h E_h \sin \theta_h \Delta \theta_h    . \nonumber
\end{eqnarray}
The summation is performed over all hadronic EFOs. For the hadronic measurement, the MC simulation yields $\Delta E_h \approx 0.8 \sqrt{E_h} \oplus 0.04 E_h$ and $\Delta \theta_h  \approx  0.07$.
The uncertainty results largely from fluctuations of the energy loss in the material ahead of CAL, and from neutrinos and muons produced in the final state.

Using the MC to estimate the errors shows that at low $W$, where $W_e$ provides a poor measurement of $W$, the value of $W_h$ is rather precise, while the opposite is true for high values of $W$. The weighting method combines the two measurements for $W$ by weighting them with the inverse of the squares of their estimated errors, $g_e = \frac{W^2}{\Delta W^2_e}$ and $g_h = \frac{W^2}{\Delta W^2_h}$:
\begin{eqnarray}
W^2_W & = & \frac{g^2_e W^2_e + g^2_h W^2_h}{g^2_e + g^2_h}. \nonumber
\end{eqnarray}
In order to arrive at reliable estimates for $g_e, g_h$, it is essential to have an estimate for $W$. This is achieved with the double-angle (DA) measurement~\cite{proc:HERA:1991:23,*add:1991:43} which relies only on the measurement of the angles of the scattered electron and of the hadronic system.

\section{Extraction of the diffractive contribution with the $M_X$ method}
\label{asec-lnmx}
In non-peripheral DIS, the incident proton is broken up and the remnant is a coloured object. This gives rise to a substantial amount of initial- and final-state radiation, populating the region between the incident proton and the current jet. The scaling of the position of the maximum and the exponential fall-off of the $\ln M^2_X$ distribution follow from the assumption of uniform, uncorrelated particle emission in rapidity (${\cal Y} = \frac{1}{2} \ln \frac{E+P_L}{E-P_L}$, where $E,P_L$ are the energy and longitudinal momentum of the particle) along the beam axis in the $\gamma^* p$ system~\cite{feynman:1972:photon,zfp:c70:391}. For an (idealized) uniform $\cal Y$ distribution between maximum and minimum rapidities of ${\cal Y}_{\rm max}$ and ${\cal Y}_{\rm min}$, respectively, the total c.m. energy $W$ is given by
\begin{eqnarray}
W^2 \approx c_0 \cdot \exp({\cal Y}_{\rm max} - {\cal Y}_{\rm min}), \nonumber
\end{eqnarray}
assuming $({\cal Y}_{\rm max} - {\cal Y}_{\rm min}) \gg 1$.
Here, $c_o$ is a constant. The mass $M_X$ of the particle system that can be observed in the detector is reduced by the loss of particles (mainly) through the forward beam hole: 
\begin{eqnarray}
M^2_X \approx c_0 \cdot \exp({\cal Y}^{\rm det}_{\rm limit} - {\cal Y}_{\rm min}) \approx W ^2 \cdot \exp({\cal Y}^{\rm det}_{\rm limit} - {\cal Y}_{\rm max}),
\label{eq:lnmx2andy}
\end{eqnarray}
where ${\cal Y}^{\rm det}_{\rm limit}$ denotes the limit of the calorimetric acceptance in the forward direction. Equation~(\ref{eq:lnmx2andy}) predicts scaling of the $\ln M^2_X$ distribution when plotted as function of\\
 $\ln (M^2_X/W^2)$, in agreement with the behaviour of the data.

The value of $M_X$ will fluctuate due to a finite probability $P(\Delta {\cal Y})$ that no particles are emitted between ${\cal Y}^{\rm det}_{\rm limit}$ and ${\cal Y}^{\rm det}_{\rm limit} - \Delta{\cal Y}$. This generates a gap of size $\Delta{\cal Y}$. The assumption of uncorrelated particle emission leads to a Poissonian rapidity gap distribution, $P(\Delta {\cal Y}) = \exp(-\lambda \Delta {\cal Y})$, resulting in an exponential fall-off of the $\ln M^2_X$ distribution,
\begin{eqnarray}
\frac{d{\cal N}^{\rm non-diff}}{d \ln M^2_X} = c \cdot \exp(b \cdot \ln M^2_X),
\end{eqnarray}
where the slope $b$ and the parameter $c$ can be determined from the data. The exponential fall-off of the $\ln M^2_X$ distribution towards small values of $\ln M^2_X$ is indeed found for models which include QCD leading-order matrix elements, parton showers and fragmentation, such as DJANGOH (see shaded area in Fig.~\ref{f:lnmxsel}). The exponential fall-off holds for $\ln M^2_X \le \ln W^2 -\eta_0$ over two units of rapidity, where $\eta_0 \approx 2$.

In the $M_X$ method, the diffractive contribution is identified as the excess of events towards small $M_X$ above the exponential fall-off of the non-diffractive contribution. In a Triple Regge model~\cite{pr:d2:2963,pr:d4:150,np:80:367}, the diffractive cross section is approximately of the form
\begin{eqnarray}
\frac{d\sigma^{\rm diff}_{\gamma^* p\to XN}}{d \ln M^2_X} & \propto & \exp[(1 + \alpha_k(0)-2\overline{\alpha_j}) \cdot \ln M^2_X]. \nonumber
\end{eqnarray}
Here, $\overline{\alpha_j}$ is the trajectory exchanged in the $t$ channel between the incoming proton and the outgoing system $N$, averaged over the $t$ distribution, as seen in Fig.~\ref{f:gptripreg}. The parameter $\alpha_k(0)$ is the intercept of the trajectory describing the production of the system $X$ by the scattering of $\gamma^*$ on a Regge-pole with $t$ averaged intercept $\overline{\alpha_j}$. For large $M_X$, $\alpha_k(0)$ is expected to be 1. Pomeron exchange in the $t$-channel with $\overline{\alpha_j} \approx 1$ leads then to $1 + \alpha_k(0)-2\overline{\alpha_j} = 0$ and to a constant $\ln M^2_X$ spectrum: 
\begin{eqnarray}
\frac{d\sigma^{\rm diff}_{\gamma^* p \to XN}}{d \ln M^2_X} & = & {\rm constant}. \nonumber
\end{eqnarray}

If, instead of the Pomeron, the highest-lying Reggeon trajectory ($\overline{\alpha_j} \approx 0.5$)  is exchanged in the $t$-channel, then the $\ln M^2_X$ spectrum for this contribution rises exponentially towards large $\ln M^2_X$:    
\begin{eqnarray}
\frac{d\sigma^{\rm diff}_{\gamma^* p \to XN}}{d \ln M^2_X} & \propto & \exp(b_{\reg} \cdot \ln M^2_X),    \nonumber
\end{eqnarray}
with $b_{\reg} = 1$. Hence, Reggeon exchange in the $t$ channel leads to an exponential rise of the $\ln M^2_X$ distribution. Note also that lower-lying Regge trajectories produce an even larger exponential slope $b_{\reg}$. Therefore, identifying the diffractive contribution as the excess of events above the exponential fall-off of the $\ln M^2_X$-distribution suppresses not only the non-diffractive contribution arising from colour exchange but also the contributions from Reggeon exchange.

\section{Reggeon contribution to the $\ln M^2_X$ spectrum}
\label{asec-lnmxreggeon}
The recent ZEUS measurement of diffraction in DIS with the LPS~\cite{desy-04-131} allows the Reggeon exchange contribution to the reaction $\gamma^*p \to Xp$ to be estimated. In the LPS analysis, the diffractive structure function of the proton, $\xpom F^{D(3)}_2(\beta, \xpom, Q^2)$, shows a rise towards small $\xpom$, and a rise towards large $\xpom$, with a minimum near $\xpom = 0.01 - 0.02$. The rise towards large $\xpom$ is indicative of a Reggeon contribution. The LPS data were used to estimate the size of the Reggeon contribution by fitting them to a sum of Pomeron and Reggeon contributions, assuming Regge factorisation: 
\begin{eqnarray}
x_{\pom} F^{D(3)}_2(\beta,x_{\pom},Q^2) = c_{\pom} \cdot x_{\pom} F^{D(3)\pom}_2(\beta,x_{\pom},Q^2) + c_{\reg} \cdot x_{\pom} F^{D(3) \reg}_2(\beta,x_{\pom},Q^2).
\label{eq:regpom}
\end{eqnarray} 
The Pomeron contribution ($\pom$) was taken to equal the result of the BEKW(mod) fit to the FPC data multiplied by the factor $c_{\pom}$ which accounts for the fact that the LPS data do not include proton dissociation. The LPS data for $\xpom < 0.005$ yielded $c_{\pom} = 0.70 \pm 0.03$. For the Reggeon contribution ($\reg$), the following ansatz was made :
\begin{eqnarray}
\xpom F^{D(3) \reg}_2(\beta,\xpom,Q^2,t) = \frac{\xpom \cdot e^{B_{\reg} t}}{\xpom^{2\alpha_{\reg}(t) -1}}F^{D(2)\reg}_2(\beta,Q^2). \nonumber
\end{eqnarray}
Defining
\begin{eqnarray}
g_{\reg}(\xpom) = \int^{t_{\rm min}}_{t_{\rm max}} dt \frac{e^{B_{\reg} t}}{\xpom^{2\alpha_{\reg}(t) -2}}, \nonumber
\end{eqnarray}
taking $t_{\rm min} = 0$, $t_{\rm max} = 1$ GeV$^2$ and $\alpha_{\reg}(t) = \alpha_{\reg}(0)+\alpha^{\prime}_{\reg}\cdot t$ \nonumber
leads to 
\begin{eqnarray}
g_{\reg}(\xpom) = \frac{1}{(B_{\reg} - 2\alpha^{\prime}_{\reg} \cdot \ln \xpom) \cdot  \xpom^{(2 \alpha_{\reg}(0) -2)}} . \nonumber
\label{eqngreg}
\end{eqnarray}
and to
\begin{eqnarray}
\xpom F^{D(3)\reg}_2(\beta,\xpom,Q^2) = g_{\reg}(\xpom) \cdot F^{D(2)\reg}_2(\beta,Q^2) . \nonumber
\end{eqnarray}
Following H1~\cite{zfp:c76:613}, the Reggeon parameters were assumed to be: $\alpha_{\reg}(0) = 0.55$, $\alpha^{\prime}_{\reg} = 0.9$ GeV$^{-2}$ and $B_{\reg} = 2$ GeV$^{-2}$. While the LPS data are the most precise information on the Reggeon contribution available, the data are still too sparse to effectively constrain the parameter $c_R$ in a fit to Eq.~\ref{eq:regpom}. In order to obtain a rough estimation needed for this study, the assumption $\xpom F^{D(3)\reg}_2 ({\rm at \; } \xpom = 0.06) \approx x_{\pom} F^{D(3)\pom}_2 ({\rm at \;} \xpom = 0.002)$ independent of $\beta$ and $Q^2$ was made (see Fig.~\ref{f:f2d3fpclps}). This allowed the determination, $c_R = 0.39$; the $\chi^2/{\rm dof}$ of the resulting description of Eq.~\ref{eq:regpom} to the LPS data was 89/78.   

The Reggeon contribution extracted from the LPS data was multiplied by a factor of $1/c_{\pom} = 1.43$. This factor accounts for the extra contribution from proton dissociation in the present analysis. The contribution from charged isovector Reggeons, which cannot contribute to the LPS data, was assumed to be negligible~\cite{pr:d56:3955}. 

The relation between $F^{D(3)\reg}_2$ and the $\ln M^2_X$ spectrum from the present analysis is given by,
\begin{eqnarray}
\frac{d\sigma^{\reg}_{\gamma^* p\to XN}} {d\ln M^2_X} = 4 \pi^2 \alpha \frac{M^2_X}{Q^2(Q^2+M^2_X)} \cdot c_{\reg}/c_{\pom} \cdot \xpom F^{D(3)\reg}(\beta,\xpom,Q^2). \nonumber
\end{eqnarray}

Figure~\ref{f:lnmxselreggeon} compares the distribution of $\ln M^2_X$ for the lowest and highest $W$ bins at low and high $Q^2$ for the data together with the expectations from Reggeon exchange which lies below the total non-diffractive contribution predicted by the fit to the $\ln M^2_X$ distributions.

} 
\vfill\eject

{
\def\bibname{\Large\bf References}
\def\refname{\Large\bf References}
\pagestyle{plain}
\ifzeusbst
  \bibliographystyle{l4z_default}
\fi
\ifzdrftbst
  \bibliographystyle{l4z_draft}
\fi
\ifzbstepj
  \bibliographystyle{l4z_epj}
\fi
\ifzbstnp
  \bibliographystyle{l4z_np}
\fi
\ifzbstpl
  \bibliographystyle{l4z_pl}
\fi
{\raggedright
\bibliography{DESY-05-011-ref.bib}}
}
\vfill\eject

\providecommand{\etal}{et al.\xspace}
\providecommand{\coll}{Coll.\xspace}
\catcode`\@=11
\def\@bibitem#1{%
\ifmc@bstsupport
  \mc@iftail{#1}%
    {;\newline\ignorespaces}%
    {\ifmc@first\else.\fi\orig@bibitem{#1}}
  \mc@firstfalse
\else
  \mc@iftail{#1}%
    {\ignorespaces}%
    {\orig@bibitem{#1}}%
\fi}%
\catcode`\@=12

\begin{table}[p]
\begin{center}
\begin{tabular}{||c|c|c|c|c|c|c|c||}
\hline
$Q^2$ (GeV$^2$) & 2.2 - 3 & 3 - 5 & 5 - 7 & 7 - 10 & 10 - 20 & 20 - 40 & 40 - 80 \\
$Q^2_{\rm ref}$ (GeV$^2$) & 2.7 & 4 & 6 & 8 & 14 & 27 & 55 \\
\hline
$W$ (GeV) & 37 - 55 & 55 - 74 & 74 - 99 & 99 - 134 & 134 - 164 & 164 - 200 & 200 - 245 \\
$W_{\rm ref}$ (GeV) & 45 & 65 & 85 & 115 & 150 & 180 & 220 \\
\hline
$M_X$ (GeV) & 0.28 - 2 & 2 - 4 & 4 - 8 & 8 - 15 & 15 - 25 & 25 - 35 & \\         $M_{X {\rm ref}}$ (GeV) & 1.2 & 3 & 6 & 11 & 20 & 30 & \\
\hline      
\end{tabular}
\caption{Binning and reference values for $M_X$, $W$ and $Q^2$.}
\label{t:binning}
\end{center}
\end{table}
\clearpage

\begin{table}[p]
\begin{center}
\begin{tabular}{||ccc|ccc||}
\hline
$Q^2$ (GeV$^2$) & $x$ &  $F_2\pm \rm stat. \pm syst.$  & $Q^2$ (GeV$^2$) & $x$ &  $F_2 \pm \rm stat. \pm syst.$ \\
\hline\hline
  2.7 & 0.001332& 0.574$\pm 0.017^{+ 0.044}_{- 0.041}$ &14.0 & 0.006869& 0.656$\pm 0.008^{+ 0.015}_{- 0.017}$\\
  2.7 & 0.000639& 0.668$\pm 0.022^{+ 0.046}_{- 0.040}$ &14.0 & 0.003303& 0.802$\pm 0.011^{+ 0.027}_{- 0.031}$\\
  2.7 & 0.000374& 0.706$\pm 0.021^{+ 0.038}_{- 0.038}$ &14.0 & 0.001934& 0.887$\pm 0.011^{+ 0.075}_{- 0.065}$\\
  2.7 & 0.000204& 0.747$\pm 0.019^{+ 0.047}_{- 0.039}$ &14.0 & 0.001058& 1.026$\pm 0.013^{+ 0.028}_{- 0.034}$\\
  2.7 & 0.000120& 0.827$\pm 0.023^{+ 0.045}_{- 0.040}$ &14.0 & 0.000622& 1.151$\pm 0.017^{+ 0.012}_{- 0.015}$\\
  2.7 & 0.000083& 0.888$\pm 0.024^{+ 0.041}_{- 0.048}$ &14.0 & 0.000432& 1.257$\pm 0.019^{+ 0.018}_{- 0.023}$\\
  2.7 & 0.000056& 0.965$\pm 0.028^{+ 0.042}_{- 0.035}$ &14.0 & 0.000289& 1.398$\pm 0.023^{+ 0.020}_{- 0.033}$\\
\end{tabular}
\caption{Proton structure function $F_2$.}
\label{t:f2tab}
\end{center}
\end{table}

\begin{table}[p]
\begin{center}
\begin{tabular}{||c|cc||}
\hline
$Q^2$ (GeV$^2$)  & $c\pm {\rm stat.\pm syst.}$ & $\alpha^{\rm tot}_{\pom}(0) \pm {\rm stat. \pm syst.}$ \\
\hline\hline
 2.7 & $0.206 \pm 0.019^{+0.025}_{-0.017}$  & $1.155 \pm 0.011^{+0.007}_{-0.011}$ \\
 4.0 & $0.209 \pm 0.010^{+0.027}_{-0.028}$  & $1.174 \pm 0.006^{+0.014}_{-0.012}$ \\
 6.0 & $0.195 \pm 0.009^{+0.022}_{-0.019}$  & $1.202 \pm 0.006^{+0.010}_{-0.011}$ \\ 
 8.0 & $0.206 \pm 0.009^{+0.016}_{-0.015}$  & $1.211 \pm 0.006^{+0.009}_{-0.009}$ \\
14.0 & $0.207 \pm 0.007^{+0.019}_{-0.017}$  & $1.233 \pm 0.005^{+0.010}_{-0.011}$ \\
27.0 & $0.215 \pm 0.014^{+0.037}_{-0.029}$  & $1.255 \pm 0.010^{+0.020}_{-0.021}$ \\
55.0 & $0.178 \pm 0.020^{+0.049}_{-0.028}$  & $1.307 \pm 0.019^{+0.027}_{-0.037}$ \\
\hline
\end{tabular}
\caption{The results of the fits of $F_2$ data for $x < 0.01$ in bins of $Q^2$ to $F_2(x,Q^2) = c \cdot x^{-\lambda}$, where $\alpha^{\rm tot}_{\pom}(0) = 1+\lambda$.}
\label{t:f2fit}
\end{center}
\end{table}

\begin{table}[p]
\begin{center}
\begin{tabular}{||ccc|ccc||}
\hline
$Q^2$ & $W$ &  $\sigma^{\rm tot}_{\gamma^*p} \pm {\rm stat. \pm syst.}$ & $Q^2$ & $W$ &  $\sigma^{\rm tot}_{\gamma^*p} \pm {\rm stat. \pm syst.}$\\
(GeV$^2$) & (GeV) & ($\rm{\mu}$b) & (GeV$^2$) & (GeV) & ($\rm{\mu}$b) \\
\hline\hline
  2.7& 45.0& 23.88$\pm  0.70^{+ 1.82}_{- 1.73}$ & 14.0& 45.0&  5.29$\pm  0.07^{+ 0.12}_{- 0.13}$\\
  2.7& 65.0& 27.75$\pm  0.93^{+ 1.92}_{- 1.67}$ & 14.0& 65.0&  6.45$\pm  0.09^{+ 0.22}_{- 0.25}$\\
  2.7& 85.0& 29.33$\pm  0.89^{+ 1.58}_{- 1.56}$ & 14.0& 85.0&  7.12$\pm  0.09^{+ 0.60}_{- 0.53}$\\
  2.7&115.0& 31.03$\pm  0.80^{+ 1.95}_{- 1.64}$ & 14.0&115.0&  8.23$\pm  0.10^{+ 0.22}_{- 0.27}$\\
  2.7&150.0& 34.38$\pm  0.97^{+ 1.86}_{- 1.67}$ & 14.0&150.0&  9.23$\pm  0.14^{+ 0.09}_{- 0.12}$\\
  2.7&180.0& 36.91$\pm  1.01^{+ 1.69}_{- 1.98}$ & 14.0&180.0& 10.08$\pm  0.15^{+ 0.14}_{- 0.18}$\\
\end{tabular}
\caption{Total $\gamma^*p$ cross section $\sigma^{\rm tot}_{\gamma^*p}$.}
\label{t:sigtottab}
\end{center}
\end{table}

\begin{table}[p]
\small
\begin{center}
\begin{tabular}{||llll|llll||}
\hline
$M_X$ & $Q^2$ & $W$ & $\frac{d\sigma^{\rm diff}_{\gamma^* p \to XN}}{dM_X}$ & $M_X$ & $Q^2$ & $W$ & $\frac{d\sigma^{\rm diff}_{\gamma^* p \to XN}}{dM_X}$ \\  
 & & & $\pm$ stat. $\pm$ syst. & & & &  $\pm$ stat. $\pm$ syst. \\
 (GeV) & (GeV$^2$) & (GeV) &  (nb/GeV) & (GeV) & (GeV$^2$) & (GeV) &  (nb/GeV)  \\
\hline\hline
  1.2&  2.7& 45.0&    773.8$\pm  76.0^{+ 109.5}_{-  79.8}$&  1.2& 14.0& 45.0&     31.6$\pm   2.3^{+   4.3}_{-   3.6}$ \\
  1.2&  2.7& 65.0&    908.4$\pm  89.8^{+ 141.2}_{- 112.6}$&  1.2& 14.0& 65.0&     30.4$\pm   2.4^{+   3.9}_{-   4.5}$ \\
  1.2&  2.7& 85.0&    952.8$\pm  89.0^{+  85.3}_{-  79.8}$&  1.2& 14.0& 85.0&     40.3$\pm   2.7^{+   5.6}_{-   6.9}$ \\
  1.2&  2.7&115.0&    958.2$\pm  73.7^{+  99.7}_{-  86.0}$&  1.2& 14.0&115.0&     39.0$\pm   2.8^{+   4.1}_{-   3.4}$ \\
  1.2&  2.7&150.0&   1010.0$\pm  86.6^{+ 113.0}_{- 115.7}$&  1.2& 14.0&150.0&     41.9$\pm   3.5^{+   3.3}_{-   4.7}$ \\
\end{tabular}
\caption{Cross section for diffractive scattering, $\gamma^*p \to XN$, $M_N < 2.3$ {\rm GeV}, for $M_X = 1.2$ {\rm GeV} in bins of $W$ and $Q^2$.}
\label{t:dsigdmx1}
\end{center}
\normalsize
\end{table}

\begin{table}[p]
\small
\begin{center}
\begin{tabular}{||llll|llll||}
\hline
$M_X$ & $Q^2$ & $W$ & $\frac{d\sigma^{\rm diff}_{\gamma^* p \to XN}}{dM_X}$ & $M_X$ & $Q^2$ & $W$ & $\frac{d\sigma^{\rm diff}_{\gamma^* p \to XN}}{dM_X}$ \\  
 & & & $\pm$ stat. $\pm$ syst. & & & &  $\pm$ stat. $\pm$ syst. \\
 (GeV) & (GeV$^2$) & (GeV) &  (nb/GeV) & (GeV) & (GeV$^2$) & (GeV) &  (nb/GeV)  \\
\hline\hline
  3.0&  2.7& 45.0&    502.7$\pm  50.8^{+  52.8}_{-  50.1}$&  3.0& 14.0& 45.0&     61.0$\pm   3.5^{+   2.8}_{-   4.2}$ \\
  3.0&  2.7& 65.0&    476.5$\pm  57.8^{+  30.7}_{-  38.4}$&  3.0& 14.0& 65.0&     67.3$\pm   3.8^{+   7.4}_{-   5.0}$ \\
  3.0&  2.7& 85.0&    501.7$\pm  58.9^{+  63.1}_{-  38.2}$&  3.0& 14.0& 85.0&     80.2$\pm   4.2^{+  13.2}_{-   9.1}$ \\
  3.0&  2.7&115.0&    555.6$\pm  58.0^{+  45.3}_{-  44.8}$&  3.0& 14.0&115.0&     88.0$\pm   4.5^{+   2.3}_{-   5.7}$ \\
  3.0&  2.7&150.0&    772.0$\pm  72.4^{+  40.4}_{-  59.4}$&  3.0& 14.0&150.0&     88.8$\pm   5.6^{+   7.3}_{-   3.6}$ \\
\end{tabular}
\caption{Cross section for diffractive scattering, $\gamma^*p \to XN$, $M_N < 2.3$ {\rm GeV}, for $M_X = 3.0$ {\rm GeV} in bins of $W$ and $Q^2$.}
\label{t:dsigdmx2}
\end{center}
\normalsize
\end{table}

\begin{table}[p]
\small
\begin{center}
\begin{tabular}{||llll|llll||}
\hline
$M_X$ & $Q^2$ & $W$ & $\frac{d\sigma^{\rm diff}_{\gamma^* p \to XN}}{dM_X}$ & $M_X$ & $Q^2$ & $W$ & $\frac{d\sigma^{\rm diff}_{\gamma^* p \to XN}}{dM_X}$ \\  
 & & & $\pm$ stat. $\pm$ syst. & & & &  $\pm$ stat. $\pm$ syst. \\
 (GeV) & (GeV$^2$) & (GeV) &  (nb/GeV) & (GeV) & (GeV$^2$) & (GeV) &  (nb/GeV)  \\
\hline\hline
   6.0&  2.7& 65.0&    221.5$\pm  26.2^{+ 11.1}_{-  11.5}$&  6.0& 14.0& 65.0&     45.6$\pm   3.3^{+   4.8}_{-   4.8}$ \\
  6.0&  2.7& 85.0&    262.7$\pm  28.4^{+  17.5}_{-  13.3}$&  6.0& 14.0& 85.0&     55.9$\pm   2.8^{+   7.1}_{-   5.5}$ \\
  6.0&  2.7&115.0&    278.8$\pm  27.8^{+  35.3}_{-  26.6}$&  6.0& 14.0&115.0&     57.3$\pm   2.8^{+   0.9}_{-   5.4}$ \\
  6.0&  2.7&150.0&    270.8$\pm  30.3^{+  23.5}_{-  30.4}$&  6.0& 14.0&150.0&     64.9$\pm   3.6^{+   2.9}_{-   1.8}$ \\
  6.0&  2.7&180.0&    313.3$\pm  30.1^{+  24.0}_{-  29.9}$&  6.0& 14.0&180.0&     73.8$\pm   3.9^{+   2.3}_{-   2.0}$ \\
\end{tabular}
\caption{Cross section for diffractive scattering, $\gamma^*p \to XN$, $M_N < 2.3$ {\rm GeV}, for $M_X = 6.0$ {\rm GeV} in bins of $W$ and $Q^2$.}
\label{t:dsigdmx3}
\end{center}
\normalsize
\end{table}

\begin{table}[p]
\small
\begin{center}
\begin{tabular}{||llll|llll||}
\hline
$M_X$ & $Q^2$ & $W$ & $\frac{d\sigma^{\rm diff}_{\gamma^* p \to XN}}{dM_X}$ & $M_X$ & $Q^2$ & $W$ & $\frac{d\sigma^{\rm diff}_{\gamma^* p \to XN}}{dM_X}$ \\  
 & & & $\pm$ stat. $\pm$ syst. & & & &  $\pm$ stat. $\pm$ syst. \\
 (GeV) & (GeV$^2$) & (GeV) &  (nb/GeV) & (GeV) & (GeV$^2$) & (GeV) &  (nb/GeV)  \\
\hline\hline
 11.0&  2.7& 85.0&    126.3$\pm  27.1^{+  24.3}_{-   8.3}$& 11.0& 14.0& 85.0&     30.5$\pm   5.7^{+   5.5}_{-   4.9}$ \\
 11.0&  2.7&115.0&    141.3$\pm  17.1^{+  14.0}_{-  12.3}$& 11.0& 14.0&115.0&     32.0$\pm   2.8^{+   1.2}_{-   2.5}$ \\
 11.0&  2.7&150.0&    124.7$\pm  16.5^{+  15.3}_{-   8.3}$& 11.0& 14.0&150.0&     34.2$\pm   2.4^{+   0.6}_{-   1.4}$ \\
 11.0&  2.7&180.0&    148.7$\pm  15.4^{+   4.5}_{-  17.9}$& 11.0& 14.0&180.0&     37.7$\pm   2.3^{+   2.0}_{-   1.6}$ \\
 11.0&  2.7&220.0&    151.4$\pm  16.4^{+  19.4}_{-   3.4}$& 11.0& 14.0&220.0&     43.1$\pm   2.7^{+   1.5}_{-   1.0}$ \\
\end{tabular}
\caption{Cross section for diffractive scattering, $\gamma^*p \to XN$, $M_N < 2.3$ {\rm GeV}, for $M_X =11.0$ {\rm GeV} in bins of $W$ and $Q^2$.}
\label{t:dsigdmx4}
\end{center}
\normalsize
\end{table}

\begin{table}[p]
\small
\begin{center}
\begin{tabular}{||llll|llll||}
\hline
$M_X$ & $Q^2$ & $W$ & $\frac{d\sigma^{\rm diff}_{\gamma^* p \to XN}}{dM_X}$ & $M_X$ & $Q^2$ & $W$ & $\frac{d\sigma^{\rm diff}_{\gamma^* p \to XN}}{dM_X}$ \\  
 & & & $\pm$ stat. $\pm$ syst. & & & &  $\pm$ stat. $\pm$ syst. \\
 (GeV) & (GeV$^2$) & (GeV) &  (nb/GeV) & (GeV) & (GeV$^2$) & (GeV) &  (nb/GeV)  \\
\hline\hline
 20.0&  2.7&150.0&     81.8$\pm  24.0^{+   8.6}_{-   4.4}$& 20.0& 14.0&150.0&     17.8$\pm   4.9^{+   2.3}_{-   1.9}$ \\
 20.0&  2.7&180.0&     68.1$\pm  13.3^{+   8.3}_{-   3.1}$& 20.0& 14.0&180.0&     21.1$\pm   3.1^{+   0.8}_{-   1.2}$ \\
 20.0&  2.7&220.0&    107.1$\pm  13.4^{+   8.8}_{-   6.7}$& 20.0& 14.0&220.0&     23.9$\pm   2.3^{+   1.0}_{-   0.9}$ \\
\hline
 20.0&  4.0&150.0&     48.0$\pm  15.2^{+   3.7}_{-   4.7}$ & & & & \\
 20.0&  4.0&180.0&     57.1$\pm   9.3^{+   4.8}_{-   3.5}$& 20.0& 27.0&180.0&     10.2$\pm   1.9^{+   0.5}_{-   0.7}$ \\
 20.0&  4.0&220.0&     66.9$\pm   6.9^{+   3.8}_{-   2.6}$& 20.0& 27.0&220.0&     10.7$\pm   1.4^{+   1.0}_{-   0.4}$ \\
\hline
 20.0&  6.0&150.0&     40.9$\pm  11.1^{+   5.2}_{-   4.0}$& 20.0& 55.0&150.0&      4.2$\pm   1.4^{+   0.9}_{-   1.0}$ \\
 20.0&  6.0&180.0&     42.5$\pm   7.0^{+   0.7}_{-   6.0}$& 20.0& 55.0&180.0&      3.7$\pm   1.0^{+   0.4}_{-   0.5}$ \\
 20.0&  6.0&220.0&     48.0$\pm   5.2^{+   3.2}_{-   4.2}$& 20.0& 55.0&220.0&      4.7$\pm   0.9^{+   1.1}_{-   0.3}$ \\
\hline
 20.0& 8.0&150.0&     31.3$\pm   8.2^{+   2.6}_{-   2.5}$ \\
 20.0& 8.0&180.0&     41.1$\pm   5.5^{+   0.9}_{-   4.3}$ \\
 20.0& 8.0&220.0&     39.6$\pm   4.2^{+   1.4}_{-   4.7}$ \\
\cline{1-4}
\end{tabular}
\caption{Cross section for diffractive scattering, $\gamma^*p \to XN$, $M_N < 2.3$ {\rm GeV}, for $M_X =20.0$ {\rm GeV} in bins of $W$ and $Q^2$.}
\label{t:dsigdmx5}
\end{center}
\normalsize
\end{table}

\begin{table}[p]
\small
\begin{center}
\begin{tabular}{||llll|llll||}
\hline
$M_X$ & $Q^2$ & $W$ & $\frac{d\sigma^{\rm diff}_{\gamma^* p \to XN}}{dM_X}$ & $M_X$ & $Q^2$ & $W$ & $\frac{d\sigma^{\rm diff}_{\gamma^* p \to XN}}{dM_X}$ \\  
 & & & $\pm$ stat. $\pm$ syst. & & & &  $\pm$ stat. $\pm$ syst. \\
 (GeV) & (GeV$^2$) & (GeV) &  (nb/GeV) & (GeV) & (GeV$^2$) & (GeV) &  (nb/GeV)  \\
\hline\hline
 30.0&  2.7&220.0&     63.3$\pm  18.6^{+   9.2}_{-   6.1}$& 30.0& 14.0&220.0&     16.9$\pm   4.5^{+   1.7}_{-   1.1}$ \\
 30.0&  4.0&220.0&     41.3$\pm  11.8^{+   1.5}_{-   5.0}$& 30.0& 27.0&220.0&      6.8$\pm   2.4^{+   0.7}_{-   0.8}$ \\
\hline
 30.0&  6.0&220.0&     29.3$\pm   8.8^{+   4.5}_{-   3.4}$ \\
  30.0& 8.0&220.0&     23.2$\pm   7.4^{+   4.6}_{-   1.1}$ \\

\cline{1-4}
\end{tabular}
\caption{Cross section for diffractive scattering, $\gamma^*p \to XN$, $M_N < 2.3$ {\rm GeV}, for $M_X =30.0$ {\rm GeV} in bins of $W$ and $Q^2$.}
\label{t:dsigdmx6}
\end{center}
\normalsize
\end{table}

\begin{table}[p]
\begin{center}
\begin{tabular}{||ccc||}
\hline
$M_X$ (GeV) & $Q^2$ (GeV$^2$)  & $\alpha^{\rm diff}_{\pom}(0) \pm {\rm stat. \pm syst.}$ \\
\hline\hline
 1.2 & 2.7 & $1.058 \pm 0.019^{+0.017}_{-0.018}$ \\
 1.2 & 4.0 & $1.085 \pm 0.012^{+0.024}_{-0.014}$ \\
 1.2 & 6.0 & $1.095 \pm 0.016^{+0.015}_{-0.012}$ \\
 1.2 & 8.0 & $1.095 \pm 0.014^{+0.015}_{-0.013}$ \\
 1.2 &14.0 & $1.074 \pm 0.015^{+0.016}_{-0.011}$ \\
 1.2 &27.0 & $1.095 \pm 0.038^{+0.024}_{-0.014}$ \\
 1.2 &55.0 & $1.087 \pm 0.113^{+0.042}_{-0.080}$ \\
\hline
 3.0 & 2.7 & $1.130 \pm 0.024^{+0.021}_{-0.010}$ \\
 3.0 & 4.0 & $1.130 \pm 0.012^{+0.015}_{-0.016}$ \\
 3.0 & 6.0 & $1.128 \pm 0.014^{+0.013}_{-0.015}$ \\
 3.0 & 8.0 & $1.103 \pm 0.011^{+0.012}_{-0.005}$ \\
 3.0 &14.0 & $1.125 \pm 0.011^{+0.015}_{-0.013}$ \\
 3.0 &27.0 & $1.201 \pm 0.019^{+0.017}_{-0.018}$ \\
 3.0 &55.0 & $1.202 \pm 0.057^{+0.013}_{-0.030}$ \\
\hline
 6.0 & 2.7 & $1.099 \pm 0.032^{+0.008}_{-0.021}$ \\
 6.0 & 4.0 & $1.131 \pm 0.017^{+0.015}_{-0.020}$ \\
 6.0 & 6.0 & $1.126 \pm 0.019^{+0.017}_{-0.010}$ \\
 6.0 & 8.0 & $1.121 \pm 0.016^{+0.015}_{-0.015}$ \\
 6.0 &14.0 & $1.125 \pm 0.014^{+0.026}_{-0.021}$ \\
 6.0 &27.0 & $1.170 \pm 0.021^{+0.020}_{-0.019}$ \\
 6.0 &55.0 & $1.208 \pm 0.038^{+0.022}_{-0.024}$ \\
\hline
11.0 & 2.7 & $1.074 \pm 0.049^{+0.019}_{-0.033}$ \\
11.0 & 4.0 & $1.098 \pm 0.041^{+0.019}_{-0.023}$ \\
11.0 & 6.0 & $1.170 \pm 0.046^{+0.022}_{-0.023}$ \\
11.0 & 8.0 & $1.097 \pm 0.038^{+0.026}_{-0.020}$ \\
11.0 &14.0 & $1.138 \pm 0.036^{+0.030}_{-0.012}$ \\
11.0 &27.0 & $1.248 \pm 0.058^{+0.023}_{-0.066}$ \\
11.0 &55.0 & $1.181 \pm 0.070^{+0.021}_{-0.022}$ \\
\hline
\end{tabular}
\caption{The value of $\alpha^{\rm diff}_{\pom}(0)$ for $M_X$ and $Q^2$ bins.}
\label{t:adiff}
\end{center}
\end{table}

\begin{table}[p]
\begin{center}
\begin{tabular}{||cc|c||}
\hline
$Q^2$ range (GeV$^2$)  & $<Q^2>$ (GeV$^2$) & $\alpha^{\rm diff}_{\pom}(0) \pm {\rm stat.\pm syst.}$  \\
\hline\hline
2.2 - 5 & 3.6 & $1.125\pm 0.009^{+0.013}_{-0.013}$ \\
5 - 10 & 7.2 &  $1.117\pm 0.007^{+0.013}_{-0.010}$ \\
10 - 20 & 14.0 & $1.126\pm 0.009^{+0.016}_{-0.013}$ \\
\hline
2.2 - 20 & 8.3 & $1.122\pm 0.005^{+0.013}_{-0.011}$ \\
\hline
20 - 80 & 34.4 & $1.193\pm 0.013^{+0.016}_{-0.019}$\\
\hline
\end{tabular}
\caption{The values of $\alpha^{\rm diff}_{\pom}(0)$ for $2 < M_X < 15$ GeV.}
\label{t:alphapomvsq2}
\end{center}
\end{table}

\begin{table}[p]
\begin{center}
\begin{tabular}{||cc|c||}
\hline
$Q^2$ range (GeV$^2$)  & $<Q^2>$ (GeV$^2$) & $\alpha^{\rm diff}_{\pom}(0) \pm {\rm stat.\pm syst.}$  \\
\hline\hline
2.2 - 5 & 3.6 & $1.127\pm 0.009^{+0.013}_{-0.013}$ \\
5 - 10 & 7.2 &  $1.115\pm 0.008^{+0.013}_{-0.010}$ \\
10 - 20 & 14.0 & $1.124\pm 0.012^{+0.020}_{-0.019}$ \\
\hline
2.2 - 20 & 8.3 & $1.121\pm 0.005^{+0.014}_{-0.012}$ \\
\hline
20 - 80 & 33.9 & $1.179\pm 0.017^{+0.020}_{-0.022}$\\
\hline
\end{tabular}
\caption{The values of $\alpha^{\rm diff}_{\pom}(0)$ for $2 < M_X < 15$ GeV and $\xpom < 0.01$.}
\label{t:alphapomvsq2lt01}
\end{center}
\end{table}
\clearpage
\clearpage

\begin{table}[p]
\small
\begin{center}
\begin{tabular}{||lrc|lrc||}
\hline
$Q^2$ & $W$ &$\frac{\int^{M_b}_{M_a} dM_X d\sigma^{\rm diff}_{\gamma^* p \to XN}/dM_X}{\sigma^{\rm tot}}$ & $Q^2$ & $W$ &$\frac{\int^{M_b}_{M_a} dM_X d\sigma^{\rm diff}_{\gamma^* p \to XN}/dM_X}{\sigma^{\rm tot}}$\\  
 & & $\pm$stat.$\pm$syst. &  &  &$\pm$stat.$\pm$syst. \\
(GeV$^2$)&(GeV)& &(GeV$^2$)&(GeV)&    \\
\hline\hline
 2.7& 45& 0.0557$\pm 0.0057^{+ 0.0079}_{- 0.0057}$&14.0& 45& 0.0103$\pm 0.0008^{+ 0.0014}_{- 0.0012}$ \\
 2.7& 65& 0.0563$\pm 0.0059^{+ 0.0088}_{- 0.0070}$&14.0& 65& 0.0081$\pm 0.0006^{+ 0.0010}_{- 0.0012}$ \\
 2.7& 85& 0.0559$\pm 0.0055^{+ 0.0050}_{- 0.0047}$&14.0& 85& 0.0097$\pm 0.0007^{+ 0.0014}_{- 0.0017}$ \\
 2.7&115& 0.0531$\pm 0.0043^{+ 0.0055}_{- 0.0048}$&14.0&115& 0.0081$\pm 0.0006^{+ 0.0009}_{- 0.0007}$ \\
 2.7&150& 0.0505$\pm 0.0046^{+ 0.0057}_{- 0.0058}$&14.0&150& 0.0078$\pm 0.0007^{+ 0.0006}_{- 0.0009}$ \\
\end{tabular}
\caption{Ratio of the cross section for diffractive scattering, $\gamma^*p \to XN$, $M_N < 2.3$ {\rm GeV}, integrated over $M_X = 0.28 - 2$ {\rm GeV}, to the total cross section.}
\label{t:rdiftot1}
\end{center}
\normalsize
\end{table}

\begin{table}[p]
\small
\begin{center}
\begin{tabular}{||lrc|lrc||}
\hline
$Q^2$ & $W$ &$\frac{\int^{M_b}_{M_a} dM_X d\sigma^{\rm diff}_{\gamma^* p \to XN}/dM_X}{\sigma^{\rm tot}}$ & $Q^2$ & $W$ &$\frac{\int^{M_b}_{M_a} dM_X d\sigma^{\rm diff}_{\gamma^* p \to XN}/dM_X}{\sigma^{\rm tot}}$\\  
 & & $\pm$stat.$\pm$syst. &  &  &$\pm$stat.$\pm$syst. \\
(GeV$^2$)&(GeV)& &(GeV$^2$)&(GeV)&    \\
\hline\hline
 2.7& 45& 0.0421$\pm 0.0044^{+ 0.0044}_{- 0.0042}$&14.0& 45& 0.0230$\pm 0.0014^{+ 0.0011}_{- 0.0016}$ \\
 2.7& 65& 0.0343$\pm 0.0043^{+ 0.0022}_{- 0.0028}$&14.0& 65& 0.0209$\pm 0.0012^{+ 0.0023}_{- 0.0015}$ \\
 2.7& 85& 0.0342$\pm 0.0042^{+ 0.0043}_{- 0.0026}$&14.0& 85& 0.0225$\pm 0.0012^{+ 0.0037}_{- 0.0026}$ \\
 2.7&115& 0.0358$\pm 0.0039^{+ 0.0029}_{- 0.0029}$&14.0&115& 0.0214$\pm 0.0011^{+ 0.0006}_{- 0.0014}$ \\
 2.7&150& 0.0449$\pm 0.0044^{+ 0.0023}_{- 0.0035}$&14.0&150& 0.0192$\pm 0.0012^{+ 0.0016}_{- 0.0008}$ \\
\end{tabular}
\caption{Ratio of the cross section for diffractive scattering, $\gamma^*p \to XN$, $M_N < 2.3$ {\rm GeV}, integrated over $M_X = 2 - 4$ {\rm GeV}, to the total cross section.}
\label{t:rdiftot2}
\end{center}
\normalsize
\end{table}

\begin{table}[p]
\small
\begin{center}
\begin{tabular}{||lrc|lrc||}
\hline
$Q^2$ & $W$ &$\frac{\int^{M_b}_{M_a} dM_X d\sigma^{\rm diff}_{\gamma^* p \to XN}/dM_X}{\sigma^{\rm tot}}$ & $Q^2$ & $W$ &$\frac{\int^{M_b}_{M_a} dM_X d\sigma^{\rm diff}_{\gamma^* p \to XN}/dM_X}{\sigma^{\rm tot}}$\\  
 & & $\pm$stat.$\pm$syst. &  &  &$\pm$stat.$\pm$syst. \\
(GeV$^2$)&(GeV)& &(GeV$^2$)&(GeV)&    \\
\hline\hline
 2.7& 65& 0.0319$\pm 0.0039^{+ 0.0016}_{- 0.0017}$&14.0& 65& 0.0283$\pm 0.0021^{+ 0.0030}_{- 0.0030}$ \\
 2.7& 85& 0.0358$\pm 0.0040^{+ 0.0024}_{- 0.0018}$&14.0& 85& 0.0314$\pm 0.0016^{+ 0.0040}_{- 0.0031}$ \\
 2.7&115& 0.0359$\pm 0.0037^{+ 0.0046}_{- 0.0034}$&14.0&115& 0.0279$\pm 0.0014^{+ 0.0004}_{- 0.0026}$ \\
 2.7&150& 0.0315$\pm 0.0036^{+ 0.0027}_{- 0.0035}$&14.0&150& 0.0281$\pm 0.0016^{+ 0.0013}_{- 0.0008}$ \\
 2.7&180& 0.0340$\pm 0.0034^{+ 0.0026}_{- 0.0032}$&14.0&180& 0.0293$\pm 0.0016^{+ 0.0009}_{- 0.0008}$ \\
\end{tabular}
\caption{Ratio of the cross section for diffractive scattering, $\gamma^*p \to XN$, $M_N < 2.3$ {\rm GeV}, integrated over $M_X = 4 - 8$ {\rm GeV}, to the total cross section.}
\label{t:rdiftot3}
\end{center}
\normalsize
\end{table}

\begin{table}[p]
\small
\begin{center}
\begin{tabular}{||lrc|lrc||}
\hline
$Q^2$ & $W$ &$\frac{\int^{M_b}_{M_a} dM_X d\sigma^{\rm diff}_{\gamma^* p \to XN}/dM_X}{\sigma^{\rm tot}}$ & $Q^2$ & $W$ &$\frac{\int^{M_b}_{M_a} dM_X d\sigma^{\rm diff}_{\gamma^* p \to XN}/dM_X}{\sigma^{\rm tot}}$\\  
 & & $\pm$stat.$\pm$syst. &  &  &$\pm$stat.$\pm$syst. \\
(GeV$^2$)&(GeV)& &(GeV$^2$)&(GeV)&    \\
\hline\hline
 2.7& 85& 0.0302$\pm 0.0065^{+ 0.0058}_{- 0.0020}$&14.0& 85& 0.0300$\pm 0.0056^{+ 0.0054}_{- 0.0048}$ \\
 2.7&115& 0.0319$\pm 0.0039^{+ 0.0032}_{- 0.0028}$&14.0&115& 0.0272$\pm 0.0024^{+ 0.0010}_{- 0.0021}$ \\
 2.7&150& 0.0254$\pm 0.0034^{+ 0.0031}_{- 0.0017}$&14.0&150& 0.0260$\pm 0.0019^{+ 0.0004}_{- 0.0011}$ \\
 2.7&180& 0.0282$\pm 0.0030^{+ 0.0009}_{- 0.0034}$&14.0&180& 0.0262$\pm 0.0017^{+ 0.0014}_{- 0.0011}$ \\
 2.7&220& 0.0264$\pm 0.0030^{+ 0.0034}_{- 0.0006}$&14.0&220& 0.0269$\pm 0.0018^{+ 0.0009}_{- 0.0006}$ \\
\end{tabular}
\caption{Ratio of the cross section for diffractive scattering, $\gamma^*p \to XN$, $M_N < 2.3$ {\rm GeV}, integrated over $M_X = 8 - 15$ {\rm GeV}, to the total cross section.}
\label{t:rdiftot4}
\end{center}
\normalsize
\end{table}

\begin{table}[p]
\small
\begin{center}
\begin{tabular}{||lrc|lrc||}
\hline
$Q^2$ & $W$ &$\frac{\int^{M_b}_{M_a} dM_X d\sigma^{\rm diff}_{\gamma^* p \to XN}/dM_X}{\sigma^{\rm tot}}$ & $Q^2$ & $W$ &$\frac{\int^{M_b}_{M_a} dM_X d\sigma^{\rm diff}_{\gamma^* p \to XN}/dM_X}{\sigma^{\rm tot}}$\\  
 & & $\pm$stat.$\pm$syst. &  &  &$\pm$stat.$\pm$syst. \\
(GeV$^2$)&(GeV)& &(GeV$^2$)&(GeV)&    \\
\hline\hline
 2.7&150& 0.0238$\pm 0.0070^{+ 0.0025}_{- 0.0013}$&14.0&150& 0.0193$\pm 0.0054^{+ 0.0025}_{- 0.0020}$ \\
 2.7&180& 0.0184$\pm 0.0036^{+ 0.0023}_{- 0.0008}$&14.0&180& 0.0210$\pm 0.0031^{+ 0.0008}_{- 0.0011}$ \\
 2.7&220& 0.0267$\pm 0.0034^{+ 0.0022}_{- 0.0017}$&14.0&220& 0.0213$\pm 0.0021^{+ 0.0009}_{- 0.0008}$ \\
\hline
 4.0&150& 0.0187$\pm 0.0059^{+ 0.0015}_{- 0.0018}$ & & & \\
 4.0&180& 0.0200$\pm 0.0033^{+ 0.0017}_{- 0.0012}$&27.0&180& 0.0190$\pm 0.0036^{+ 0.0010}_{- 0.0014}$ \\
 4.0&220& 0.0218$\pm 0.0023^{+ 0.0012}_{- 0.0008}$&27.0&220& 0.0176$\pm 0.0024^{+ 0.0016}_{- 0.0007}$ \\
\hline
 6.0&150& 0.0214$\pm 0.0058^{+ 0.0027}_{- 0.0021}$&55.0&150& 0.0188$\pm 0.0065^{+ 0.0039}_{- 0.0047}$ \\
 6.0&180& 0.0206$\pm 0.0034^{+ 0.0004}_{- 0.0029}$&55.0&180& 0.0142$\pm 0.0038^{+ 0.0016}_{- 0.0019}$ \\
 6.0&220& 0.0208$\pm 0.0023^{+ 0.0014}_{- 0.0018}$&55.0&220& 0.0157$\pm 0.0029^{+ 0.0036}_{- 0.0011}$ \\
\hline
 8.0&150& 0.0205$\pm 0.0054^{+ 0.0017}_{- 0.0016}$ \\
 8.0&180& 0.0245$\pm 0.0033^{+ 0.0005}_{- 0.0026}$ \\
 8.0&220& 0.0216$\pm 0.0023^{+ 0.0007}_{- 0.0026}$ \\
\cline{1-3}
\end{tabular}
\caption{Ratio of the cross section for diffractive scattering, $\gamma^*p \to XN$, $M_N < 2.3$ {\rm GeV}, integrated over $M_X = 15 - 25$ {\rm GeV}, to the total cross section.}
\label{t:rdiftot5}
\end{center}
\normalsize
\end{table}

\begin{table}[p]
\small
\begin{center}
\begin{tabular}{||lrc|lrc||}
\hline
$Q^2$ & $W$ &$\frac{\int^{M_b}_{M_a} dM_X d\sigma^{\rm diff}_{\gamma^* p \to XN}/dM_X}{\sigma^{\rm tot}}$  & $Q^2$ & $W$ &$\frac{\int^{M_b}_{M_a}dM_X d\sigma^{\rm diff}_{\gamma^* p \to XN}/dM_X}{\sigma^{\rm tot}}$\\  
 & & $\pm$stat.$\pm$syst. &  &  &$\pm$stat.$\pm$syst. \\
(GeV$^2$)&(GeV)& &(GeV$^2$)&(GeV)&    \\
\hline\hline
 2.7&220& 0.0158$\pm 0.0047^{+ 0.0023}_{- 0.0015}$&14.0&220& 0.0151$\pm 0.0040^{+ 0.0015}_{- 0.0010}$ \\
 4.0&220& 0.0135$\pm 0.0039^{+ 0.0005}_{- 0.0016}$&27.0&220& 0.0111$\pm 0.0039^{+ 0.0012}_{- 0.0014}$ \\
\hline
 6.0&220& 0.0127$\pm 0.0038^{+ 0.0020}_{- 0.0015}$ \\
 8.0&220& 0.0126$\pm 0.0040^{+ 0.0025}_{- 0.0006}$ \\
\cline{1-3}
\end{tabular}
\caption{Ratio of the cross section for diffractive scattering, $\gamma^*p \to XN$, $M_N < 2.3$ {\rm GeV}, integrated over $M_X = 25 - 35$ {\rm GeV}, to the total cross section.}
\label{t:rdiftot6}
\end{center}
\normalsize
\end{table}

\begin{table}[p]
\begin{center}
\begin{tabular}{||cc||}
\hline
$Q^2$ (GeV$^2$)  & $\sigma^{\rm diff}/\sigma^{\rm tot}\pm {\rm stat.\pm syst.}$ \\
\hline\hline
4.0 & $0.158\pm 0.007^{+0.009}_{-0.007}$ \\
6.0 & $0.149\pm 0.007^{+0.005}_{-0.005}$ \\
8.0 & $0.134\pm 0.006^{+0.005}_{-0.004}$ \\
14.0 &$0.118\pm 0.005^{+0.003}_{-0.002}$ \\
27.0 &$0.096\pm 0.006^{+0.003}_{-0.004}$ \\
\hline
\end{tabular}
\caption{Ratio of the total diffractive cross section observed to the total cross section, $\sigma^{\rm diff}(0.28 < M_X < 35 {\rm GeV}, M_N < 2.3 {\rm GeV})/\sigma^{\rm tot}$, for $200 < W < 245$ \GeV, at different values of $Q^2$.}
\label{t:sumrdiftot}
\end{center}
\end{table}

\begin{table}[p]
\small
\begin{center}
\begin{tabular}{||llll|llll||}
\hline
$\beta$ & $\xpom$ & $Q^2$ & $\xpom F^{D(3)}_2$ & $\beta$ & $\xpom$& $Q^2$ & $\xpom F^{D(3)}_2$  \\
& & & $\pm$ stat. $\pm$ syst. & & & & $\pm$ stat. $\pm$ syst.   \\
& & \hspace*{-0.4cm}(GeV$^2$) & & & & \hspace*{-0.4cm}(GeV$^2$)  & \\
\hline\hline
0.6522&0.00204&  2.7& 0.0321$\pm0.0032^{+0.0045}_{-0.0033}$&0.7353&0.00268&  4.0& 0.0314$\pm0.0018^{+0.0032}_{-0.0044}$ \\
0.6522&0.00098&  2.7& 0.0377$\pm0.0037^{+0.0059}_{-0.0047}$&0.7353&0.00129&  4.0& 0.0342$\pm0.0022^{+0.0046}_{-0.0051}$ \\
0.6522&0.00057&  2.7& 0.0396$\pm0.0037^{+0.0035}_{-0.0033}$&0.7353&0.00075&  4.0& 0.0392$\pm0.0025^{+0.0056}_{-0.0052}$ \\
0.6522&0.00031&  2.7& 0.0398$\pm0.0031^{+0.0041}_{-0.0036}$&0.7353&0.00041&  4.0& 0.0456$\pm0.0026^{+0.0049}_{-0.0045}$ \\
0.6522&0.00018&  2.7& 0.0419$\pm0.0036^{+0.0047}_{-0.0048}$&0.7353&0.00024&  4.0& 0.0389$\pm0.0027^{+0.0052}_{-0.0039}$ \\
\hline
\end{tabular}
\caption{The diffractive structure function multiplied by $\xpom$, $\xpom F^{D(3)}_2(\beta,\xpom,Q^2)$ for diffractive scattering, $\gamma^*p \to XN$, $M_N < 2.3$ {\rm GeV}, for $Q^2 = 2.7$ and 4.0 \GeV$^2$, in bins of $\beta$ and $\xpom$.} 
\label{t:f2d301}
\end{center}
\normalsize
\end{table}
\clearpage

\scriptsize
\begin{table}[p]
\small
\begin{center}
\begin{tabular}{||llll|llll||}
\hline
$\beta$ & $\xpom$ & $Q^2$ & $\xpom F^{D(3)}_2$ & $\beta$ & $\xpom$ & $Q^2$ & $\xpom F^{D(3)}_2$  \\
& & & $\pm$ stat. $\pm$ syst. & & & & $\pm$ stat. $\pm$ syst.   \\
& & \hspace*{-0.4cm}(GeV$^2$) & & & & \hspace*{-0.4cm}(GeV$^2$)  & \\
\hline\hline
0.8065&0.00366&  6.0& 0.0272$\pm0.0023^{+0.0029}_{-0.0035}$&0.8475&0.00464&  8.0& 0.0297$\pm0.0023^{+0.0031}_{-0.0034}$ \\
0.8065&0.00176&  6.0& 0.0319$\pm0.0029^{+0.0072}_{-0.0033}$&0.8475&0.00223&  8.0& 0.0338$\pm0.0025^{+0.0047}_{-0.0043}$ \\
0.8065&0.00103&  6.0& 0.0321$\pm0.0029^{+0.0036}_{-0.0037}$&0.8475&0.00131&  8.0& 0.0353$\pm0.0026^{+0.0055}_{-0.0043}$ \\
0.8065&0.00056&  6.0& 0.0350$\pm0.0029^{+0.0034}_{-0.0048}$&0.8475&0.00071&  8.0& 0.0400$\pm0.0029^{+0.0043}_{-0.0040}$ \\
0.8065&0.00033&  6.0& 0.0425$\pm0.0037^{+0.0046}_{-0.0041}$&0.8475&0.00042&  8.0& 0.0426$\pm0.0035^{+0.0036}_{-0.0037}$ \\
\hline
\end{tabular}
\caption{The diffractive structure function multiplied by $\xpom$, $\xpom F^{D(3)}_2(\beta,\xpom,Q^2)$ for diffractive scattering, $\gamma^*p \to XN$, $M_N < 2.3$ {\rm GeV}, for $Q^2 = 6.0$ and 8.0 {\rm GeV$^2$}, in bins of $\beta$ and $\xpom$.}
\label{t:f2d302}
\end{center}
\normalsize
\end{table}
\clearpage

\begin{table}[p]
\small
\begin{center}
\begin{tabular}{||llll|llll||}
\hline
$\beta$ & $\xpom$ & $Q^2$ & $\xpom F^{D(3)}_2$ & $\beta$ & $\xpom$ & $Q^2$ & $\xpom F^{D(3)}_2$  \\
& & & $\pm$ stat. $\pm$ syst. & & & & $\pm$ stat. $\pm$ syst.   \\
& & \hspace*{-0.4cm}(GeV$^2$) & & & & \hspace*{-0.4cm}(GeV$^2$)  & \\
\hline\hline
0.9067&0.00757& 14.0& 0.0253$\pm0.0018^{+0.0034}_{-0.0029}$&0.9494&0.01386& 27.0& 0.0181$\pm0.0029^{+0.0016}_{-0.0035}$ \\
0.9067&0.00364& 14.0& 0.0244$\pm0.0019^{+0.0031}_{-0.0036}$&0.9494&0.00669& 27.0& 0.0154$\pm0.0031^{+0.0029}_{-0.0017}$ \\
0.9067&0.00213& 14.0& 0.0324$\pm0.0022^{+0.0045}_{-0.0055}$&0.9494&0.00392& 27.0& 0.0251$\pm0.0037^{+0.0039}_{-0.0045}$ \\
0.9067&0.00117& 14.0& 0.0313$\pm0.0022^{+0.0033}_{-0.0027}$&0.9494&0.00215& 27.0& 0.0133$\pm0.0032^{+0.0026}_{-0.0024}$ \\
0.9067&0.00069& 14.0& 0.0336$\pm0.0028^{+0.0026}_{-0.0038}$&0.9494&0.00126& 27.0& 0.0343$\pm0.0062^{+0.0025}_{-0.0107}$ \\
\hline
\hline
\end{tabular}
\caption{The diffractive structure function multiplied by $\xpom$, $\xpom F^{D(3)}_2(\beta,\xpom,Q^2)$ for diffractive scattering, $\gamma^*p \to XN$, $M_N < 2.3$ {\rm GeV}, for $Q^2 = 14.0$ and 27.0 {\rm GeV$^2$}, in bins of $\beta$ and $\xpom$.}
\label{t:f2d303}
\end{center}
\normalsize
\end{table}
\clearpage

\begin{table}[p]
\small
\begin{center}
\begin{tabular}{||llll||}
\hline
$\beta$ & $\xpom$ & $Q^2$ & $\xpom F^{D(3)}_2$ \\
& &  & $\pm$ stat. $\pm$ syst. \\
& & \hspace*{-0.4cm}(GeV$^2$) &  \\
\hline\hline
0.9745&0.02713& 55.0& 0.0121$\pm0.0045^{+0.0040}_{-0.0046}$ \\
0.9745&0.01319& 55.0& 0.0148$\pm0.0067^{+0.0026}_{-0.0027}$ \\
0.9745&0.00775& 55.0& 0.0062$\pm0.0035^{+0.0058}_{-0.0016}$ \\
0.9745&0.00425& 55.0& 0.0136$\pm0.0047^{+0.0016}_{-0.0027}$ \\
0.9745&0.00250& 55.0& 0.0171$\pm0.0108^{+0.0111}_{-0.0032}$ \\
0.9745&0.00117& 55.0& 0.0163$\pm0.0086^{+0.0024}_{-0.0099}$ \\
\hline
0.8594&0.01495& 55.0& 0.0131$\pm0.0037^{+0.0026}_{-0.0007}$ \\
0.8594&0.00879& 55.0& 0.0208$\pm0.0042^{+0.0018}_{-0.0024}$ \\
0.8594&0.00482& 55.0& 0.0211$\pm0.0045^{+0.0050}_{-0.0019}$ \\
0.8594&0.00284& 55.0& 0.0212$\pm0.0060^{+0.0080}_{-0.0040}$ \\
0.8594&0.00197& 55.0& 0.0313$\pm0.0063^{+0.0031}_{-0.0080}$ \\
0.8594&0.00132& 55.0& 0.0345$\pm0.0075^{+0.0076}_{-0.0053}$ \\
\hline
0.6044&0.02126& 55.0& 0.0155$\pm0.0037^{+0.0025}_{-0.0019}$ \\
0.6044&0.01250& 55.0& 0.0218$\pm0.0034^{+0.0064}_{-0.0033}$ \\
0.6044&0.00685& 55.0& 0.0296$\pm0.0034^{+0.0015}_{-0.0040}$ \\
0.6044&0.00404& 55.0& 0.0248$\pm0.0041^{+0.0066}_{-0.0030}$ \\
0.6044&0.00280& 55.0& 0.0459$\pm0.0053^{+0.0016}_{-0.0054}$ \\
0.6044&0.00188& 55.0& 0.0388$\pm0.0051^{+0.0026}_{-0.0035}$ \\
\hline
0.3125&0.01325& 55.0& 0.0281$\pm0.0039^{+0.0015}_{-0.0027}$ \\
0.3125&0.00780& 55.0& 0.0279$\pm0.0038^{+0.0033}_{-0.0023}$ \\
0.3125&0.00542& 55.0& 0.0345$\pm0.0039^{+0.0031}_{-0.0005}$ \\
0.3125&0.00363& 55.0& 0.0398$\pm0.0045^{+0.0016}_{-0.0025}$ \\
\hline
0.1209&0.02017& 55.0& 0.0232$\pm0.0080^{+0.0049}_{-0.0058}$ \\
0.1209&0.01402& 55.0& 0.0205$\pm0.0054^{+0.0023}_{-0.0027}$ \\
0.1209&0.00939& 55.0& 0.0261$\pm0.0049^{+0.0060}_{-0.0018}$ \\
\hline
\end{tabular}
\caption{The diffractive structure function multiplied by $\xpom$, $\xpom F^{D(3)}_2(\beta,\xpom,Q^2)$ for diffractive scattering, $\gamma^*p \to XN$, $M_N < 2.3$ {\rm GeV}, for $Q^2 = 55.0$ {\rm GeV$^2$}, in bins of $\beta$ and $\xpom$. }
\label{t:f2d304}
\end{center}
\normalsize
\end{table}
\clearpage

\begin{table}[p]
\begin{center}
\begin{tabular}{||c|ll||}
\hline
$Q^2$ (GeV$^2$)  &  $\beta$ &  $x_0F^{D(3)}_2 \pm {\rm stat.} \pm {\rm syst.}$\\
\hline\hline
  2.7&  0.2308&  0.0219$\pm 0.0022^{+ 0.0023}_{- 0.0022}$\\
  2.7&  0.0698&  0.0178$\pm 0.0014^{+ 0.0010}_{- 0.0009}$\\
  2.7&  0.0218&  0.0171$\pm 0.0015^{+ 0.0019}_{- 0.0013}$\\
  2.7&  0.0067&  0.0214$\pm 0.0023^{+ 0.0021}_{- 0.0012}$\\
  4.0&  0.3077&  0.0239$\pm 0.0015^{+ 0.0021}_{- 0.0012}$\\
  4.0&  0.1000&  0.0190$\pm 0.0011^{+ 0.0025}_{- 0.0018}$\\
  4.0&  0.0320&  0.0189$\pm 0.0013^{+ 0.0012}_{- 0.0017}$\\
\hline
\end{tabular}
\caption{The diffractive structure function of the proton multiplied by $\xpom$ at the point $\xpom \equiv x_0 = 0.01$, $x_0F^{D(3)}_2(\beta,x_0,Q^2)$, in bins of $Q^2$ and $\beta$.}
\label{t:f2d2tab}
\end{center}
\end{table}

\begin{figure}[p]
\begin{center}
\vfill
\vspace*{-1.5cm} \epsfig{file=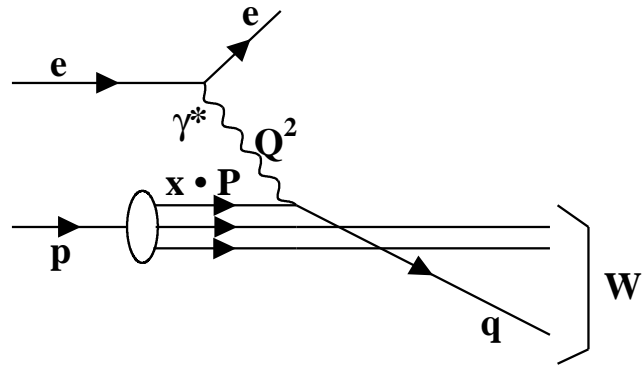,width=11.5cm,clip=}
\end{center}
\vspace*{-1.cm}
\caption{Non-peripheral deep inelastic scattering.} 
\label{f:nondifdiag}
\vfill
\end{figure}

\begin{figure}[p]
\vfill
\vspace*{-4.cm}
\hspace*{4.cm}
\epsfig{file=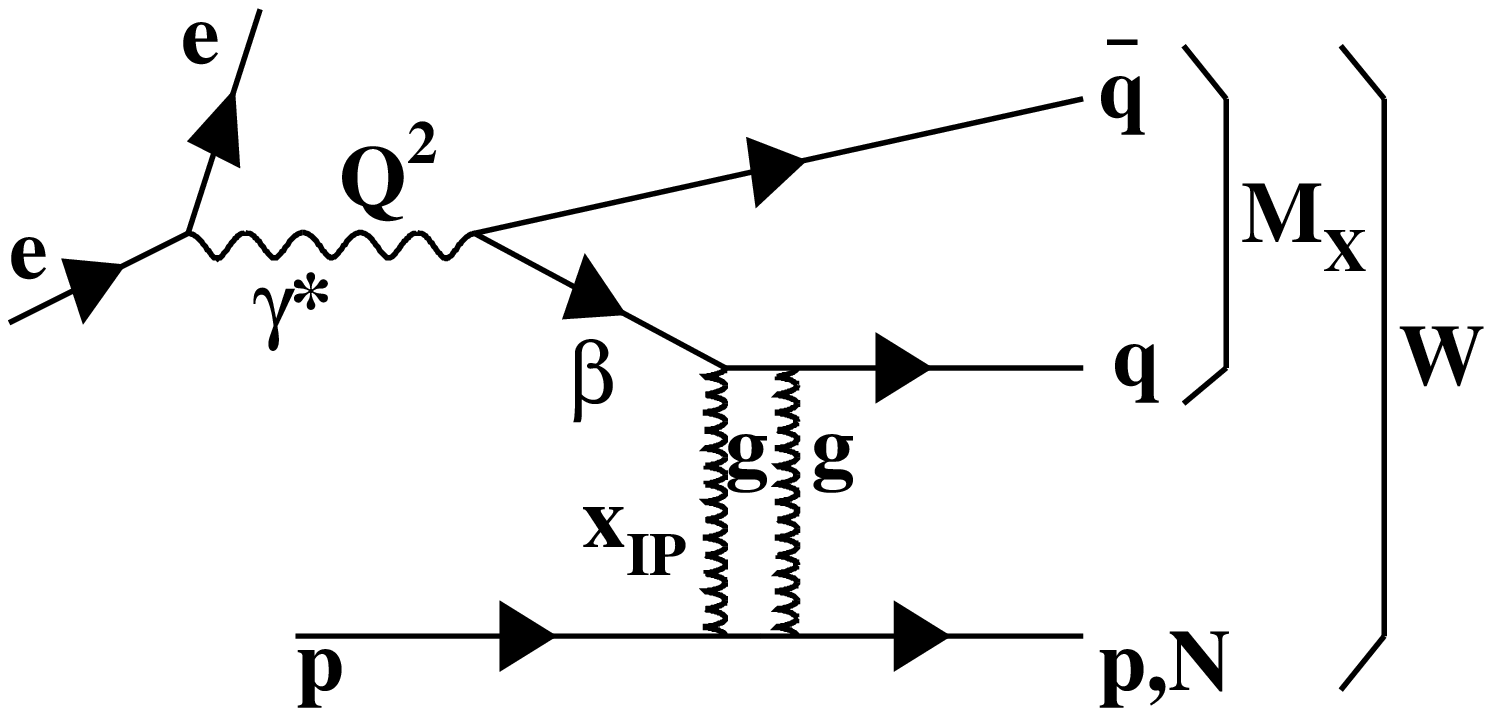,width=7.cm,clip=}
\vfill
\vspace*{0.5cm}
\hspace*{1cm}
\epsfig{file=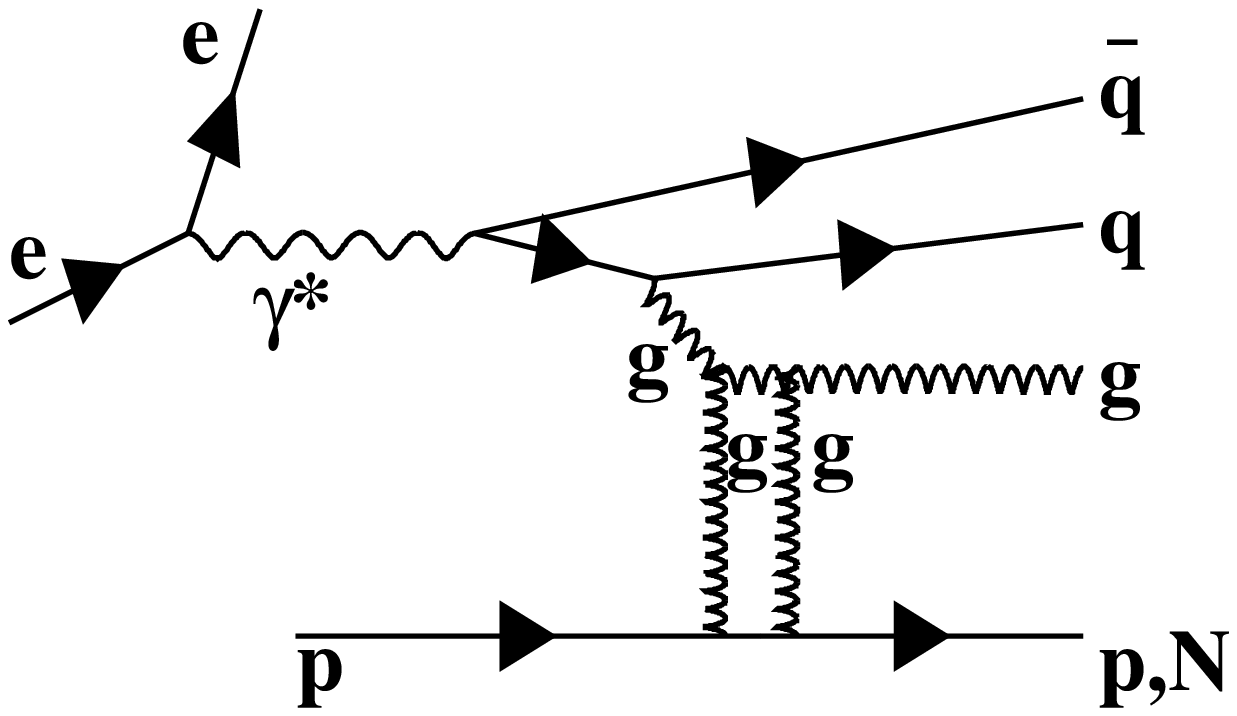,width=7.cm,clip=}
\hspace*{0.5cm}
\epsfig{file=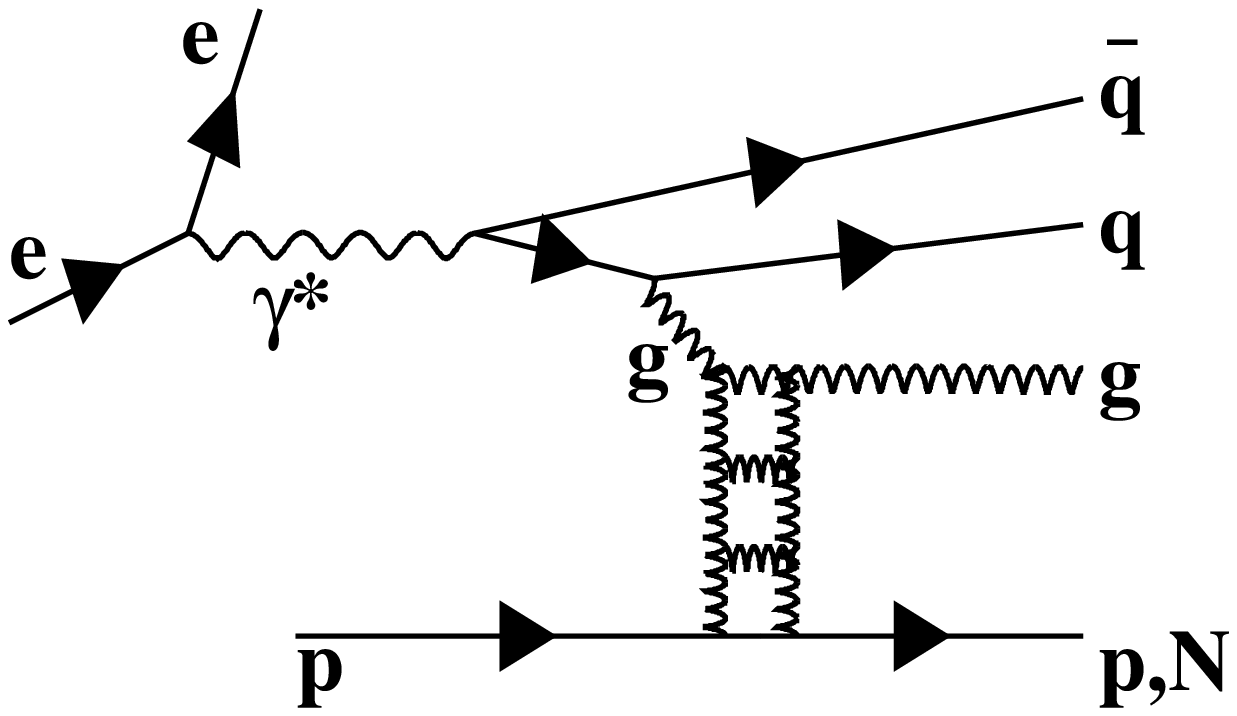,width=7.cm,clip=}
\caption{Diffractive deep inelastic scattering, $e p \to e X N$.} 
\label{f:diffdiag}
\vfill
\end{figure}
\clearpage
\begin{figure}[p]
\vfill
{\epsfig{file=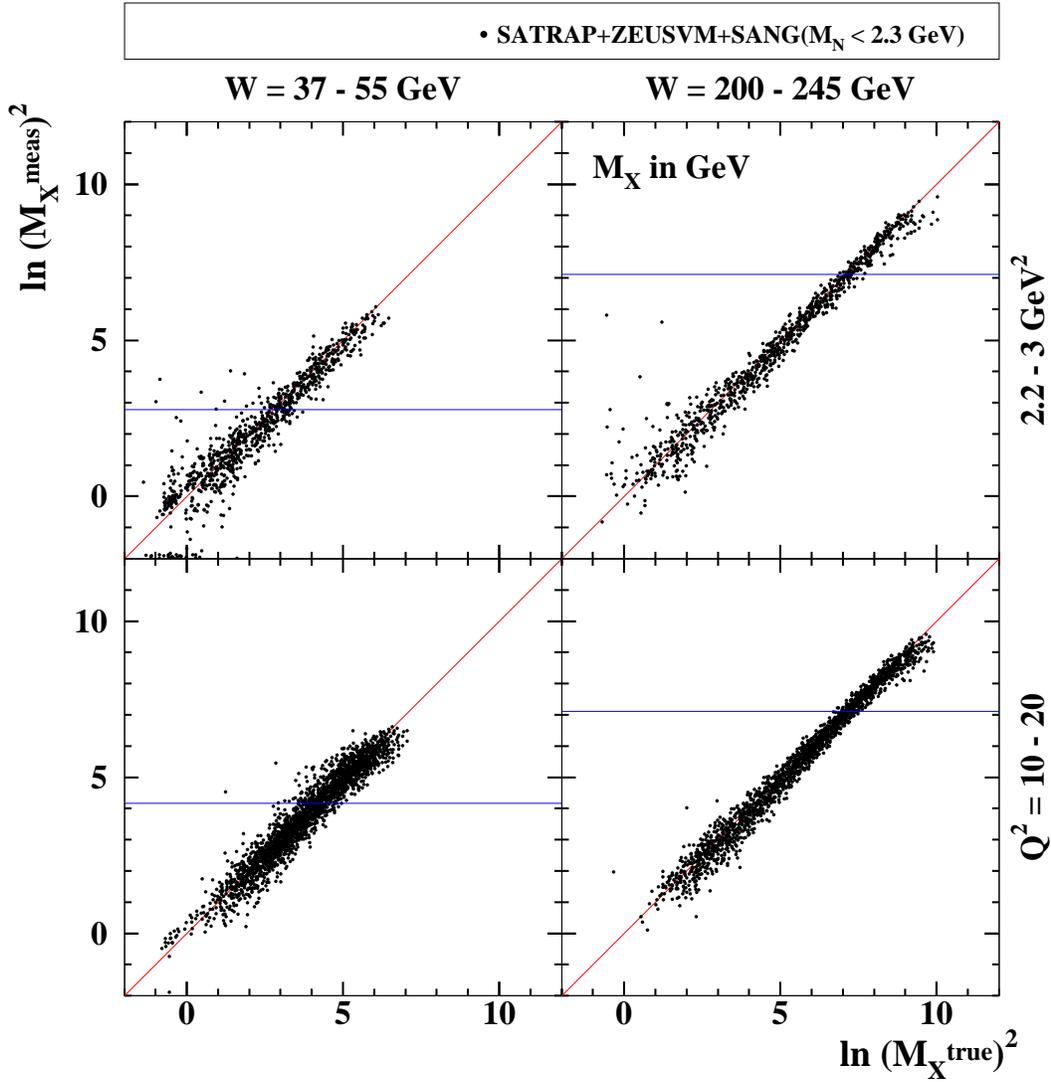,width=15cm}}
\caption{The measured versus the generated $\ln M^2_X$ values for the lowest and highest $W$ bins at low and high $Q^2$ as determined by Monte Carlo simulation. The horizontal lines give the maximum values of $\ln M^2_X$ for which the diffractive contribution was determined. Lines are drawn at $\ln (M^{\rm meas}_X)^2 = \ln (M^{\rm true}_X)^2$ to guide the eye.}
\label{f:mxresol}
\vfill
\end{figure}
\clearpage

\begin{figure}[p]
\vfill
{\epsfig{file=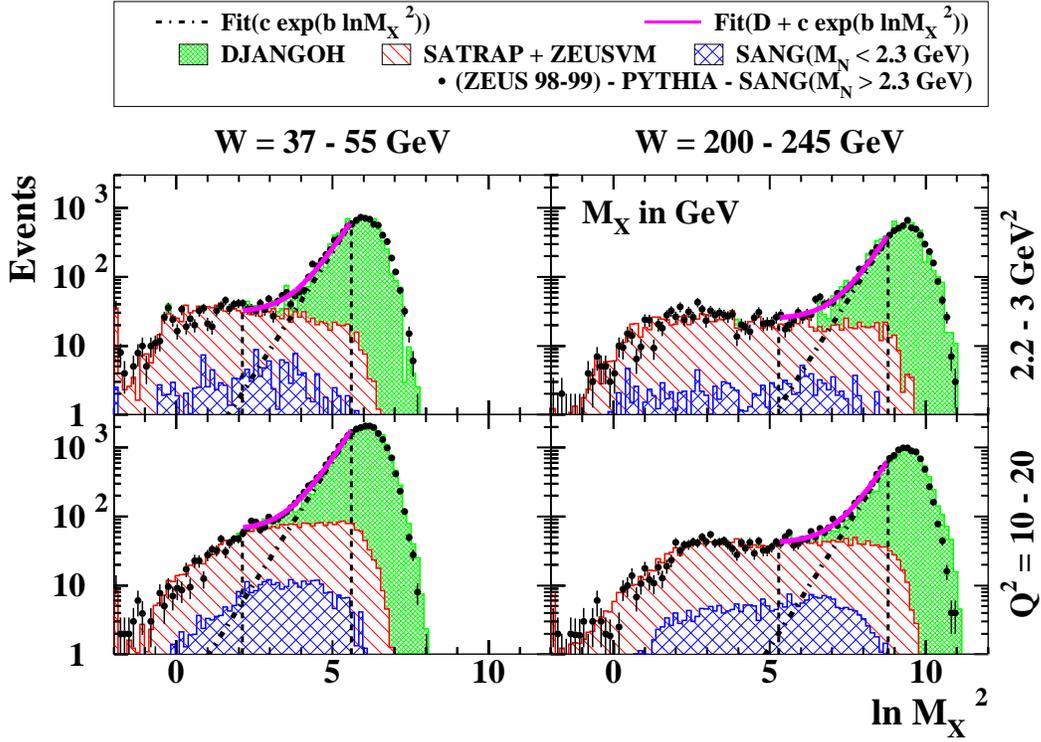,width=15cm}}
\caption{Distributions of $\ln M^2_X$ ($M_X$ in units of \GeV)  at the 
detector level for different ($W$, $Q^2$) bins. The points with error bars 
show the data. The shaded areas show the non-peripheral contributions as 
predicted by DJANGOH. The diffractive contributions from $\gamma^* p \to Xp$ 
($\gamma^* p \to XN$, $M_N < 2.3$ \GeV) as predicted by SATRAP+ZEUSVM (SANG) are shown as hatched (cross-hatched) areas. The dash-dotted lines show 
the results for the non-diffractive contribution from fitting the data 
in the $\ln M^2 _X$ range delimited by the two vertical dashed lines.}
\label{f:lnmxsel}
\vfill
\end{figure}%

%
\begin{figure}[p]
\begin{center}
\vfill
  \vspace*{-3.5cm}
{\epsfig{file=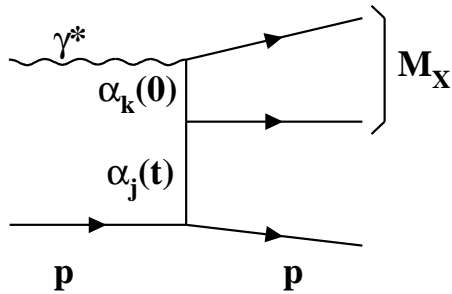,width=11.cm,clip=}}
\end{center}
  \vspace*{-1.cm}
\caption{The reaction $\gamma^* p \to Xp$ proceeding via the exchange of a Reggeon $\alpha_j$ in the $t$-channel. The system $X$ is produced by the scattering of the virtual photon on the Reggeon via the exchange of the pole $\alpha_k$.}
\label{f:gptripreg}
\vfill
\end{figure}%
\clearpage

\begin{figure}[p]
\vfill
{\epsfig{file=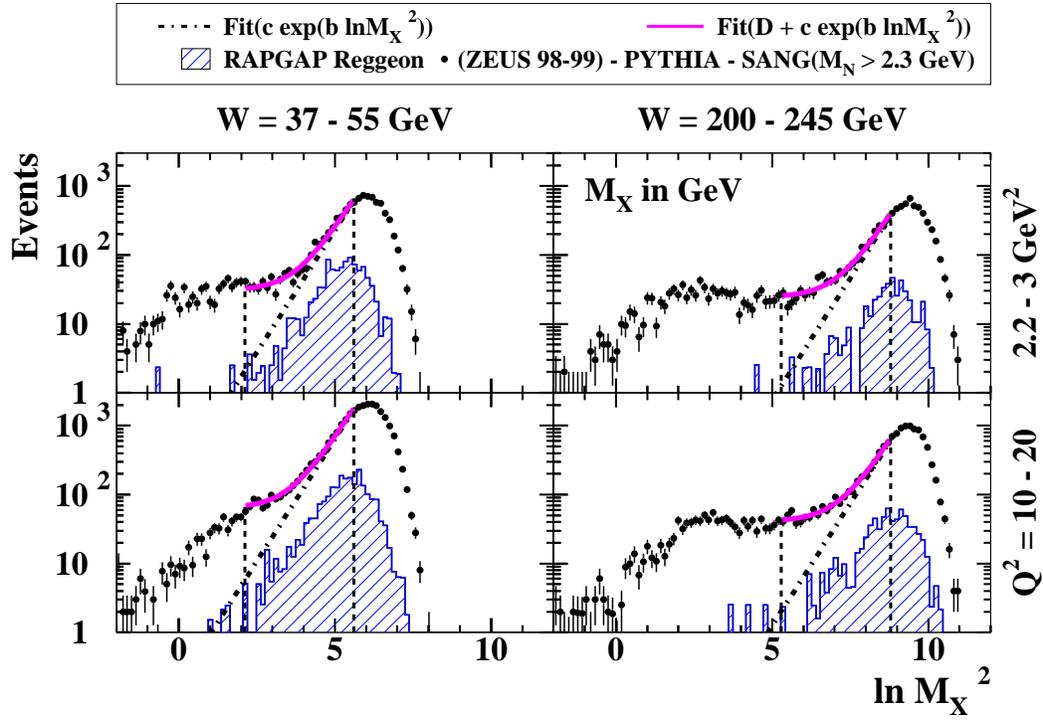,width=15cm}}
\caption{Distributions of $\ln M^2_X$ ($M_X$ in units of \GeV) at the detector level for different ($W$, $Q^2$) bins. The points with error bars show the data. The hatched histograms show the contributions predicted by the exchange of the $\rho$-Reggeon trajectory. The dash-dotted lines show the results for the non-diffractive contribution from fitting the sum of the diffractive and non-diffractive contributions in the $\ln M^2_X$ range delimited by the two vertical dashed lines.}
\label{f:lnmxselreggeon}
\vfill
\end{figure}%

\begin{figure}[p]
\vfill
{\epsfig{file=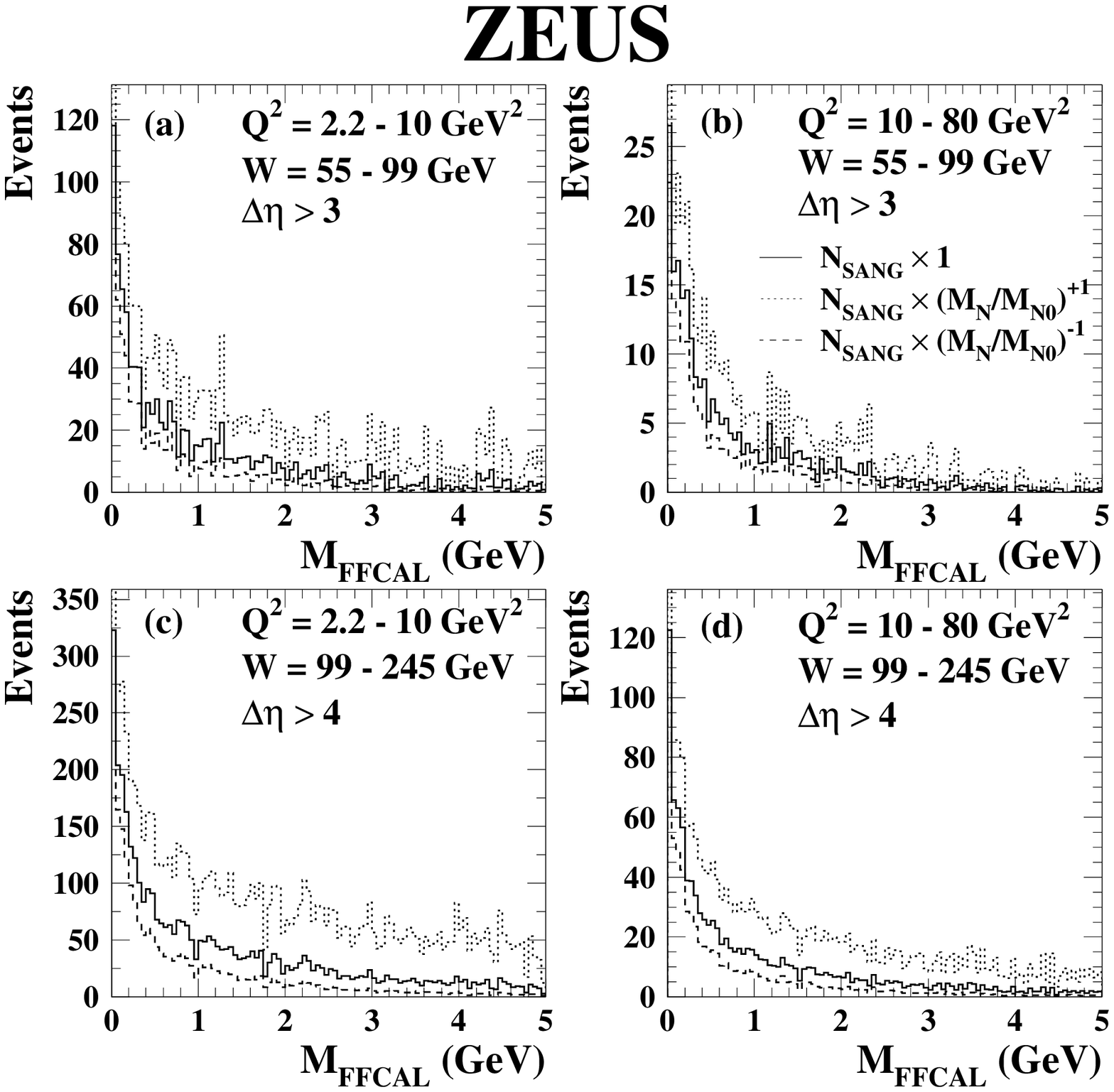,height=14cm,width=16cm}}
\caption
{The $M_{\rm FFCAL}$ distributions predicted by SANG (solid histograms), weighted by an extra factor $(\frac{M_N}{M_{N0}})^{-1}$ (dashed histograms) or by $(\frac{M_N}{M_{N0}})^{+1}$ (dotted histograms), where $M_{N0} = 2.3$ \GeV.  The $M_{\rm FFCAL}$ distributions are shown for four different regions of $W, Q^2$ and $\Delta \eta$.}
\label{f:mffcalsang}
\vfill
\end{figure}
\clearpage

\begin{figure}[p]
\vfill
{\epsfig{file=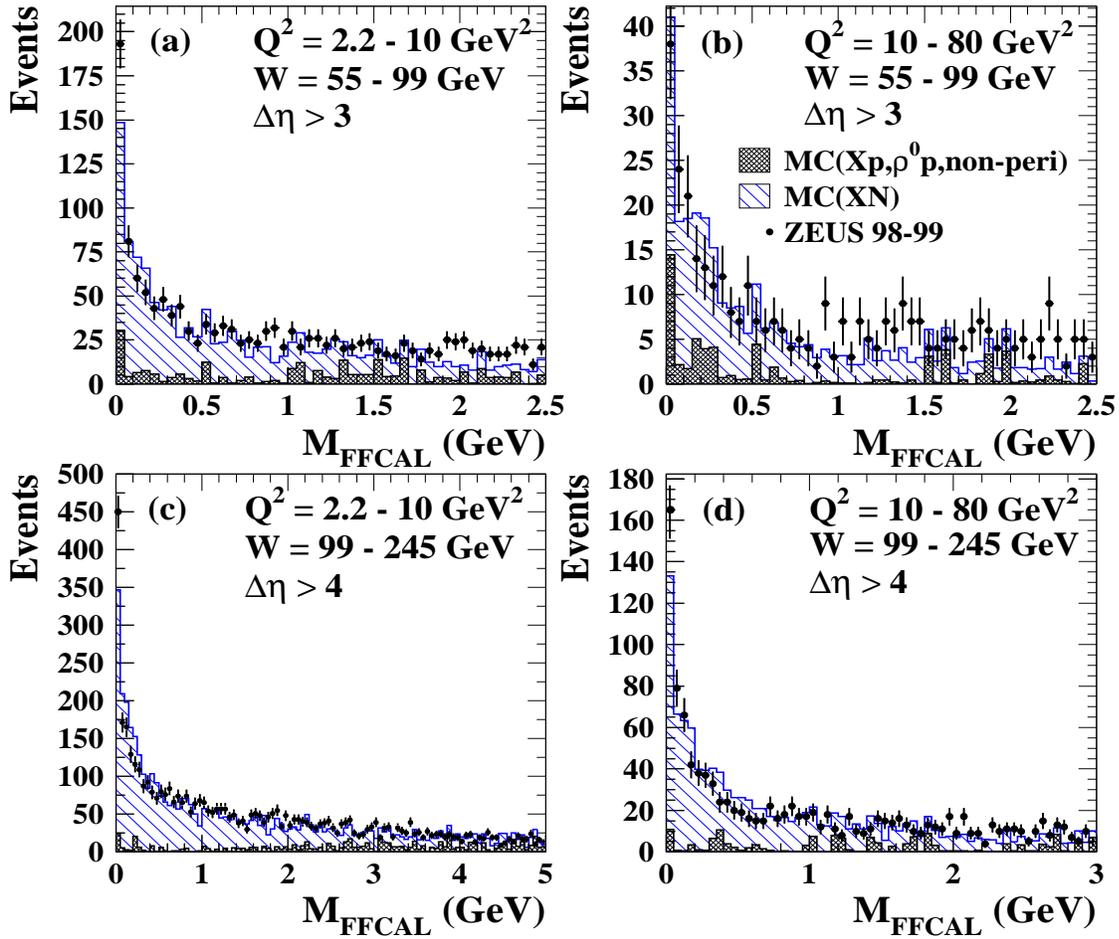,height=15cm,width=16cm}}
\caption
{Distributions of $M_{\rm FFCAL}$ for four different ($W,Q^2,\Delta \eta$) 
regions. The points with error bars show the data. The hatched  histograms 
show the MC predictions for the contribution from diffractive double 
dissociation ($XN$); the cross-hatched histograms show the sum of the 
contributions from $Xp$, $\rho^0 p$ and non-peripheral processes.}
\label{f:mffcal}
\vfill
\end{figure}
\clearpage

\begin{figure}[p]
\vfill
{\epsfig{file=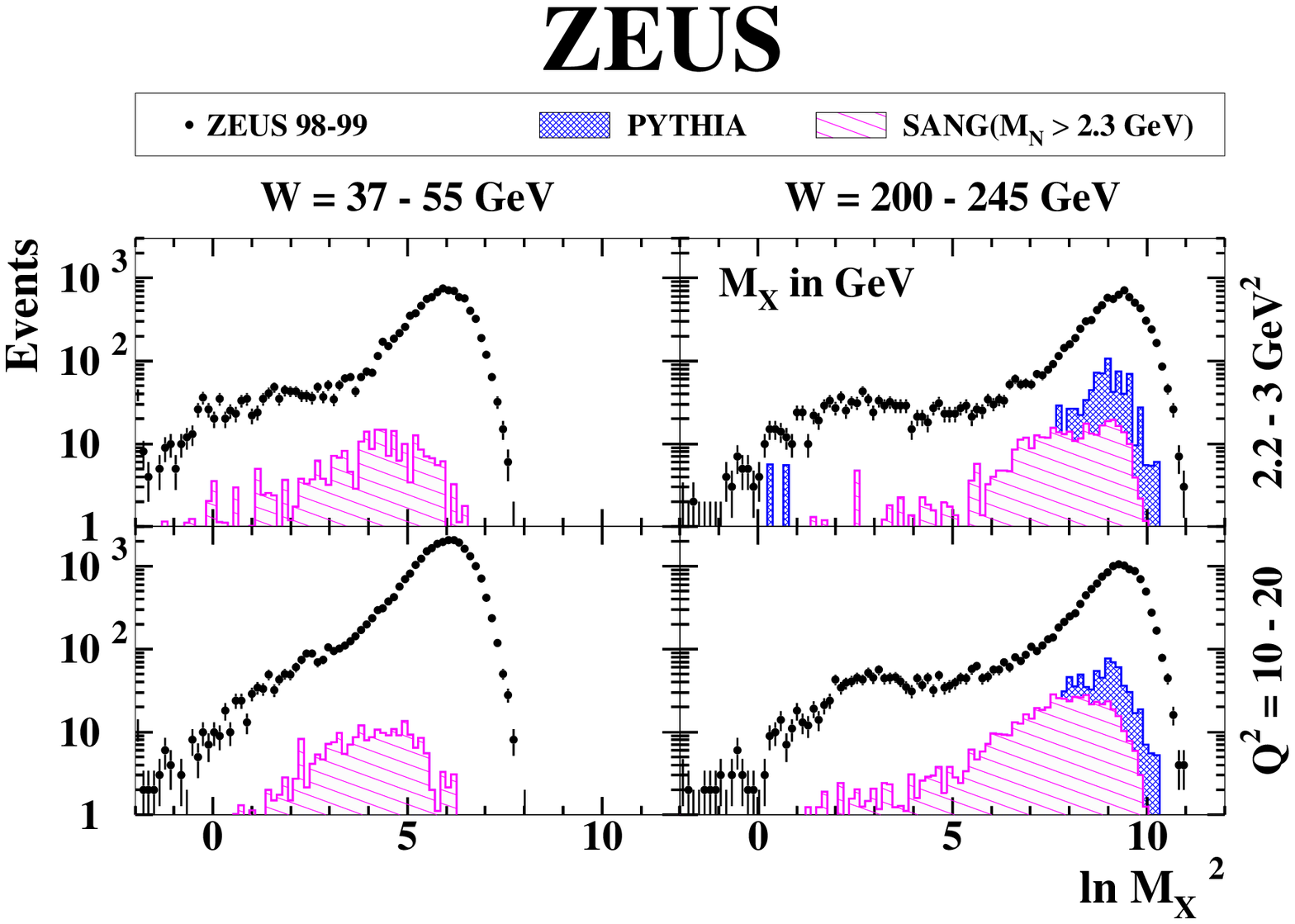,width=15cm}}
\caption{
Distributions of $\ln M^2_X$ ($M_X$ in units of \GeV) at the detector level for different ($W$, $Q^2$) bins. The points with error bars show the data. The hatched histograms show the expectation for the contribution from
diffractive double dissociation, $\gamma^* p \to XN$, where $N$ has a mass
more than 2.3 \GeV, as predicted by SANG. The cross-hatched histograms
show the background expected from photoproduction determined with PYTHIA.}
\label{f:lnmxxn}
\vfill
\end{figure}
\clearpage

%
\begin{figure}[p]
\vfill
 \vspace*{-0.5cm} \hspace*{0.7cm}
{\epsfig{file=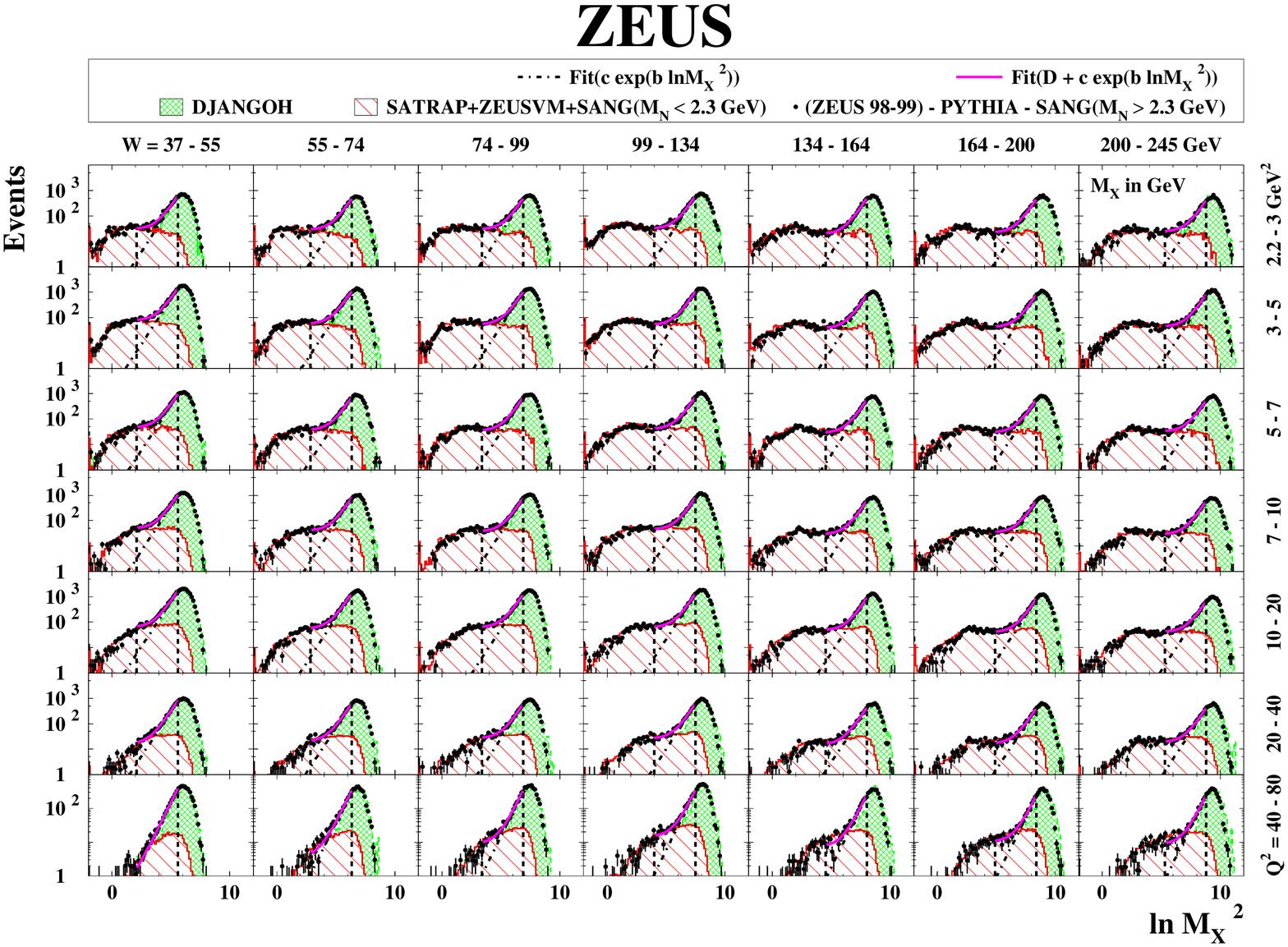,width=19.5cm,angle=90}}
 \vspace*{-0.5cm}
\caption{Distributions of $\ln M^2_X$ ($M_X$ in units of \GeV) at the detector level for different ($W$, $Q^2$) bins. The full points with error bars show the data. The diffractive contributions from $\gamma^* p \to XN$, $M_N < 2.3$ \GeV, as predicted by SATRAP+ZEUSVM+SANG are shown by the hatched histograms.
The cross-hatched histograms show the non-peripheral contributions as 
predicted by DJANGOH. The dash-dotted lines show the results for the 
non-diffractive contribution from fitting the data in the $\ln M_X ^2$ 
range delimited by the two vertical dashed lines.}
\label{f:lnmxall}
\vfill
\end{figure}
\clearpage
\begin{figure}[p]
\vfill
{\epsfig{file=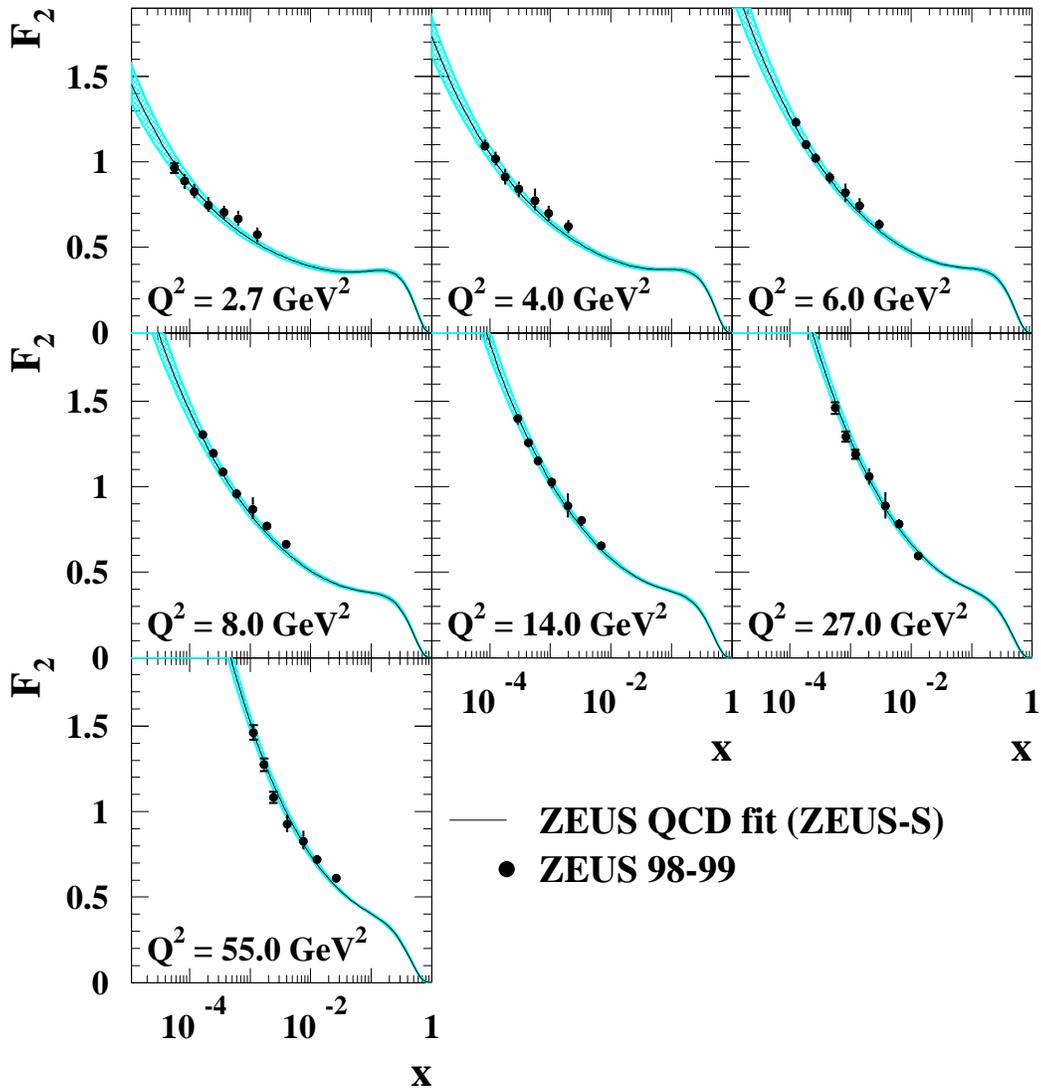,width=16cm}}
\caption{The proton structure function $F_2$ determined in this analysis for the $Q^2$ values indicated. The inner error bars show the statistical uncertainties and the full bars the statistical and systematic systematic uncertainties added in quadrature. The line shows the result of ZEUS QCD fit with its uncertainty band.}
\label{f:fpcf2}
\vfill
\end{figure}
\clearpage

\begin{figure}[p]
\vfill
\begin{center}
{\epsfig{file=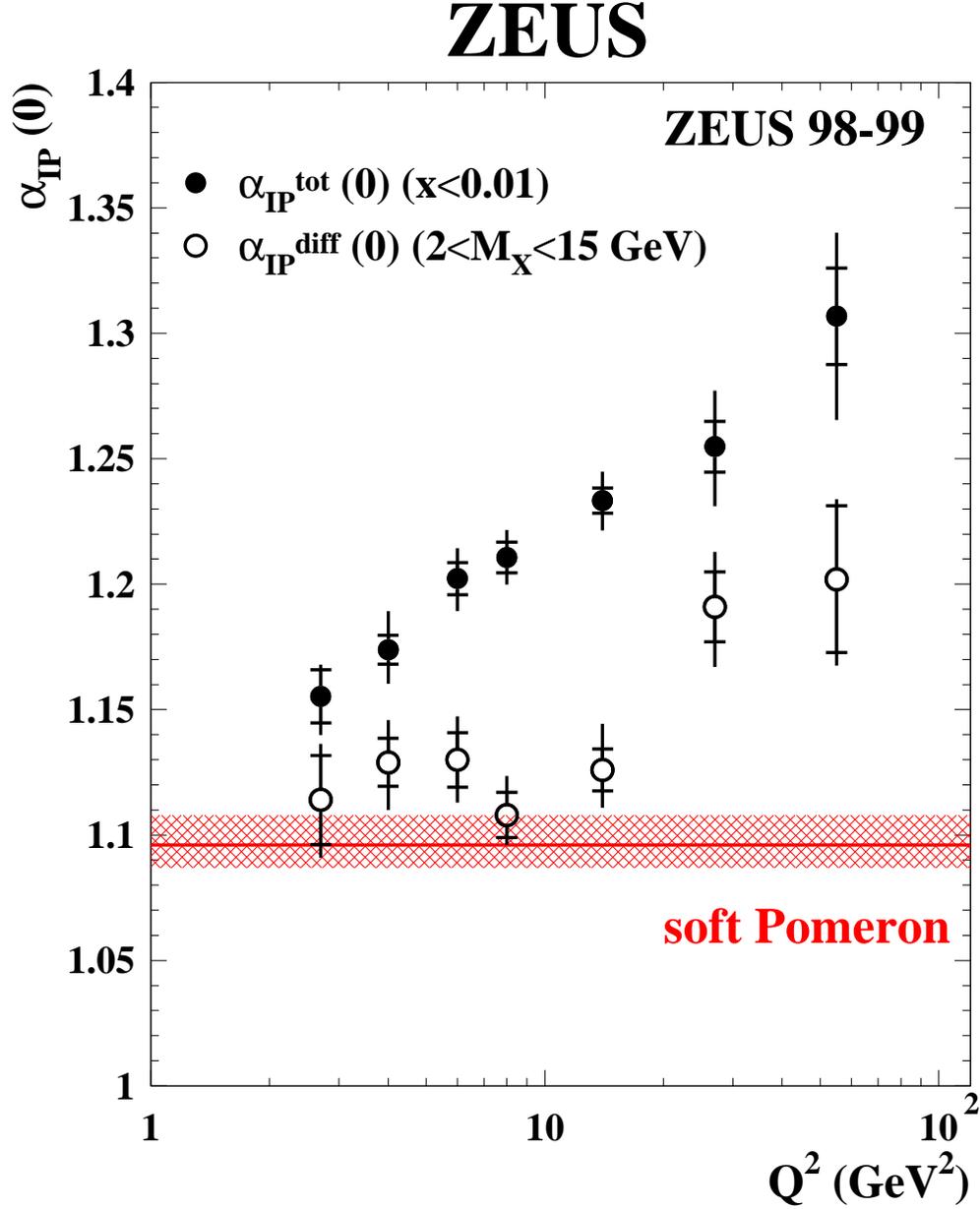,width=15cm}}
\end{center}
\caption{The intercepts of the Pomeron trajectory, $\alpha^{\rm tot}_{\pom}(0)$
and $\alpha^{\rm diff}_{\pom}(0)$, as a function of $Q^2$, obtained from the 
$W$ dependences of the total $\gamma^*p$ cross section and of the 
diffractive cross section, $d\sigma^{\rm diff}_{\gamma^* p \to XN}/dM_X$ for 
$2 < M_X < 15$ \GeV. The inner error bars show the statistical uncertainties and the full bars the combined statistical and systematic systematic uncertainties. The $\alpha_{\pom}(0)$ values for the latter were corrected for the $t$ dependence of $\alpha^{\rm diff}_{\pom}$. The cross-hatched band shows the expectation for the soft Pomeron.}
\label{f:slopef2dif}
\vfill
\end{figure}

\begin{figure}[p]
\vfill
{\epsfig{file=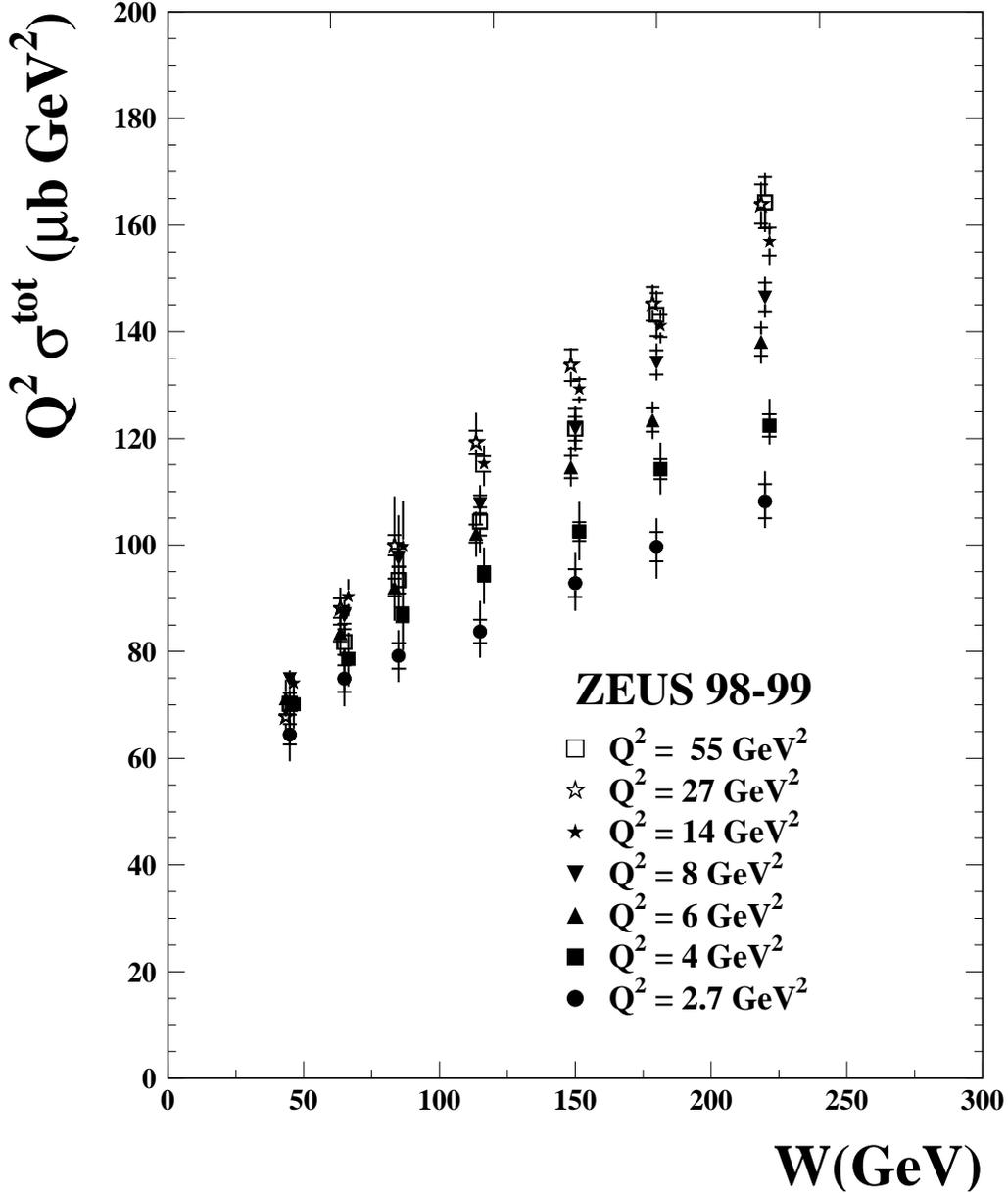,width=14.5cm}}
\caption{The total virtual photon-proton cross section, $\sigma^{\rm tot}_{\gamma^{\ast} p}$, multiplied by $Q^2$, as a function of $W$, for the $Q^2$ intervals indicated. The inner error bars show the statistical uncertainties and the full bars the statistical and systematic systematic uncertainties added in quadrature. For better visibility, the points for adjacent values of $Q^2$ were shifted in $W$ by zero, +1.5 \GeV or -1.5 \GeV.}
\label{f:sigtot}
\vfill
\end{figure}

\begin{figure}[p]
\vfill
{\epsfig{file=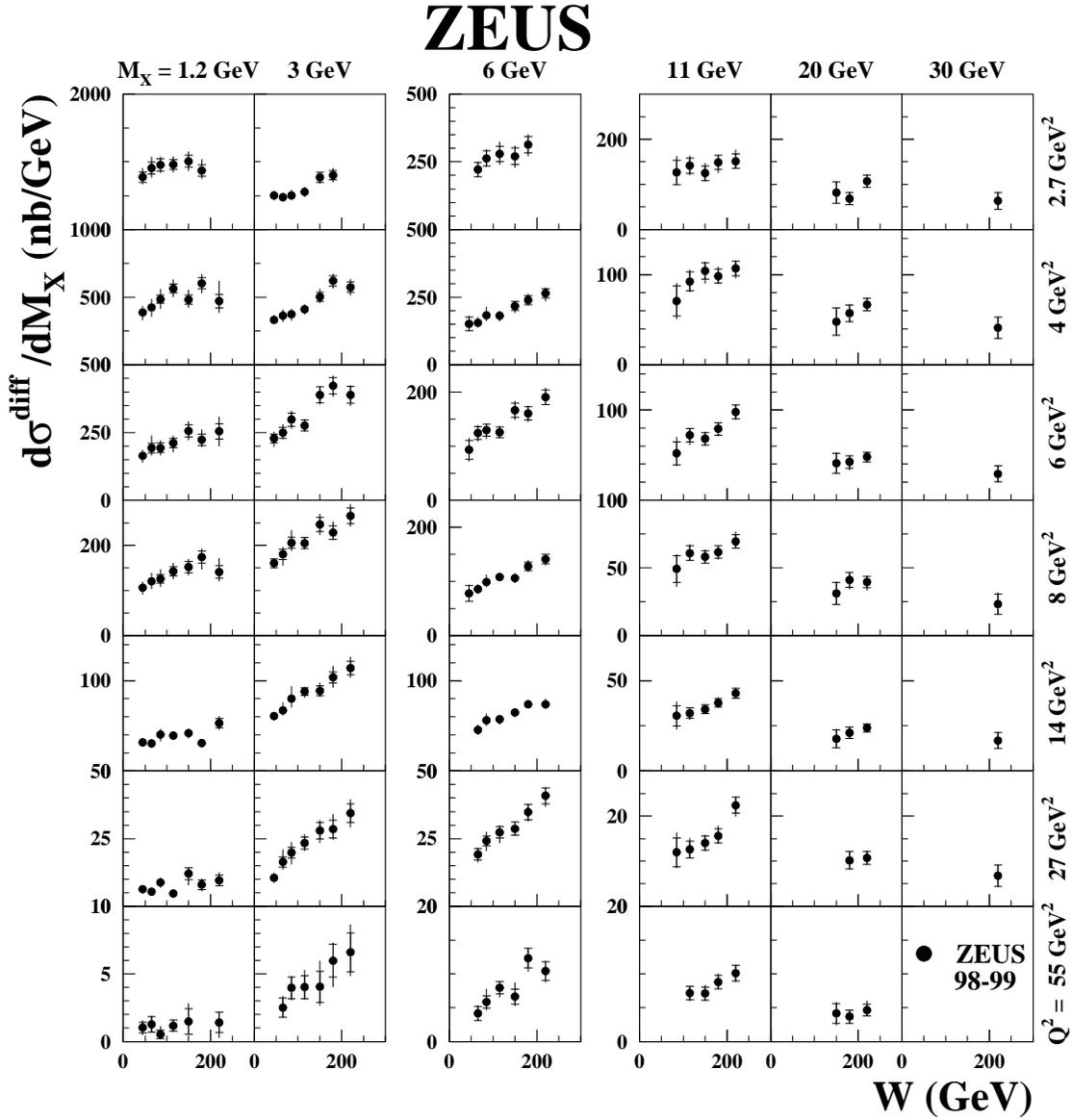,height=15.4cm,clip=}}
\caption{The differential cross sections, $d\sigma^{\rm diff}_{\gamma^*p \to XN}/dM_X$, $M_N < 2.3$ \GeV, as a function of $W$ for bins of $M_X$ and $Q^2$. The inner error bars show the statistical uncertainties and the full bars the statistical and systematic uncertainties added in quadrature.}
\label{f:dsigdmx}
\vfill
\end{figure}
%

\begin{figure}[p]
\vfill
{\epsfig{file=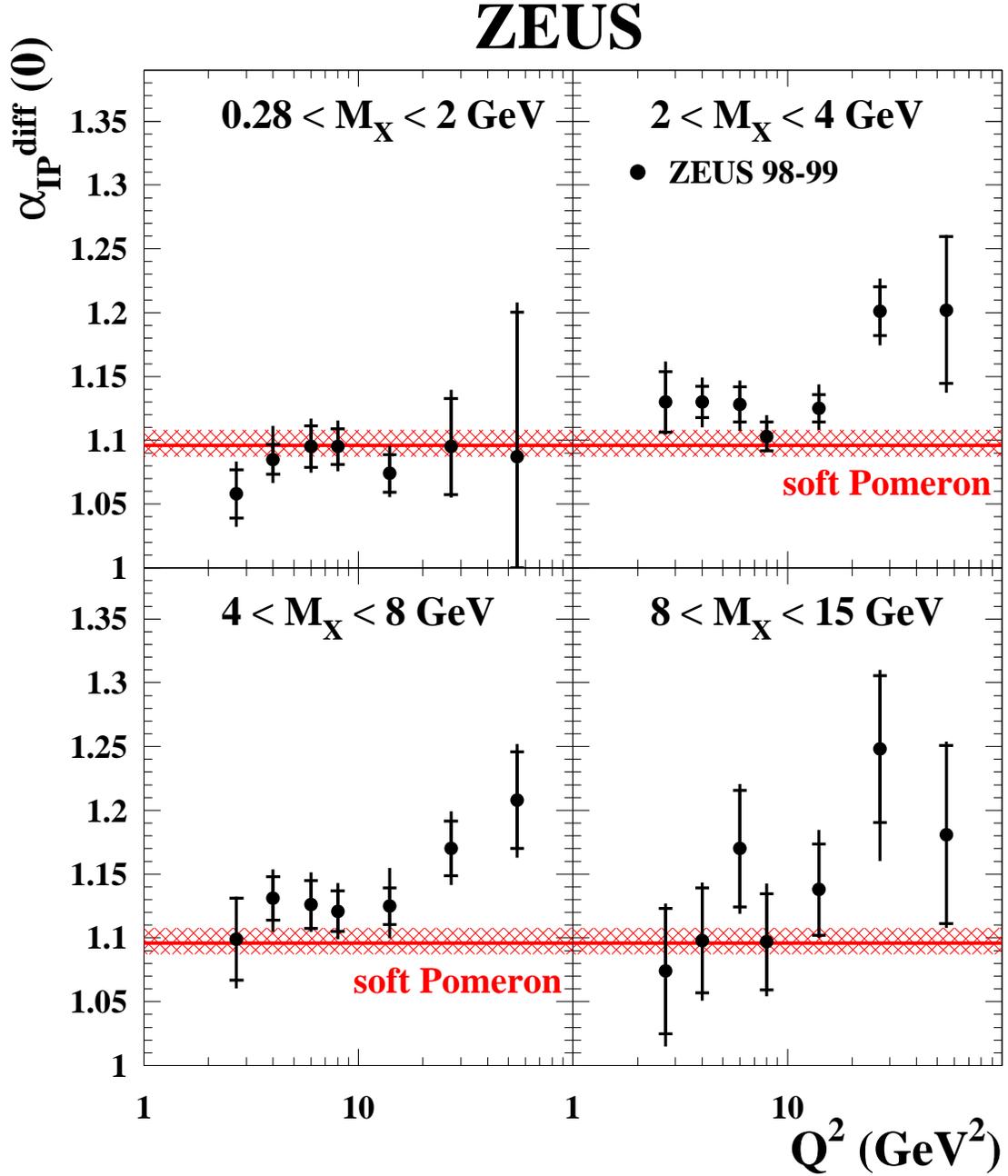,width=17cm}}
\caption{The intercept $\alpha^{\rm diff}_{\pom}(0)$, obtained from fitting the diffractive cross section, $d\sigma^{\rm diff}_{\gamma^* p \to XN}(M_X,W, Q^2)/dM_X$, as a function of $Q^2$ for the different $M_X$ bins. The inner error bars show the statistical uncertainties and the full bars the combined statistical and systematic systematic uncertainties. The cross-hatched band shows the expectation for the soft Pomeron, see text.}
\label{f:difslope}
\vfill
\end{figure}
%

%
\begin{figure}[p]
\vfill
{\epsfig{file=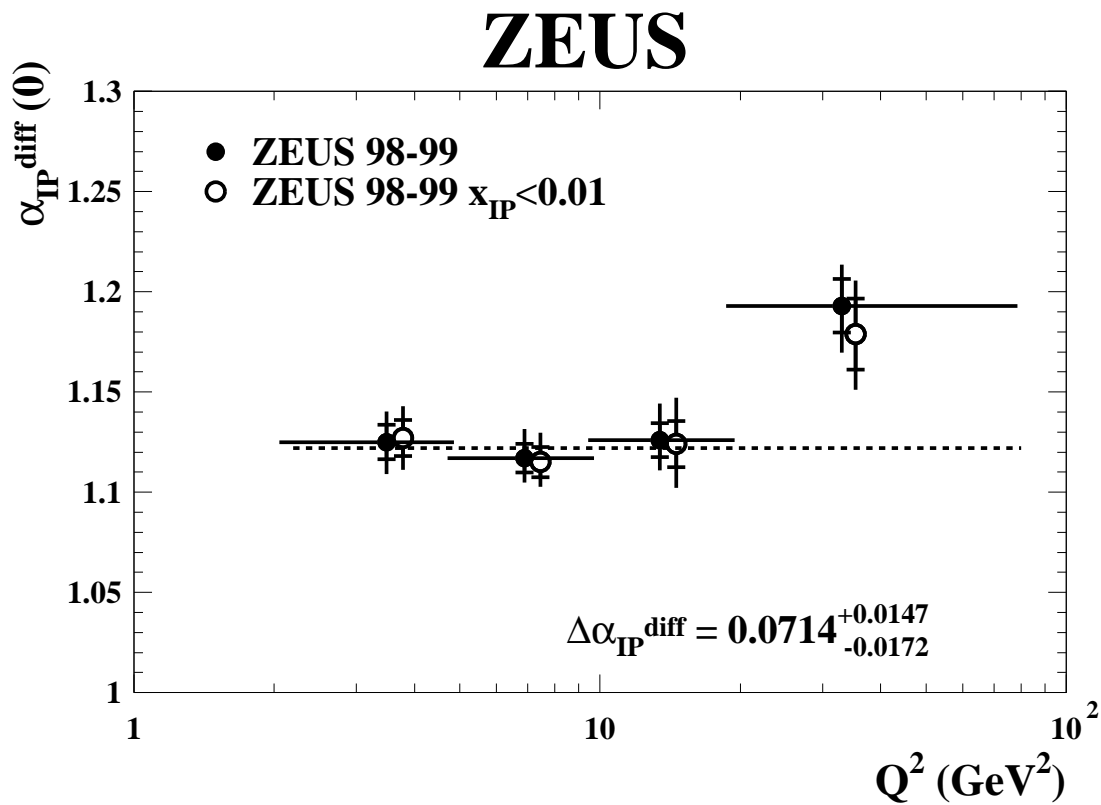,width=16cm,clip=}}
\caption{The $Q^2$ dependence of $\alpha^{\rm diff}_{\pom}(0)$ as determined 
from fitting the $W$ dependence of the diffractive cross section for 
$2 < M_X < 15$ \GeV. The inner error bars show the statistical uncertainties and the full bars the combined statistical and systematic systematic uncertainties. The horizontal bars represent the bin widths. The open symbol indicates the results which were determined using the diffractive cross section
with $x_{\pom} < 0.01$. For better visibility, the points obtained from the 
same $Q^2$ range were shifted in $Q^2$ by 4 $\%$ and -4 $\%$. 
The dashed line shows $\alpha^{\rm diff}_{\pom}(0)$ averaged over 
$2.2 < Q^2 < 20$ \GeV$^2$. The difference $\Delta \alpha^{\rm diff}_{\pom}$ is defined as 
$ \alpha^{\rm diff}_{\pom}(0,\langle Q^2 \rangle = 34.6 \; {\rm GeV}^2) - \alpha^{\rm diff}_{\pom}(0, \langle Q^2 \rangle  = 7.8 \; {\rm GeV}^2)$.}
\label{f:alpomvsq2}
\vfill
\end{figure}

%
\begin{figure}[p]
\vfill
{\epsfig{file=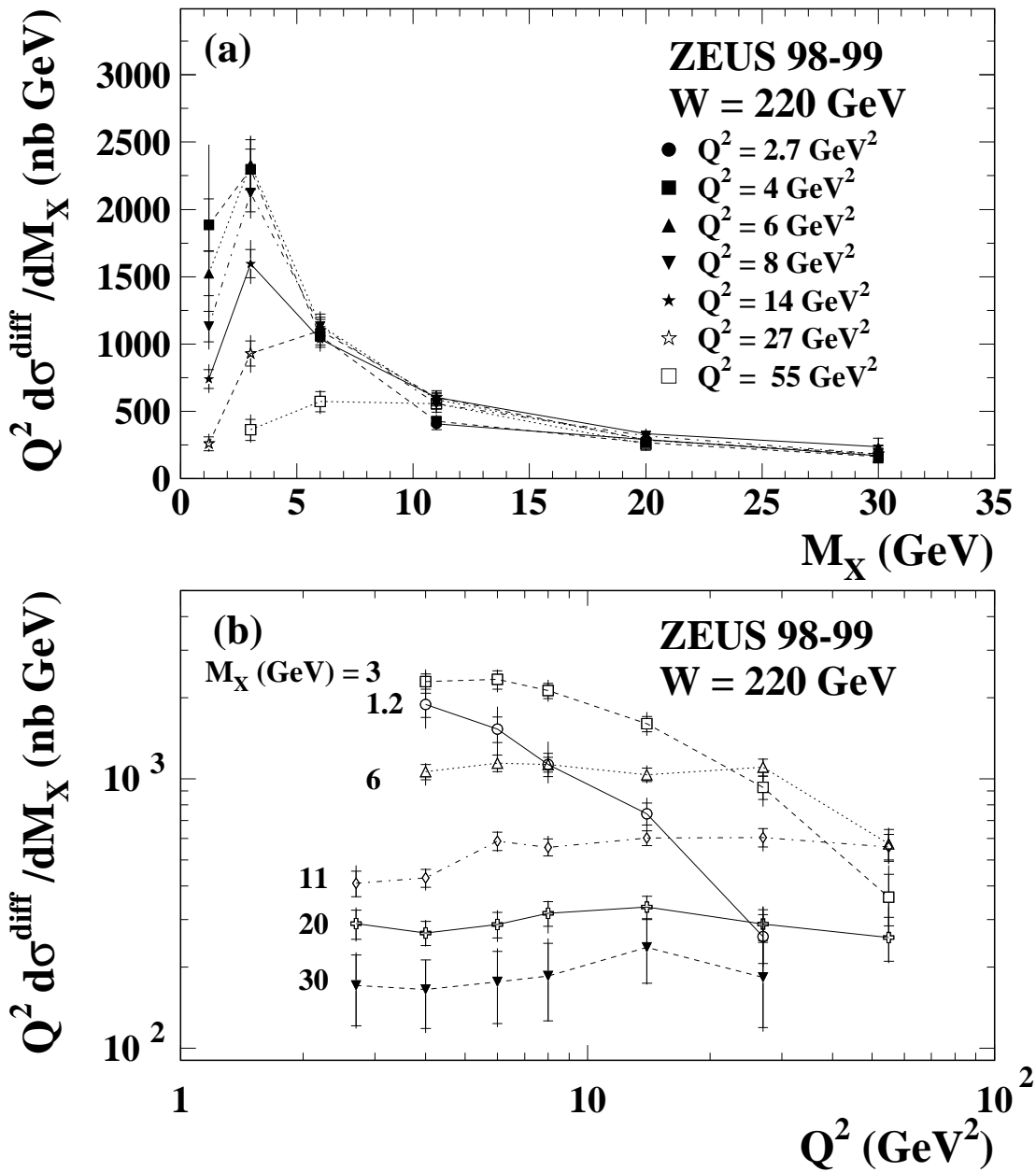,width=15.2cm,clip=}}
\caption{The diffractive cross section multiplied by $Q^2$,  $Q^2 d\sigma^{\rm diff}_{\gamma^*p \to XN}/dM_X$, $M_N < 2.3$\GeV, for $W = 220$\GeV:(a) as a function of $M_X$ for the $Q^2$ intervals indicated; the lines connect the points measured for the same value of $Q^2$; (b) as a function of $Q^2$ for the $M_X$ intervals indicated; the lines connect the points measured for the same value of $M_X$. The inner error bars show the statistical uncertainties and the full bars the statistical and systematic uncertainties added in quadrature. }
\label{f:dsigdmxvsmx}
\vfill
\end{figure}
\clearpage

%
\begin{figure}[p]
\vfill
{\epsfig{file=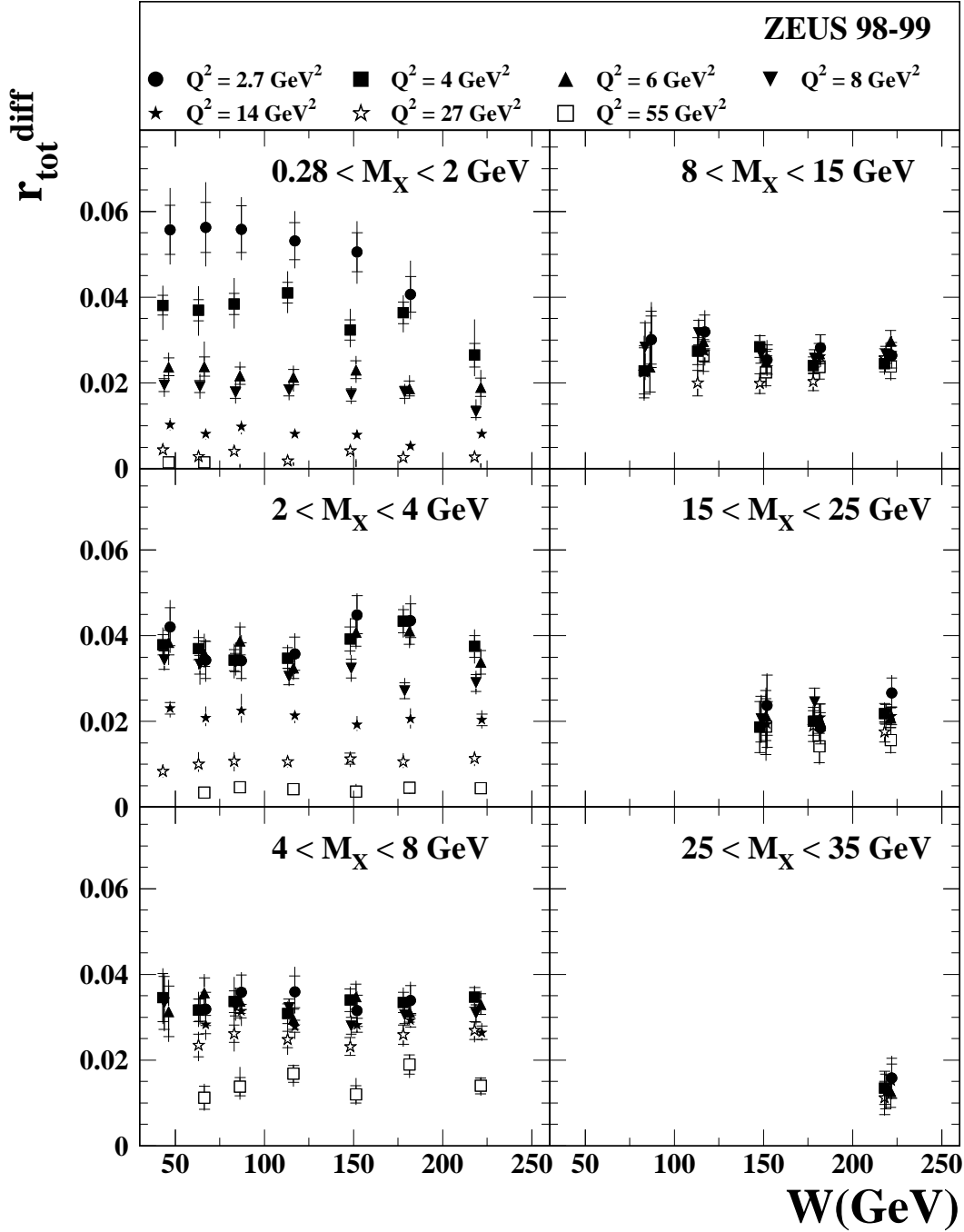,width=15.2cm,clip=}}
\caption{The ratio of the diffractive cross section, integrated over the $M_X$ intervals indicated, $\int^{M_b}_{M_a} dM_X d\sigma^{\rm diff}_{\gamma^* p \to XN, M_N < 2.3 {\rm GeV}}/dM_X$, to the total $\gamma ^{\ast}p$ cross section, $r^{\rm diff}_{\rm tot} = \sigma^{\rm diff}/\sigma^{\rm tot}_{\gamma^*p}$, as a function of $W$ for the $M_X$ and $Q^2$ intervals indicated. The inner error bars show the statistical uncertainties and the full bars the statistical and systematic uncertainties added in quadrature.}
\label{f:rdiftot}
\vfill
\end{figure}

%
\begin{figure}[p]
\centerline
{\epsfig{file=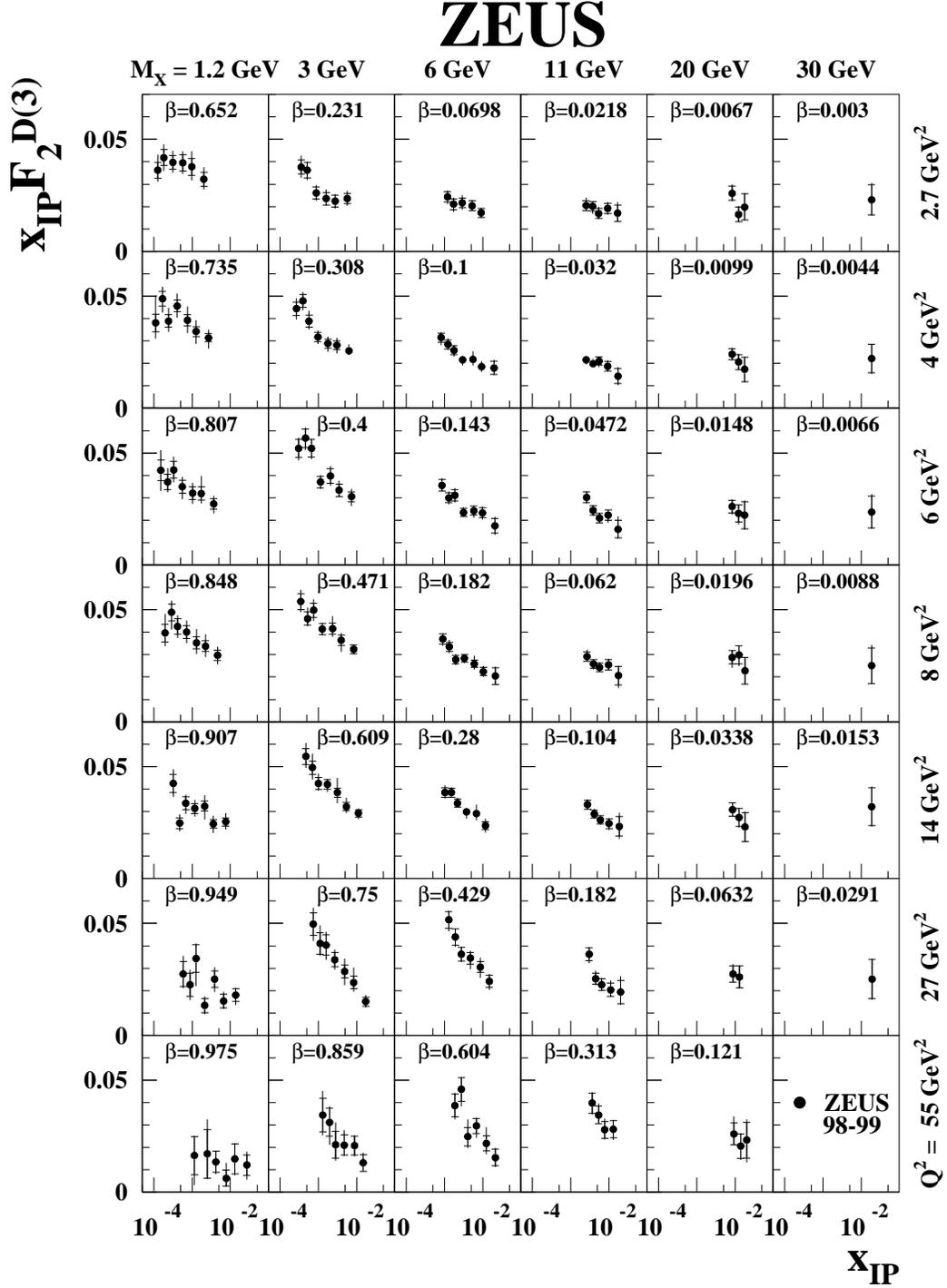,width=14.3cm}}
\caption{The diffractive structure function of the proton multiplied by $\xpom$, $\xpom F^{D(3)}_2$, for $\gamma^*p \to XN$, $M_N < 2.3$ \GeV as a function of $\xpom$ for different regions of $\beta$ and $Q^2$. The inner error bars show the statistical uncertainties and the full bars the statistical and systematic uncertainties added in quadrature.}
\label{f:f2d3}
\end{figure}
\clearpage

%
\begin{figure}[p]
\centerline
{\epsfig{file=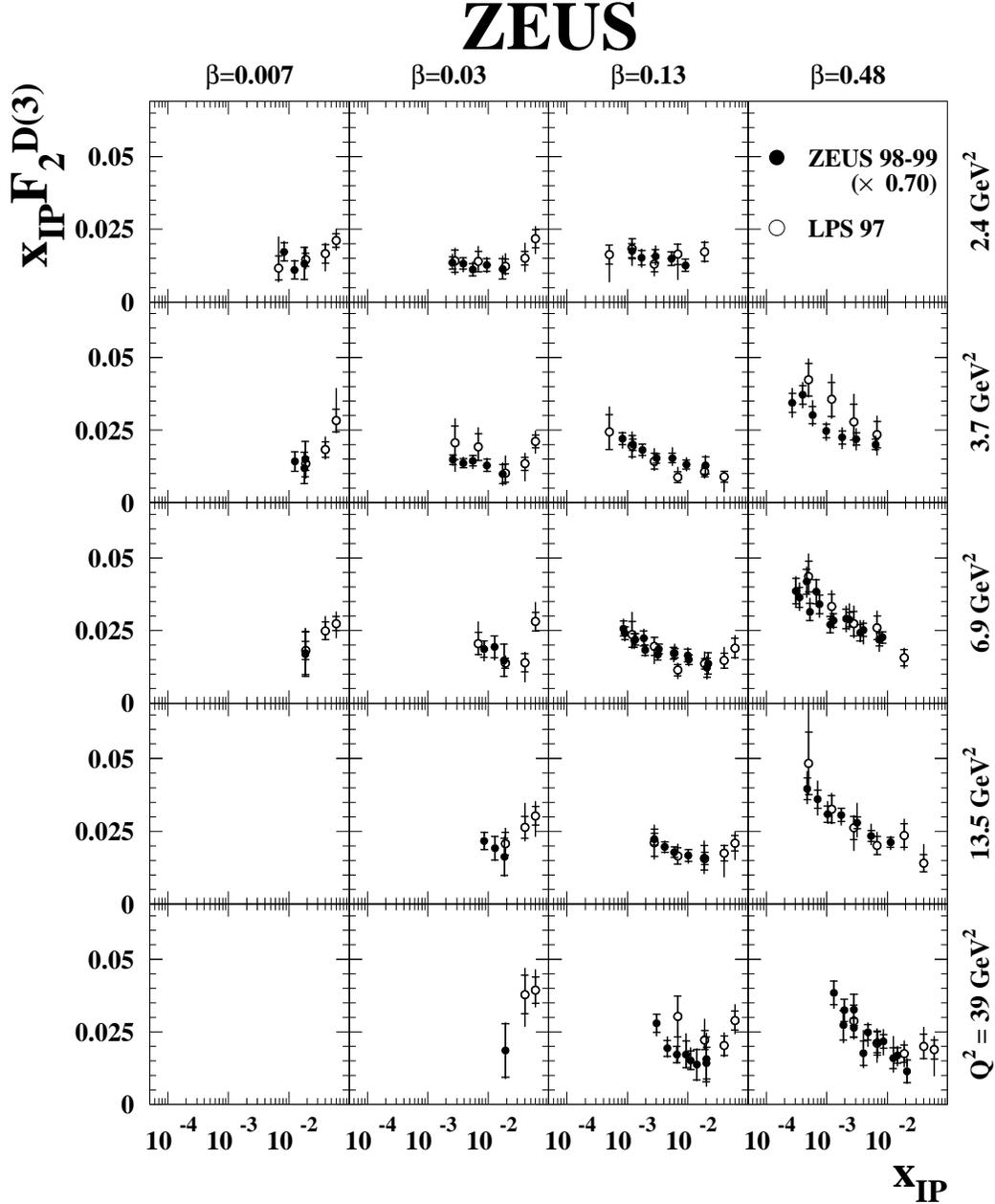,width=14.3cm,clip=}}
\caption{The diffractive structure function of the proton multiplied by 
$\xpom$, $\xpom F^{D(3)}_2$, from this analysis multiplied by a factor 
0.70 to account for the contribution from diffractive nucleon dissociation 
with $M_N < 2.3$ \GeV (solid points), compared with the results from the ZEUS LPS measurement. The inner error bars show the statistical uncertainties and the full bars the statistical and systematic systematic uncertainties added in quadrature. For ease of comparison, the results from this analysis have 
been transported to the ($\beta, Q^2$) values used by the ZEUS LPS analysis 
(open points). Only bins of ($\beta, Q^2$) in which both analyses have results are shown.}
\label{f:f2d3fpclps}
\end{figure}
\clearpage

%
\begin{figure}[p]
\centerline
{\epsfig{file=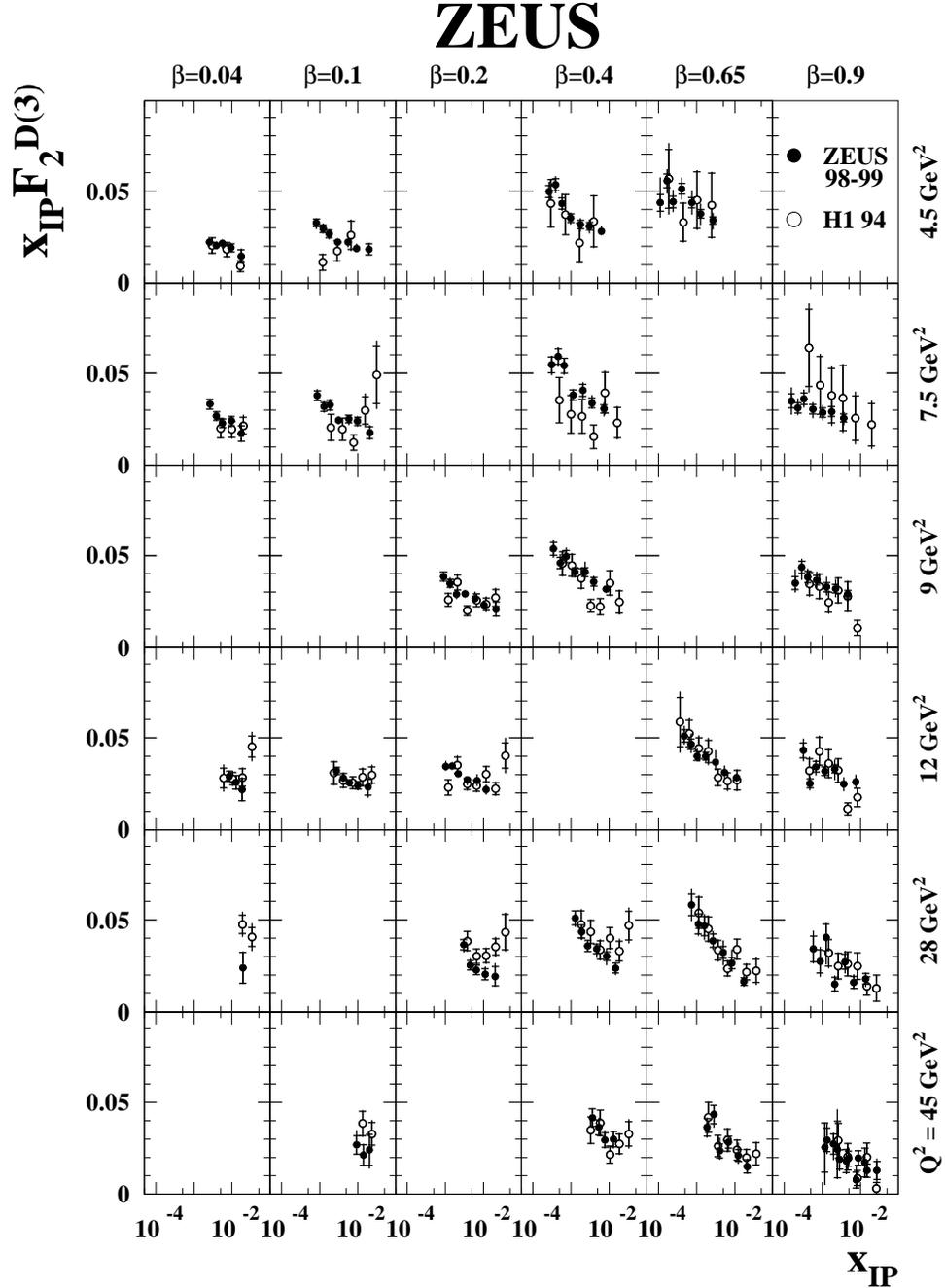,width=13.3cm}}
\caption{The diffractive structure function of the proton multiplied by 
$\xpom$, $\xpom F^{D(3)}_2$, from this analysis for 
$\gamma^* p \to XN, M_N < 2.3$ \GeV (solid points) compared with the H1 
results given for $M_N < 1.6$ \GeV (open points). The inner error bars show the statistical uncertainties and the full bars the statistical and systematic systematic uncertainties added in quadrature. For ease of comparison, 
the results from 
this analysis have been transpoted to ($\beta, Q^2$) values used by the H1 
analysis. Only bins of ($\beta, Q^2$) in which both analyses have results are shown.}
\label{f:f2d3fpch1}
\end{figure}
\clearpage

\begin{figure}[p]
\centerline
{\epsfig{file=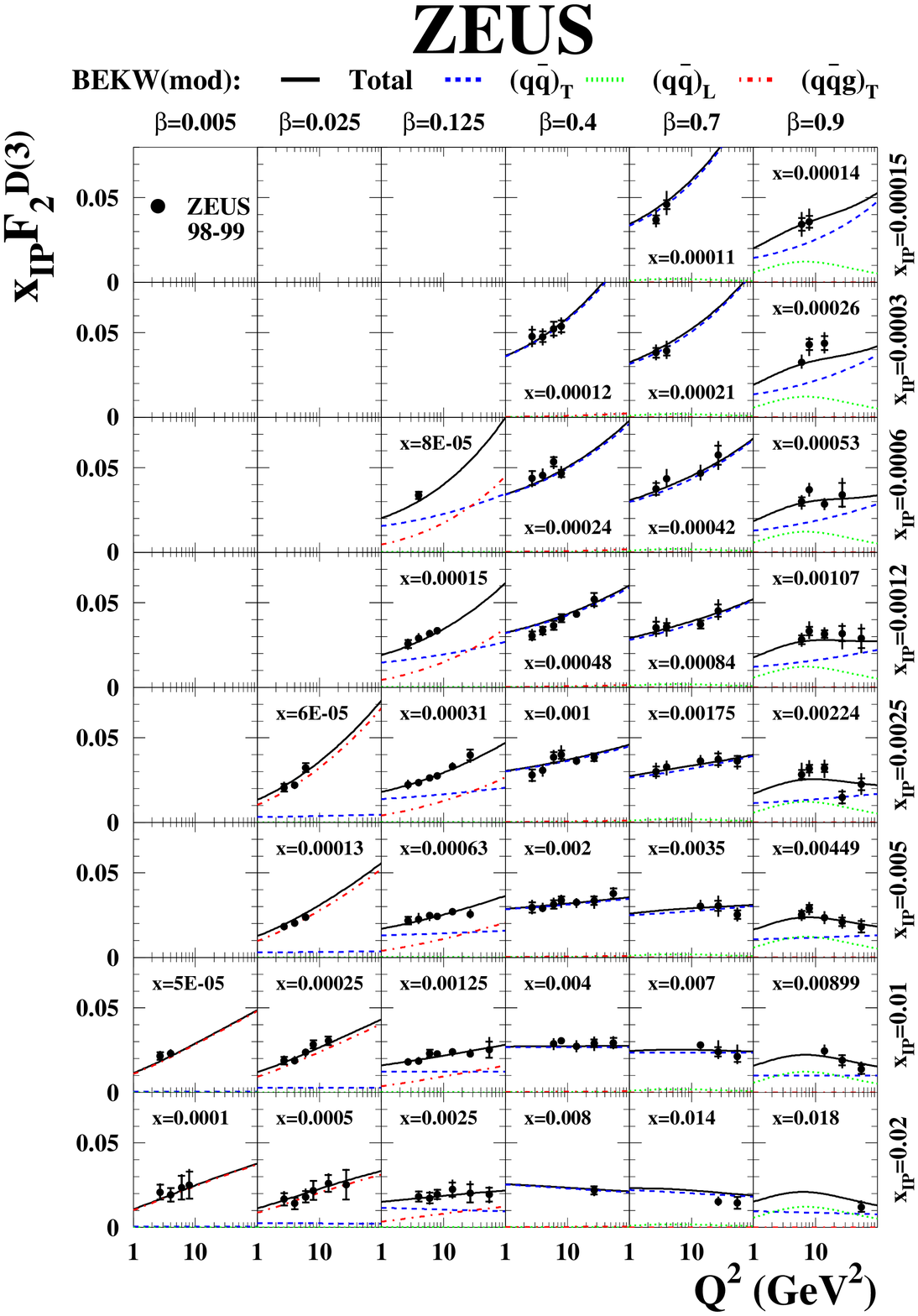,width=14.cm}}
\vspace*{-0.5cm}
\caption{The diffractive structure function of the proton multiplied by $\xpom$, $\xpom F^{D(3)}_2$, as a function of $Q^2$ for different regions of $\beta$ and $\xpom$.  The inner error bars show the statistical uncertainties and the full bars the statistical and systematic uncertainties added in quadrature. The curves show the result of the BEKW(mod) fit for the contributions from ($q\overline{q}$) for transverse (dashed) and longitudinal photons (dotted) and for the ($q\overline{q} g$) contribution for transverse photons (dashed-dotted) together with the sum of all contributions (solid).}
\label{f:f2d3vsq2}
\end{figure}
\clearpage
%

\begin{figure}[p]
\vfill
\centering
{\epsfig{file=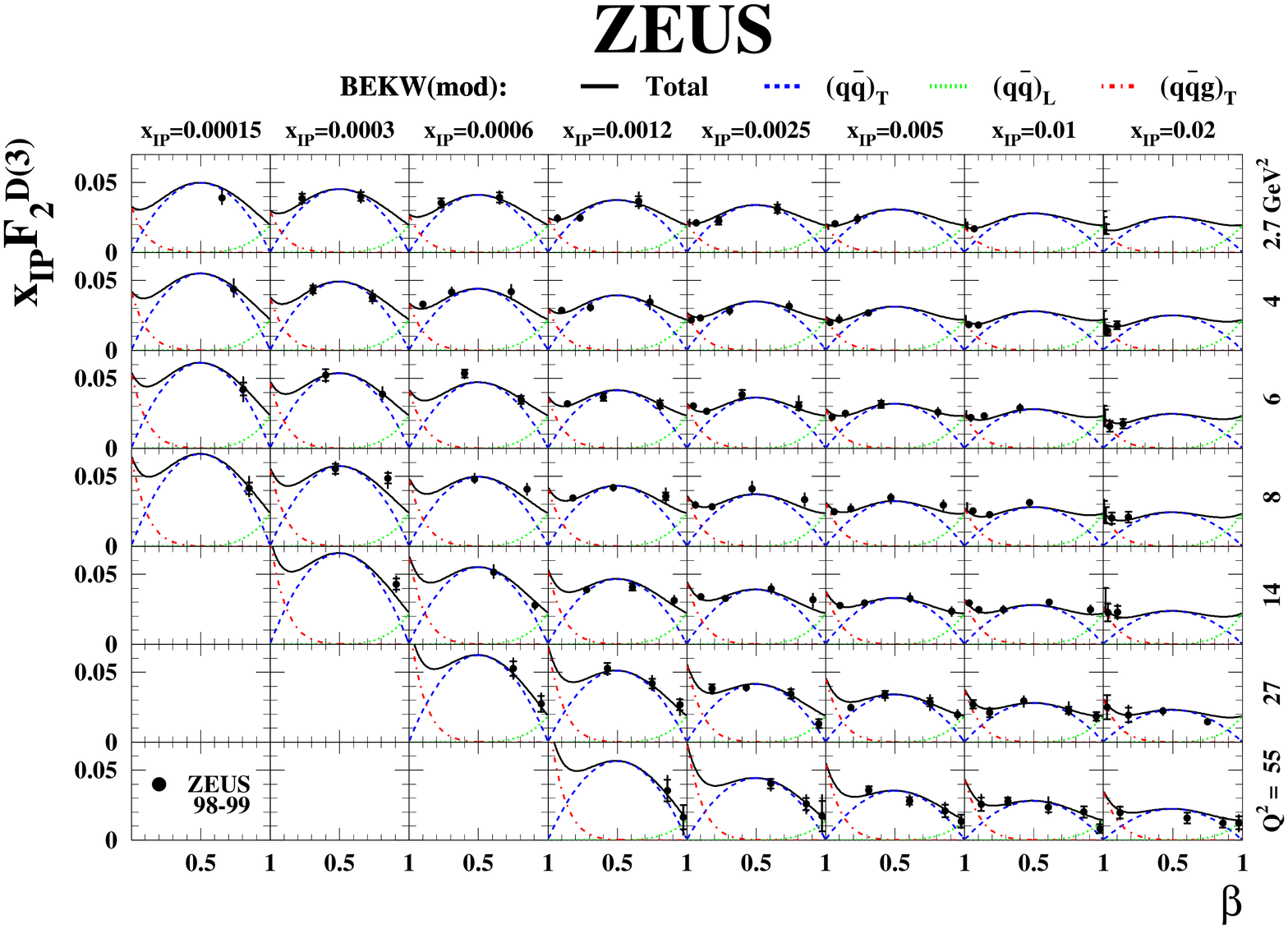,width=20.cm,angle=90}}
  \vspace*{-0.9cm}
\caption{The diffractive structure function of the proton multiplied by
   $\xpom$, $\xpom F^{D(3)}_2$, as a function of $\beta$ for different
   regions of $Q^2$ and $\xpom$. The inner error bars show the statistical 
   uncertainties and the full bars the statistical and systematic systematic 
   uncertainties added in quadrature.  The curves show the result of the 
   BEKW(mod) fit for the contributions from ($q\overline{q}$) for 
   transverse (dashed) and longitudinal photons (dotted) and for the 
   ($q\overline{q} g$) contribution for transverse photons (dashed-dotted)
   together with the sum of all contributions (solid).}
\label{f:f2d3vsbeta}
\vfill
\end{figure}
\clearpage

\begin{figure}[p]
{\epsfig{file=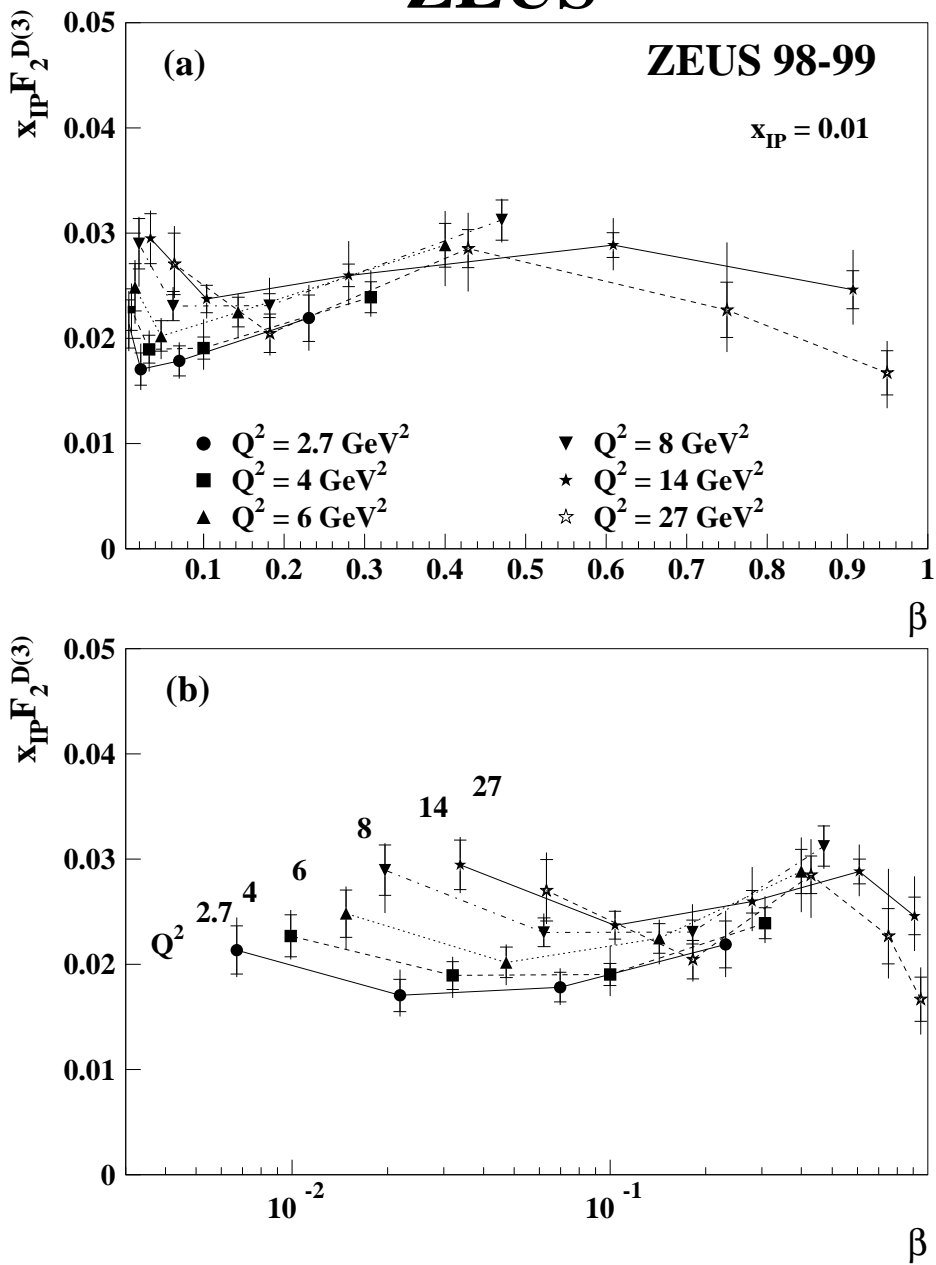,width=14cm}}
\caption{The diffractive structure function of the proton multiplied by $\xpom$, $\xpom F^{D(3)}_2$, for $x_{\pom}=0.01$ as a function of $\beta$ for the $Q^2$ intervals indicated, (a) on a linear scale, (b) on a logarithmic scale. The inner error bars show the statistical uncertainties and the full bars the statistical and systematic systematic uncertainties added in quadrature.  
The lines connect measurement at the same value of $Q^2$.}
\label{f:f2d201}
\end{figure}
\clearpage

\begin{figure}[p]
{\epsfig{file=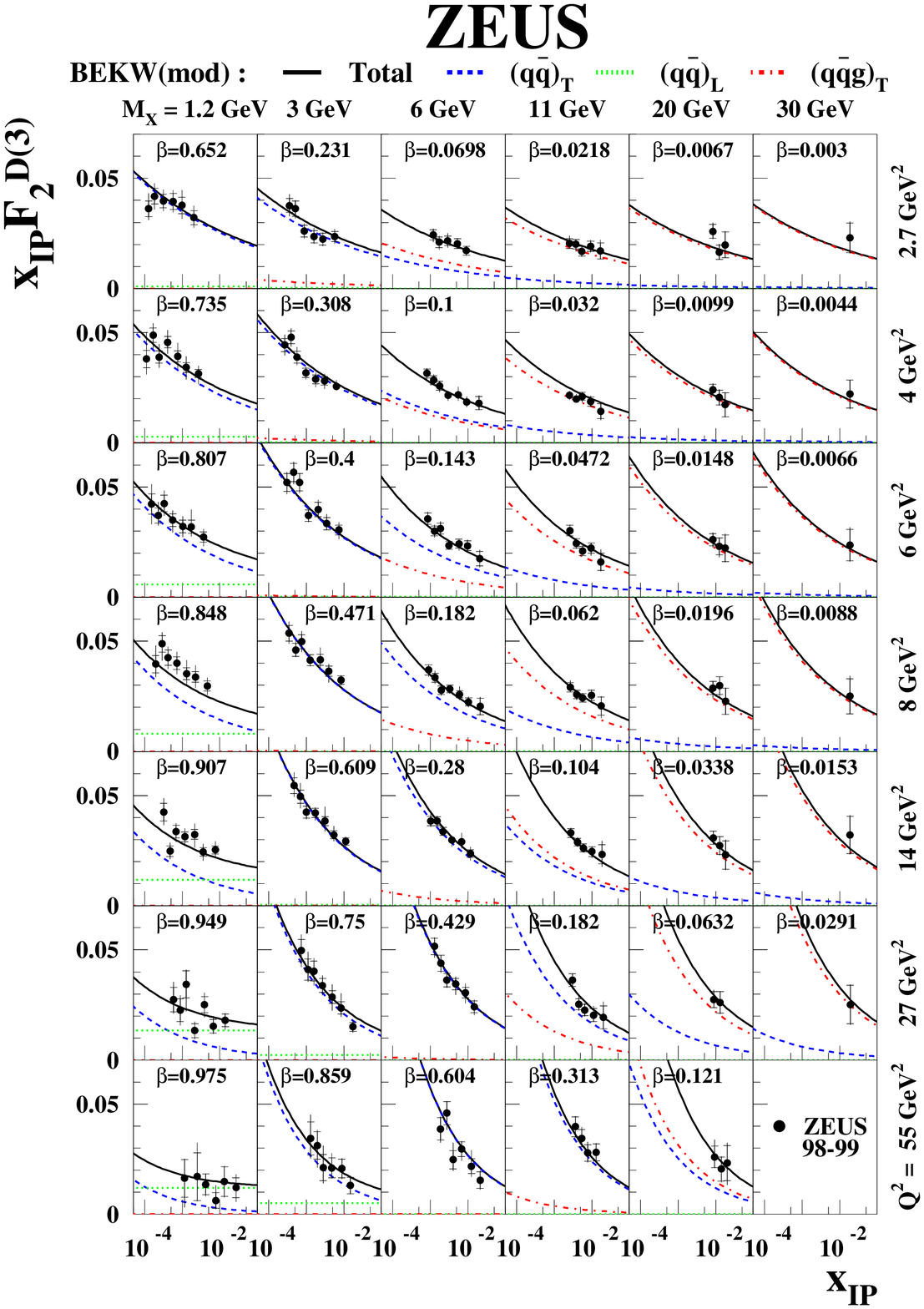,width=13.3cm}}
\caption{The diffractive structure function of the proton multiplied by $\xpom$, $\xpom F^{D(3)}_2$, as a function of $\xpom$ for different regions of $\beta$ and $Q^2$. The inner error bars show the statistical uncertainties and the full bars the statistical and systematic systematic uncertainties added in quadrature. The curves show the result of the BEKW fit for the contributions from ($q\overline{q}$) for transverse (dashed) and longitudinal photons (dotted) and for the ($q\overline{q} g$) contribution for transverse photons (dashed-dotted) together with the sum of all contributions (solid).}
\label{f:f2d3.bekw}
\end{figure}
\clearpage

%
%
\end{document}